\documentclass[11pt,a4paper]{article}
\pdfoutput=1

\usepackage{jheppub}
\usepackage{color}
\usepackage[dvipsnames]{xcolor}
\usepackage{graphicx}
\usepackage{wrapfig}
\usepackage{verbatim}
\usepackage{amsmath}
\usepackage{amssymb}
\usepackage{url}
\usepackage{bbold}
\usepackage{xspace}
\usepackage{slashed}
\usepackage{multirow}
\usepackage{threeparttable}
\usepackage{paralist}
\usepackage{subcaption}
\usepackage{floatrow}
\usepackage{multirow}
\usepackage{soul}
\usepackage{natbib}

\usepackage{tikz}
\usetikzlibrary{trees}
\usetikzlibrary{decorations.pathmorphing}
\usetikzlibrary{decorations.markings}
\usetikzlibrary{arrows}

\usepackage{tikz}
\usetikzlibrary{arrows,shapes}
\usetikzlibrary{trees}
\usetikzlibrary{matrix,arrows} 				
\usetikzlibrary{calc} 
\usetikzlibrary{positioning}				
\usetikzlibrary{calc,through}				
\usetikzlibrary{decorations.pathreplacing}  
\usepackage[tikz]{bclogo} 					
\usepackage{pgffor}							

\usetikzlibrary{decorations.pathmorphing}	
\usetikzlibrary{decorations.markings}
\usetikzlibrary{intersections}				
\tikzset{
    vector/.style={decorate, decoration={snake, pre length=.5mm, post length=.5mm}, draw},
    fermion/.style={postaction={decorate}, decoration={markings,mark=at position .55 with {\arrow{>}}}},
    fermionbar/.style={draw, postaction={decorate}, decoration={markings,mark=at position .55 with {\arrow{<}}}},
    fermionnoarrow/.style={},
    gluon/.style={decorate, decoration={coil,amplitude=4pt, segment length=5pt}},
    scalar/.style={dashed, postaction={decorate}, decoration={markings,mark=at position .55 with {\arrow{>}}}},
    scalarbar/.style={dashed, postaction={decorate}, decoration={markings,mark=at position .55 with {\arrow{<}}}},
    scalarnoarrow/.style={dashed,draw},
    graviton/.style={decorate, decoration={snake, amplitude=1.0mm, segment length=2.7mm, pre length=.5mm, post length=.5mm}, double}
}

\newcommand{\TeV}{\,\mathrm{TeV}}

\newcommand{\an}[1]{\langle #1 \rangle}
\newcommand{\sq}[1]{[ #1 ]}

\newcommand{\uv}{\text{\tiny UV}}
\newcommand{\ir}{\text{\tiny IR}}
\newcommand{\eft}{\text{\tiny EFT}}
\newcommand{\gr}{\text{\tiny GR}}
\newcommand{\nm}{\text{\tiny NM}}
\newcommand{\tree}{\text{\tiny tree}}
\renewcommand{\loop}{\text{\tiny 1-loop}}
\newcommand{\ct}{\text{\tiny CT}}

\newcommand{\mpl}{M_{\rm Pl}}
\newcommand{\coff}{M}
\newcommand{\Er}{\mathcal{E}}

\newcommand{\beq}{\begin{equation}}
\newcommand{\eeq}{\end{equation}}
\newcommand{\bea}{\begin{eqnarray}}
\newcommand{\eea}{\end{eqnarray}}

\newcommand{\amp}{\mathcal{A}}

\DeclareRobustCommand{\Sec}[1]{Sec.~\ref{#1}}

\DeclareRobustCommand{\App}[1]{App.~\ref{#1}}

\DeclareRobustCommand{\Fig}[1]{Fig.~\ref{#1}}

\DeclareRobustCommand{\Eq}[1]{Eq.~(\ref{#1})}
\DeclareRobustCommand{\Eqs}[2]{Eqs.~(\ref{#1}), (\ref{#2})}
\DeclareRobustCommand{\EqsRange}[2]{Eqs.~(\ref{#1})--(\ref{#2})}

\begin{document}

\title{Negative running of gravitational positivity}

\author[1]{Jaime Fernandez,}
\author[2]{Maximilian Ruhdorfer,}
\author[1]{and Javi Serra}

\affiliation[1]{Instituto de F\'isica Te\'orica UAM/CSIC, Madrid 28049, Spain}
\affiliation[2]{Leinweber Institute for Theoretical Physics at Stanford, Department of Physics, Stanford University, Stanford, CA 94305, USA \\}

\emailAdd{jaime.fernandez@csic.es}
\emailAdd{m.ruhdorfer@stanford.edu}
\emailAdd{javi.serra@ift.csic.es}

\date{\today}

\abstract{We investigate the one-loop renormalization group evolution in four dimensions of the leading operators in the effective field theories of shift-symmetric scalars, photons, and gravitons. We show that certain non-minimal three-point interactions induce a negative running of the corresponding Wilson coefficients, with beta-functions suppressed by the Planck scale. The decrease of the coefficients toward the infrared prompts us to revisit their dispersive bounds, in particular accounting for graviton loops. Gravitational interactions generate positive infrared contributions which, after smearing over the momentum transfer, are argued to dominate over the negative running, provided the number of non-minimally coupled particles is bounded from above according to the species bound.}

\preprint{IFT-UAM/CSIC-26-28}

\maketitle

\newpage

\section{Introduction} \label{sec:intro}

Understanding which new dynamics or principles complete quantum gravity at short distances remains one of the central open problems in theoretical particle physics.
While it bears some resemblance to the pre-LHC question of what completed the Standard Model at energies above the electroweak scale, its most compelling solution, string theory, is undoubtedly more complex and richer than the Higgs boson, a seemingly fundamental scalar.
Yet perhaps the most significant difference between these two problems is that gaining experimental access to a UV-complete gravitational theory has proven notoriously difficult, in contrast to probing electroweak physics with a collider operating at center-of-mass energies above $4 \pi v \approx 3 \TeV$.

This last fact is one of the main original motivations behind the string theory swampland program \cite{Vafa:2005ui}.
At the same time, it provides a strong rationale for a low-energy approach to modifications of general relativity (GR), where information about potential UV completions is encoded in the Wilson coefficients of an effective field theory (EFT).
These two perspectives, one top-down and the other bottom-up, have found common ground in the $S$-matrix theoretical constraints: bounds on EFTs derived from dispersion relations whose only UV input consists of the principles of unitarity, locality, and causality \cite{Adams:2006sv}.
A prototypical example is the EFT of a single shift-symmetric scalar, whose leading self-interaction takes the form $c_2 (\partial \phi)^4$. In the absence of gravity, dispersion relations require the coefficient $c_2$ to be positive, and EFTs with a negative $c_2$ are said to belong to the swampland.

Dispersion relations therefore provide a way to theoretically test gravitational theories, which is worth pursuing because, as Weinberg put it \cite{Weinberg:1965nx}, ``something might go wrong, and that would be interesting.''
Indeed, many things can go wrong in gravitational EFTs, and ensuring their compliance with dispersive bounds has already led to numerous interesting novel results, such as extremal black holes being the required charged states by the weak gravity conjecture \cite{Cheung:2014ega,Hamada:2018dde,Bellazzini:2019xts,Charles:2019qqt,Alberte:2020jsk,Arkani-Hamed:2021ajd,Henriksson:2022oeu} the species bound \cite{Caron-Huot:2024lbf}, 
the inconsistency of a massive graviton \cite{Cheung:2016yqr,Bellazzini:2017fep,deRham:2018qqo,Bellazzini:2023nqj}, and the limited range of validity of modifications of GR of phenomenological relevance for black hole physics \cite{Camanho:2014apa,Endlich:2017tqa,Caron-Huot:2022ugt,Serra:2022pzl}.
In general, these theoretical constraints help delineate the space of gravitational EFTs that any unitary and causal UV completion must fall into \cite{Bellazzini:2015cra,Cheung:2016wjt,Tokuda:2020mlf,Herrero-Valea:2020wxz,Arkani-Hamed:2020blm,Caron-Huot:2021rmr,Bern:2021ppb,Bellazzini:2021shn,Chiang:2022jep,Caron-Huot:2022jli,Bellazzini:2022wzv,deRham:2022gfe,Hong:2023zgm,McPeak:2023wmq,Caron-Huot:2024tsk,Beadle:2024hqg,Alviani:2024sxx,Dong:2024omo,Bellazzini:2025shd} 
(see also the review \cite{deRham:2022hpx} and references therein for other interesting developments). 

In this paper, we conduct another low-energy test of gravitational EFTs. Specifically, we systematically analyze the properties in the deep IR of the least irrelevant EFT operators involving shift-symmetric scalars, photons, or gravitons, of the form $(\partial \phi)^4$, $F^4$, and $C^4$, respectively ($F$ the field strength and $C$ the Weyl tensor). At asymptotically low energies, and in theories with a small separation between the EFT cutoff $\coff$ and $\mpl$, the leading contribution to the corresponding EFT coefficients is dominated by the logarithmic running induced by loops of massless particles within the EFT. These include dynamical gravitons, as well as any particle that can consistently be taken to be massless (i.e., particles with spin $\leqslant 2$), with both minimal gravitational couplings and generic non-minimal couplings.

In this regard, while the analysis in \cite{Baratella:2021guc} showed that minimally coupled SM-like matter leads to coefficients increasing toward the IR (i.e., positive running or negative $\beta$-function), \cite{Arkani-Hamed:2021ajd} found that a negative running for photons 
is possible in theories with a large number $N> 136$ 
of non-minimally coupled fermions via $\bar\chi F \chi$ operators, or a large number $N > 45$ 
of scalar-vector pairs with $\sigma F^2$ non-minimal couplings.
In addition to reproducing and extending these results, in this paper we show that a positive $\beta$-function for scalars 
is induced by $N > 16$ massless vectors non-minimally coupled via $\phi V \tilde{V}$ operators, and that whenever there are scalars with $\sigma C^2$ non-minimal couplings, the coefficient of $C^4$ decreases toward lower energies.

We will devote special attention to understanding why all such positive $\beta$-functions are necessarily suppressed by $\mpl$. 
When dynamical gravitons are decoupled, any contribution to the renormalization group evolution (RGE) will be shown to be manifestly positive and directly determined by a finite cross section. Consequently, in the scalar and photon cases, $\beta > 0$ can only arise when the non-minimal three-point interactions are comparable in magnitude to the minimal gravitational coupling.
In the graviton case, reaching the same conclusion requires taking into account that the coefficient of the non-minimal interaction of a scalar to two gravitons is bounded by causality arguments, as derived in \cite{Serra:2022pzl}.

The finding that non-minimal interactions can drive the running of $(\partial \phi)^4$, $F^4$, and $C^4$ operators to negative values raises the question of whether these theories remain consistent with positivity constraints. 
The observation that $\beta \to 0$ in the decoupling limit $\mpl \to \infty$ for all cases exhibiting negative running indicates that gravitational contributions to the dispersion relations are crucial.
Indeed, it is well known that tree-level graviton exchange contributes to the (spin-2 subtracted) dispersion relations that single out at tree level the coefficients of the leading EFT operators for scalars and photons.
This term actually dominates at low exchanged momentum, precluding the derivation of a positivity constraint. The approach proposed in \cite{Caron-Huot:2021rmr,Caron-Huot:2022ugt} deals with the Coulomb singularity at tree level by constructing non-forward dispersion relations in impact parameter space,
yielding the result that the coefficients of $(\partial \phi)^4$ and $F^4$ are allowed to be negative, by an amount $(\coff \mpl)^{-2}$.%
\footnote{Times an infrared $1/\epsilon$ divergence in $d = 4 - 2 \epsilon$ dimensional regularization. Such a negative lower bound is qualitatively equivalent to the universal gravitational contribution to the time delay \cite{Adams:2006sv,Camanho:2014apa,Bellazzini:2021shn,Arkani-Hamed:2021ajd,Caron-Huot:2022ugt,Serra:2022pzl}, where the $1/\epsilon$ is traded by a (large) reference impact parameter. 
As recently shown in \cite{Bellazzini:2025bay}, $1/\epsilon$ divergences in dispersive bounds should, and in fact can, be eliminated to yield physically interpretable bounds in $d = 4$.
Therefore, our bounds strictly apply only for $\epsilon \neq 0$, or alternatively be interpreted qualitatively by keeping in mind that $1/\epsilon$ can effectively be traded by a finite logarithm; see \App{sec:IRsafe} for further details.}
We will show that such a degree of negativity is parametrically consistent with the maximal amount of negative running we obtain, provided that $N \lesssim (4\pi\mpl/\coff)^2$, with the cutoff $\coff$ thus identified with the species scale.

In contrast, even with dynamical gravity, tree-level dispersion relations still require the coefficient of $C^4$ to be positive \cite{Bellazzini:2015cra}, since the graviton $1/t$ pole does not contribute to the corresponding (spin-4 subtracted) dispersion relation.
This fact motivates us to examine in detail graviton exchange diagrams at one loop, as these are expected to allow for negativity as shown in \cite{Chang:2025cxc,Beadle:2025cdx} for scalar scattering.
Remarkably, albeit not unexpectedly \cite{Caron-Huot:2022ugt,Bellazzini:2022wzv}, we find that the dispersive bound on the coefficient of $C^4$ is modified to $g_4 \gtrsim - \coff^{-4}/(4\pi)^2$, restoring parametric consistency with the negative running we find provided that the $\sigma C^2$ coupling is bounded as $|c_{\sigma C^2}| \lesssim (4\pi/\sqrt{N})\mpl/\coff^2$, or equivalently $N \lesssim (4\pi\mpl/c_{\sigma C^2}\coff^2)^2$.\\

This paper is organized as follows. In \Sec{sec:EFTs} we identify and present the relevant three- and four-point amplitudes. \Sec{sec:betas} is devoted to the calculation of the anomalous dimensions, together with a discussion of their general negativity in the gravitational decoupling limit. In \Sec{sec:dispersions} we construct the non-forward dispersion relations from which we derive positivity bounds. Finally, we present our conclusions in \Sec{sec:conclusions}.

Several appendices collect additional results. In \App{sec:sugra} we compute tree-level amplitudes and anomalous dimensions in supergravity. The basis of scalar integrals and their double cuts is presented in \App{sec:bubble}. A discussion of infrared divergences in minimally and non-minimally coupled theories is given in  \App{sec:irdivs}. In \App{sec:loop} we present our results for the one-loop four-point scalar, photon, and graviton amplitudes. \App{sec:details} provides additional details on our smearing procedure. Finally, in \App{sec:IRsafe} we rederive the positivity bounds in terms of detector amplitudes.

\section{EFT amplitudes, gravity, and non-minimal couplings}\label{sec:EFTs}

The objects of our study are the four-point elastic scattering amplitudes of a single real scalar $\phi$ endowed with a shift symmetry $\phi \to \phi + b$, a single photon $\gamma^{\pm}$, and a graviton $h^{\pm}$. 
For energies below a cutoff $\coff$, these admit an EFT expansion, 
which at tree level reads
\bea
\amp^\eft(1_\phi,2_\phi,3_\phi,4_\phi) &=& c_2 (s^2 +t^2 + u^2) + c_3 stu + c_4 (s^2 +t^2 + u^2)^2 + \dots \,, 
\label{eq:AEFTtreephi} \\
\amp^\eft(1_{\gamma^+},2_{\gamma^-},3_{\gamma^-},4_{\gamma^+}) &=& \an{23}^2 \sq{14}^2 \left( \alpha_2 + \alpha_3 u + \alpha_4 u^2 + \alpha_{4'} (st)^2 + \dots \right) \,,
\label{eq:AEFTtreeg} \\
\amp^\eft(1_{h^+},2_{h^-},3_{h^-},4_{h^+}) &=& \an{23}^4 \sq{14}^4 \frac{1}{\mpl^4} \left(\frac{g_3^2}{\mpl^2} \frac{st}{u} + g_4 + g_5 u 
+ \dots \right) \,,
\label{eq:AEFTtreeh}
\eea
where $s = (p_1 + p_2)^2$, $t = (p_1 + p_3)^2$, and $u = (p_1 + p_4)^2$ are the Mandelstam variables, and $c_i$, $\alpha_i$, and $g_i$ are Wilson coefficients, which scale with appropriate powers of the cutoff and a (generic) coupling $g_*$: $c_i, \, \alpha_i \sim (g_*/\coff^{i})^2$, and $g_{i+2} \sim 1/(g_*\coff^{i})^2$. 
Each term in these expansions can be mapped to an independent Lagrangian operator. Throughout this work we choose the convention that all particles are incoming.

All the amplitudes above receive a tree-level contribution from graviton exchange,
\bea
\amp^\gr(1_\phi,2_\phi,3_\phi,4_\phi) &=& \frac{1}{\mpl^2} \left(\frac{tu}{s} + \frac{su}{t} + \frac{st}{u} \right) \,,
\label{eq:AGRtreephi}\\
\amp^\gr(1_{\gamma^+},2_{\gamma^-},3_{\gamma^-},4_{\gamma^+}) &=& -\an{23}^2 \sq{14}^2 \frac{1}{\mpl^2} \left( \frac{1}{s} + \frac{1}{t} \right) \,,
\label{eq:AGRtreegam}\\
\amp^\gr(1_{h^+},2_{h^-},3_{h^-},4_{h^+}) &=& \an{23}^4 \sq{14}^4 \frac{1}{\mpl^2} \frac{1}{stu} \,. 
\label{eq:AGRtreeh}
\eea
These amplitudes can be constructed from three-point amplitudes in GR via factorization on the poles for complex kinematics.%
\footnote{We adopt the convention $|\!-p\rangle = i |p\rangle$ and $|\!-p] = i |p]$. This convention implies that, upon factorization of a four-point amplitude, $\lim_{s_{12} \to 0} s_{12} \amp(1,2,3,4) = (-i)^{F[\Phi]} i \amp_L(1,2,-\ell_{\bar{\Phi}}) i \amp_R(\ell_{\Phi},3,4)$. If the amplitude also has poles in other channels, additional minus signs must be included from the reordering of the particles. For instance, if both 2 and 3 are fermions, $\lim_{s_{13} \to 0} s_{13} \amp(1,2,3,4) = -\lim_{s_{13} \to 0} s_{13} \amp(1,3,2,4) = -(-i)^{F[\Phi]} i \amp_L(1,3,-\ell_{\bar{\Phi}}) i \amp_R(\ell_{\Phi},2,4)$. Likewise, there is a minus sign in $s_{14}$-channel factorization if 4 and either 2 or 3 are fermions. In addition, $\amp(1_{\Phi_1}, \dots, m_{\Phi_m}) = (\mp i)^{F[\Phi_1,\dots,\Phi_m]} \amp(-m_{\bar \Phi_m}, \dots, -1_{\bar \Phi_1})^*$ in our convention, where $F[\Phi_1, \dots, \Phi_m]$ counts the number of fermions in the list $\{ \Phi_1, \dots, \Phi_m \}$. \label{fn:conventions}}
Since all of them are elastic in the $t$ channel, they feature a physical $1/t$ pole. 

In \Sec{sec:betas}, we will systematically study the renormalization group evolution at one loop of the leading contact terms in each of the EFT amplitudes \EqsRange{eq:AEFTtreephi}{eq:AEFTtreeh}, 
associated with 
$c_2$, $\alpha_2$, and $g_4$, respectively.%
\footnote{The corresponding operators are
$\tfrac{1}{2}c_2(\partial_\mu \phi)^4$, $\tfrac{1}{16} \alpha_2 [ (F_{\mu\nu} F^{\mu\nu})^2 + (F_{\mu\nu} \tilde{F}^{\mu\nu})^2]$, and $\tfrac{1}{16} g_4 [ (C_{\mu\nu\rho\sigma} C^{\mu\nu\rho\sigma})^2 +(C_{\mu\nu\rho\sigma} \tilde{C}^{\mu\nu\rho\sigma})^2]$.
}
Out of the amplitudes in \EqsRange{eq:AEFTtreephi}{eq:AGRtreeh}, only $\amp^\gr$ will be relevant in the computation of the corresponding anomalous dimensions, as directly implied by naive dimensional analysis (NDA). 
Instead, several three- and four-point amplitudes involving other massless particles are less irrelevant (i.e., less suppressed by powers of $\coff$, $\mpl$) than $\amp^\eft$ and can therefore contribute to the running of $c_2$, $\alpha_2$, and $g_4$. Based on non-renormalization theorems in extended supergravity theories, some of these amplitudes were identified in \cite{Arkani-Hamed:2021ajd} and shown to give rise to a positive $\beta$-function for $\alpha_2$.
Here, we take a more comprehensive approach and identify all possible interactions that can contribute to the running based on NDA \cite{Ruhdorfer:2019qmk,Durieux:2019siw}.
We will assume that supersymmetry is broken and, consequently, that no massless gravitini are present. Results in extended supersymmetric theories are provided in \App{sec:sugra}.
We reiterate that the only minimal couplings considered are gravitational.

Logarithmic running at one loop can be efficiently computed via two-particle unitarity cuts, which yield phase-space integrals of products of two four-point tree-level amplitudes (see \Sec{sec:betas}).
In the following, we present all four-point amplitudes relevant for the RGEs, omitting their conjugates. Some are constructed from non-minimal three-point amplitudes via factorization, while others arise directly as contact terms.

\subsection{Scalar} \label{sec:scalar}

The only relevant non-minimal three-point amplitude of a shift-symmetric scalar with other massless fields is
\beq
\amp(1_{\phi},2_{V_i^+},3_{V_j^+}) = i c_{\phi V^2}^{ij} \sq{23}^2 \,,
\label{eq:phiV2}
\eeq
where $V_i^{\pm}$ is a massless U$_i$(1) vector with real $c_{\phi V^2}^{ij} = c_{\phi V^2}^{ji} \sim g_*/\coff$, while the only relevant contact four-point amplitude is
\beq
\amp^\ct(1_{\phi},2_{\phi},3_{\sigma_i},4_{\sigma_j}) = c_{\phi^2 \sigma^2}^{ij} s \,,
\label{eq:phi2S2}
\eeq
where $\sigma_i$ is a real scalar of flavor $i$ with $c_{\phi^2 \sigma^2}^{ij} = c_{\phi^2 \sigma^2}^{ji} \sim g_*^2/\coff^2$.

Besides \Eq{eq:phi2S2}, the four-point amplitude constructed via factorization from \Eq{eq:phiV2} is also relevant for the RGE of the leading four-scalar EFT amplitude, 
\beq
\amp^\nm(1_{\phi},2_{\phi},3_{V_i^+},4_{V_j^-}) = c_{\phi V^2}^{ik} c_{\phi V^2}^{kj} 
[3|1|4\rangle^2 \left(\frac{1}{t}+\frac{1}{u}\right) \,,
\label{eq:phi2V2}
\eeq
with a sum over $k$ implicit, as well as amplitudes mediated by minimally-coupled graviton exchange
\bea
\amp^\gr(1_{\phi},2_{\phi},3_{\chi_i^+},4_{\chi_i^-}) &=&  \frac{1}{2\mpl^2} 
[3|1|4\rangle \frac{t-u}{s} \,, 
\label{eq:phi2chi2}\\
\amp^\gr(1_{\phi},2_{\phi},3_{\Phi},4_{\bar{\Phi}}) &=& \frac{1}{\mpl^2} \left( 
\frac{[3|1|4\rangle}{[4|1|3\rangle}
\right)^{h_\Phi} \frac{tu}{s} \,,
\label{eq:phi2Phi2}
\eea
where $\chi_i^\pm$ is a (Weyl) fermion of flavor $i$ and $\Phi = \sigma_i, \, V_i^\pm$ a scalar and a vector, respectively. Note that \Eq{eq:phi2Phi2} holds as well for a graviton, $\Phi = h^\pm$.

Other three- and four-point amplitudes, such as $\amp(1_{\phi},2_{h^+},3_{h^+})$ and $\amp(1_{\phi},2_{\phi},3_{\chi_i^+},4_{\chi_j^+})$, are suppressed by too many powers of $\coff$, $\mpl$ to contribute to the RGE. Note that we ignore contributions from operators that do not respect the scalar shift-symmetry, like $\phi^2 F^2$.

\subsection{Photon} \label{sec:photon}

As in the scalar case, the set of non-minimal interactions that contribute to the RGE of the leading four-photon (MHV) EFT amplitude is rather limited. The relevant three-point amplitudes correspond to dipole moments,
\beq
\amp(1_{\gamma^+},2_{\chi_i^+},3_{\chi_j^+}) = c_{F\chi^2}^{ij} \sq{12} \sq{13} \,,
\label{eq:dipole}
\eeq
where $\chi_{i}^\pm$ is a fermion with complex and antisymmetric couplings $c_{F\chi^2}^{ij} = - c_{F\chi^2}^{ji} \sim g_*/\coff$,
and to dilaton/axion couplings,
\beq
\amp(1_{V_i^+},2_{V_j^+},3_{\sigma_k}) = c_{\sigma F^2}^{ij;k} \sq{12}^2 \,,
\label{eq:dilaton}
\eeq
where $V_i^{\pm}$ is a U$_i$(1) vector and $\sigma_i$ a scalar, with
$c_{\sigma F^2}^{ij;k} = c_{\sigma F^2}^{ji;k} \sim g_*/\coff$.
These are complex in general, whereas $c_{\sigma F^2}^{ij;k}$ are real for a dilaton (i.e., a CP-even scalar), and purely imaginary for an axion (CP-odd).
In the following we will be identifying $V_0 = \gamma$, with $c_{\sigma F^2}^{0j;k} \equiv c_{\sigma F^2}^{j;k}$ for $j \neq 0$ and $c_{\sigma F^2}^{00;k} \equiv c_{\sigma F^2}^{k}$. 
The non-minimal couplings give rise to the following tree-level four-point amplitudes with two opposite-helicity photons
\bea
\amp^\nm(1_{\gamma^+},2_{\gamma^-},3_{\chi_i^+},4_{\chi_j^-}) &=& -c_{F\chi^2}^{ik} c_{F\chi^2}^{kj\,*} \sq{13}^2 \an{23} \an{24}\frac{1}{t} \,, 
\label{eq:g2chi2} \\
\amp^\nm(1_{\gamma^+},2_{\gamma^-},3_{V_i^+},4_{V_j^-}) &=& - c_{\sigma F^2}^{i;k}  c_{\sigma F^2}^{j;k\,*} \sq{13}^2 \an{24}^2 \frac{1}{t} \,, 
\label{eq:g2V2} \\
\amp^\nm(1_{\gamma^+},2_{\gamma^-},3_{\sigma_i},4_{\sigma_j}) &=& 
[1|3|2\rangle^2 \left( \frac{c_{\sigma F^2}^{k;i} c_{\sigma F^2}^{k;j\,*}}{t} + \frac{c_{\sigma F^2}^{k;i\,*} c_{\sigma F^2}^{k;j}}{u} \right) \,,
\label{eq:g2sigma2}
\eea
as well as the same-helicity amplitudes
\bea
\amp^\nm(1_{\gamma^+},2_{\gamma^+},3_{V_i^-},4_{V_j^-}) &=& - c_{\sigma F^2}^{k} c_{\sigma F^2}^{ij;k\,*} \sq{12}^2 \an{34}^2 \frac{1}{s} \,, 
\label{eq:g2VpVm}\\
\amp^\nm(1_{\gamma^+},2_{\gamma^+},3_{V_i^+},4_{V_j^+}) &=& 
\left( c_{\sigma F^2}^{i;k}  c_{\sigma F^2}^{j;k} - c_{\sigma F^2}^{k} c_{\sigma F^2}^{ij;k} \right) 
\sq{12}^2 \sq{34}^2 
\frac{1}{s} \,. 
\label{eq:gp2Vp2}
\eea
All of these are suppressed by as many powers of the cutoff as the four-point GR amplitudes are suppressed by the Planck scale,
\bea
\amp^\gr(1_{\gamma^+},2_{\gamma^-},3_{\chi_i^+},4_{\chi_i^-}) &=& 
- \frac{1}{\mpl^2} [1|3|2\rangle \sq{13} \an{24} \frac{1}{s} \,, 
\label{eq:gam2chi2}\\
\amp^\gr(1_{\gamma^+},2_{\gamma^-},3_{V_i^+},4_{V_i^-}) &=& - \frac{1}{\mpl^2} \sq{13}^2 \an{24}^2 \frac{1}{s} \,, 
\label{eq:gam2V2}\\
\amp^\gr(1_{\gamma^+},2_{\gamma^-},3_{\sigma_i},4_{\sigma_i}) &=& \frac{1}{\mpl^2} 
[1|3|2\rangle^2 \frac{1}{s} \,, 
\label{eq:gam2sigma2} \\
\amp^\gr(1_{\gamma^+},2_{\gamma^-},3_{h^+},4_{h^-}) &=& -\frac{1}{\mpl^2} [3|2|4\rangle^2 \sq{13}^2 \an{24}^2 \frac{1}{stu} \,.
\label{eq:gam2h2}
\eea
Let us note that there is no GR counterpart to the same-helicity amplitudes in \Eqs{eq:g2VpVm}{eq:gp2Vp2} even for $i = j$. This is with the exception of $i, j = 0$ in \Eq{eq:g2VpVm}, which constitutes a tree-level contribution to the the four-photon amplitude of interest, explicitly
\beq
\amp^\nm(1_{\gamma^+},2_{\gamma^-},3_{\gamma^-},4_{\gamma^+}) = -|c_{\sigma F^2}^{k}|^2 \an{23}^2 \sq{14}^2 \frac{1}{u} \,.
\eeq
This amplitude also follows from \Eq{eq:g2V2} for $i, j = 0$.
There is a mixed GR-NM four-point amplitude to consider when the interaction in \Eq{eq:dilaton} is non-vanishing for $i, j = 0$,
\beq
\amp^\nm(1_{\gamma^+},2_{\gamma^+},3_{h^-},4_{\sigma_i}) = - \frac{c_{\sigma F^2}^{i}}{\mpl} [1|4|3\rangle^2 [2|4|3\rangle^2 \frac{1}{stu} \,.
\label{eq:gp2h}
\eeq

There exists only one contact four-point amplitude with the correct scaling dimension to contribute to the RGE,
\beq
\amp^\ct(1_{\gamma^+},2_{\gamma^+},3_{\sigma_i},4_{\sigma_j}) = c_{\sigma^2 F^2}^{ij} \sq{12}^2 \, ,
\label{eq:gp2sigma2}
\eeq
where $c_{\sigma^2F^2}^{ij} = c_{\sigma^2 F^2}^{ji} \sim g_*^2/\coff^2$.

Other three- and four-point amplitudes, such as $\amp(1_{\gamma^+},2_{V_i^+},3_{V_j^+})$, $\amp(1_{\gamma^+},2_{\gamma^+},3_{h^+})$ or a contact term in $\amp(1_{\gamma^+},2_{\gamma^+},3_{V_i^+},4_{V_j^+})$, feature too many powers of $\coff$, $\mpl$ to contribute.%
\footnote{We have identified several instances of non-gravitational four-point amplitudes, generated from the tree-level exchange of mediators with minimal couplings other than gravitational and non-minimal couplings to photons (gravitons), which lead to a negative running of the leading four-photon (four-graviton) coefficients. Since these lie outside of the gravitational context of this work, they will be discussed elsewhere \cite{Fernandez2026}.}

\subsection{Graviton} \label{sec:graviton}

The three-point amplitudes with appropriate scaling dimension to contribute to the RGE of the leading four-graviton (MHV) EFT amplitude correspond to scalar couplings to the Gauss-Bonnet or Chern-Simons terms (see, e.g. \cite{Alviani:2025xvf}),
\beq
\amp(1_{h^+},2_{h^+},3_{\sigma_i}) = \frac{c_{\sigma C^2}^i}{\mpl^2} \sq{12}^4 \,,
\label{eq:phiC2}
\eeq
with $c_{\sigma C^2}^i \sim 1/g_*\coff$,
\footnote{This admits a supersymmetric version, $\amp(1_{h^+},2_{\psi^+},3_{\chi_i^+}) = - (c_{\sigma C^2}^i/\mpl^2) \sq{12}^3 \sq{13}$ with $\psi^{\pm}$ a gravitino, which was overlooked in \cite{McGady:2013sga}; see \App{sec:sugra} for further details on supersymmetric amplitudes.}
and to non-minimal graviton-photon interactions,
\beq
\amp(1_{h^+},2_{V_i^+},3_{V_j^+}) = \frac{c_{CV^2}^{ij}}{\mpl} \sq{12}^2 \sq{13}^2 \,,
\label{eq:hV2}
\eeq
with complex $c_{CV^2}^{ij} = c_{CV^2}^{ji} \sim 1/\coff^2$. 
The relevant amplitudes generated by these couplings are
\bea
\amp^\nm(1_{h^+},2_{h^-},3_{\sigma_i},4_{\sigma_j}) &=& -\frac{1}{\mpl^4} [1|3|2\rangle^4 \left( \frac{c_{\sigma C^2}^i c_{\sigma C^2}^{j\,*}}{t} + \frac{c_{\sigma C^2}^{i\,*} c_{\sigma C^2}^j}{u} \right) \,, 
\label{eq:h2SiSj} \\
\amp^\nm(1_{h^+},2_{h^-},3_{h^+},4_{\sigma_i}) &=& -\frac{c_{\sigma C^2}^i}{\mpl^3} \sq{13}^6 \an{12}^2 \an{23}^2 \frac{1}{stu} \,, 
\label{eq:h3sigma}\\
\amp^\nm(1_{h^+},2_{h^-},3_{h^-},4_{h^+}) &=& -\frac{|c_{\sigma C^2}^k|^2}{\mpl^4} \an{23}^4 \sq{14}^4 \frac{1}{u} \,,
\label{eq:ANMtreeh}
\eea
as well as
\bea
\amp^\nm(1_{h^+},2_{h^-},3_{V_i^+},4_{V_j^-}) &=& \frac{c_{CV^2}^{ik} c_{CV^2}^{kj\,*}}{\mpl^2}  [1|3|2\rangle^2 \sq{13}^2 \an{24}^2 \frac{1}{t} \,, 
\label{eq:h2ViVj} \\
\amp^\nm(1_{h^+},2_{h^-},3_{V_i^+},4_{V_j^+}) &=& - \frac{c_{CV^2}^{ij}}{\mpl^2} \an{2|1|3|2}^2 \sq{13}^2 \sq{14}^2 \frac{1}{stu} \,. \label{eq:h2Vp2}
\eea
Note that \Eq{eq:ANMtreeh} constitutes a tree-level contribution to the four-graviton MHV amplitude. Several of the four-point amplitudes above are also generated from graviton exchange, in particular
\bea
\amp^\gr(1_{h^+},2_{h^-},3_{\sigma_i},4_{\sigma_j}) &=& \frac{\delta_{ij}}{\mpl^2} [1|3|2\rangle^4 \frac{1}{stu} \,, 
\label{eq:h2sigma2}\\
\amp^\gr(1_{h^+},2_{h^-},3_{V_i^+},4_{V_j^-}) &=& - \frac{\delta_{ij}}{\mpl^2} [1|3|2\rangle^2 \sq{13}^2 \an{24}^2 \frac{1}{stu} \,, \label{eq:h2V2}
\eea
as well as \Eq{eq:AGRtreeh}.

Another non-minimal three-point amplitude is also relevant for the RGE,
\beq
\amp(1_{\sigma_i}, 2_{V^+_j}, 3_{V^+_k}) = c_{\sigma V^2}^{jk;i} \sq{23}^2 \,, 
\label{eq:sigmaV2}
\eeq
where $c_{\sigma V^2}^{jk;i}$ is in fact the same coupling appearing in \Eq{eq:dilaton}. This interaction generates amplitudes with gravitons when considered in combination with \Eq{eq:phiC2},
\bea
\amp^\nm(1_{h^+},2_{h^+},3_{V^+_i}, 4_{V^+_j}) &=& - \frac{c_{\sigma C^2}^k c_{\sigma V^2}^{ij;k}}{\mpl^2} \sq{12}^4 \sq{34}^2 \frac{1}{s} \,, 
\label{eq:hp2Vp2} \\
\amp^\nm(1_{h^+},2_{h^+},3_{V^-_i}, 4_{V^-_j}) &=& - \frac{c_{\sigma C^2}^k c_{\sigma V^2}^{ij;k\,*}}{\mpl^2} \sq{12}^4 \an{34}^2 \frac{1}{s} \,. 
\label{eq:hp2Vm2}
\eea

As in the photon case, there exists only one contact four-point amplitude with the correct scaling dimension to contribute to the RGE,
\beq
\amp^\ct(1_{h^+},2_{h^+},3_{\sigma_i},4_{\sigma_j}) = \frac{c_{\sigma^2 C^2}^{ij}}{\mpl^2} \sq{12}^4 \,, 
\label{eq:hp2sigma2}
\eeq
with $c_{\sigma^2 C^2}^{ij} = c_{\sigma^2 C^2}^{ji} \sim 1/\coff^2$.

Other four-point amplitudes are suppressed by too many powers of $\coff$, $\mpl$ to contribute to the RGE, except for $\amp(1_{h^+},2_{h^-},3_{h^+},4_{h^+})$ generated by a single insertion of the non-minimal three-graviton coupling, $\amp(1_{h^+},2_{h^+},3_{h^+}) = g_3 \sq{12}^2 \sq{23}^2 \sq{13}^2/\mpl^3$, and $\amp(1_{h^+},2_{h^+},3_{V^+}, 4_{V^+}) = c_{V^2C^2}/\mpl^2$ with $c_{V^2C^2} \sim 1/\coff^4$. While these two amplitudes have a small enough scaling dimension to contribute to the RGE in combination with a GR amplitude, they do not because of helicity selection rules \cite{Baratella:2021guc}.

\section{Beta-functions}\label{sec:betas}

One-loop amplitudes can be decomposed, following the Passarino-Veltman reduction, into a basis of scalar integrals consisting of bubbles, triangles, and boxes \cite{tHooft:1978jhc,Passarino:1978jh,Bern:1993kr}. Anomalous dimensions, to be associated with $\beta$-functions of corresponding running EFT coefficients, can be obtained from the coefficients of bubble integrals, as these are the only ones that are UV divergent. In turn, bubble coefficients can be efficiently extracted using (generalized) unitarity methods, specifically via double cuts, reducing the computation of anomalous dimensions to simple phase-space integrals of two tree-level amplitudes \cite{Bern:1994cg,Bern:1994zx,Britto:2004nc,Forde:2007mi,Arkani-Hamed:2008owk,Huang:2012aq,Caron-Huot:2016cwu,EliasMiro:2020tdv,Baratella:2020lzz,Jiang:2020mhe}; see \App{sec:bubble} for more details. 
As an example, the $s$-channel bubble coefficient of the four-scalar amplitude is given by
\beq
b_s = \frac{2}{\pi} \sum_{\Phi,\Phi'} \sigma_{\Phi,\Phi'} \mathcal{R} \int d{\rm LIPS} \, \amp _L(1_\phi,2_\phi,-\ell_{\bar{\Phi}},-\ell'_{\bar{\Phi'}}) \amp_R(\ell'_{\Phi'},\ell_{\Phi},3_\phi,4_\phi) \,,
\label{eq:bs}
\eeq
where $\int d{\rm LIPS} = \int d^4\ell d^4\ell' \delta^+(\ell^2) \delta^+(\ell'^2) \delta^{(4)}(\ell+\ell'+p_3+p_4)= \pi/2$, the sum is over all the internal states $\Phi, \Phi'$ with momenta $\ell, \ell'$ respectively, and $\sigma_{\Phi, \Phi'} = (-i)^{F[\Phi, \Phi']}$ with $F[\Phi, \Phi']$ counting the number of fermions in the list $\{\Phi, \Phi'\}$.
When $\Phi$ and $\Phi'$ are indistinguishable, a factor $1/2$ must be included.
The operator $\mathcal{R}$ acts on the integral (at the integrand level) to project out contributions to the double cut from triangles and boxes.
As we will see in \Sec{sec:positivity}, if there are no IR divergences, the $d$LIPS integral can be performed just as well without acting with $\mathcal{R}$, with the contributions from triangles and boxes identified via the resulting logarithmic terms.
Note that we do not consider massless bubbles, as these are associated with collinear divergences, which are absent in the theories relevant for this work, gravity in particular \cite{Weinberg:1965nx,Dunbar:1995ed,Akhoury:2011kq,Beneke:2012xa}; see also \App{sec:irdivs} for a discussion on IR divergences. The advantages of computing anomalous dimensions from double cuts, as in \Eq{eq:bs}, include non-renormalization theorems through the application of helicity and angular momentum selection rules, as discussed in, e.g., \cite{Cheung:2015aba,Bern:2019wie,Jiang:2020rwz,Baratella:2020dvw,Baratella:2021guc}.

All the coefficients of the Passarino-Veltman decomposition are rational functions of the Mandelstam invariants, times a rational function of $\sq{ij}$ and $\an{ij}$ with the appropriate helicity weight if the scattered particles carry spin.
The anomalous dimensions are obtained from the sum of bubble coefficients across all channels, which is proportional to a sum of tree-level amplitudes due to the locality of the associated counterterms.
For instance, the UV part of the one-loop four-scalar EFT amplitude reads, in dimensional regularization,
\beq
\amp^\eft|_\uv = - \frac{1}{2\epsilon} \frac{\gamma_i \amp_i}{c_i} - \frac{1}{(4\pi)^2} \left( b_s \log\Big(\frac{-s}{\mu^2}\Big) + b_t \log\Big(\frac{-t}{\mu^2}\Big) + b_u \log\Big(\frac{-u}{\mu^2}\Big) \right) \,,
\eeq
where the anomalous dimensions, $\gamma_i = dc_i/d\log\mu$, follow from $\gamma_i \amp_i/c_i = - (b_s + b_t + b_u)/8\pi^2$, with $\amp_i$ the term associated with the coefficient $c_i$ in \Eq{eq:AEFTtreephi}. In the scalar case, the bubble coefficients in the different channels are related by crossing symmetry. Instead, for photons and gravitons, crossing only relates the $s$- and $t$-channel coefficients.

\subsection{Scalar} \label{sec:scalarbeta}

We are now ready to present our results for the most general one-loop running of the leading EFT coefficient of the four-scalar amplitude, $c_2$ in \Eq{eq:AEFTtreephi}. 
For clarity, we organize the different contributions to its $\beta$-function based on the interactions involved: gravitational ($1/\mpl$), non-minimal three-point and contact terms ($c_{\phi V^2}, c_{\phi^2\sigma^2}$), and their combination.
Since we are particularly interested in the possibility of $c_2$ decreasing toward lower energies, we also highlight the specific combination of these contributions that can lead to a positive $\beta$-function (or anomalous dimension), $a_2 = dc_2/d\log\mu$.

Pure GR contributions to the running have been computed in \cite{Baratella:2021guc}. These are associated with double cuts of one-loop amplitudes that leave on both sides of the cut either a tree-level amplitude mediated by a minimally-coupled graviton involving four scalars, \Eq{eq:AGRtreephi}, or two scalars and either two fermions, two distinct scalars, or two vectors of the same flavor, \Eq{eq:phi2chi2} and \Eq{eq:phi2Phi2} with $\Phi = \sigma_i, \, V_i^\pm$, or the two-scalar-two-graviton amplitude in \Eq{eq:phi2Phi2} ($\Phi = h^\pm$).%
\footnote{Supersymmetric contributions, in particular those involving a gravitino, are discussed in \App{sec:sugra}.}
The sum of these contributions is given by
\beq
a_2^\gr = 
- \frac{1}{4\pi^2} \frac{1}{\mpl^4} \left(\frac{203}{80} + \frac{N_\sigma}{80} + \frac{N_\chi}{60} + \frac{N_V}{15} \right) \,,
\label{eq:betac2GR}
\eeq
where the first term in brackets is the contribution from cut gravitons and (shift-symmetric) scalars, and $N_\sigma, N_\chi, N_V$ are the number of (distinct) scalars, fermions, and vectors involved in the double cut, respectively.

\begin{figure}[t]
\centering
\begin{tikzpicture}[line width=1.1 pt, scale=.50, baseline=(current bounding box.center)]
	
	\begin{scope}[shift={(-1,0)}]
	
	\begin{scope}[shift={(-.5,0)}]
	\draw[dashed] (-12,5) -- (-10,3) ;
	\draw[] (-8,5) -- (-10,3) ;
	\draw[dashed] (-12,1) -- (-10,3) ;
	\draw[] (-8,1) -- (-10,3) ;
	\draw[fill=gray!10] (-10,3) circle (1.2cm) ;
	\node at (-10,3) {$\amp^\gr$\,};
	\node at (-12.6,5) {$\phi$};
	\node at (-12.6,1) {$\phi$};
	\end{scope}
	
	\draw[dotted] (-7,5.5) -- (-7,.5) ;
	\node at (-6.2,5.) {$\bar{\Phi}$};
	\node at (-6.2,1.) {$\Phi$};
	\node at (-7.8,5.) {$\Phi$};
	\node at (-7.8,1.) {$\bar{\Phi}$};
	
	\begin{scope}[shift={(.5,0)}]
	\draw[] (-6,5) -- (-4,3) ;
	\draw[dashed] (-2,5) -- (-4,3) ;
	\draw[] (-6,1) -- (-4,3) ;
	\draw[dashed] (-2,1) -- (-4,3) ;
	\draw[fill=gray!10] (-4,3) circle (1.2cm) ;
	\node at (-1.4,5) {$\phi$};
	\node at (-1.4,1) {$\phi$};
	\node at (-4,3) {$\amp^\gr$\,};
	\end{scope}

	\end{scope}


	\begin{scope}[shift={(14,0)}]

	\begin{scope}[shift={(-.5,0)}]
	\draw[dashed] (-12,5) -- (-10,3) ;
	\draw[] (-8,5) -- (-10,3) ;
	\draw[dashed] (-12,1) -- (-10,3) ;
	\draw[] (-8,1) -- (-10,3) ;
	\draw[fill=gray!10] (-10,3) circle (1.2cm) ;
	\node at (-9.9,3) {$\amp^{\nm}$\,};
	\node at (-12.6,5) {$\phi$};
	\node at (-12.6,1) {$\phi$};
	\end{scope}
	
	\draw[dotted] (-7,5.5) -- (-7,.5) ;
	\node at (-6.2,5.) {$V_i^-$};
	\node at (-6.2,1.) {$V_j^+$};
	\node at (-7.7,5.) {$V_i^+$};
	\node at (-7.7,1.) {$V_j^-$};
	
	\begin{scope}[shift={(.5,0)}]
	\draw[] (-6,5) -- (-4,3) ;
	\draw[dashed] (-2,5) -- (-4,3) ;
	\draw[] (-6,1) -- (-4,3) ;
	\draw[dashed] (-2,1) -- (-4,3) ;
	\draw[fill=gray!10] (-4,3) circle (1.2cm) ;
	\node at (-1.4,5) {$\phi$};
	\node at (-1.4,1) {$\phi$};
	\node at (-3.9,3) {$\amp^{\nm}$\,};
	\end{scope}

	\end{scope}


	\begin{scope}[shift={(-1,-7)}]

	\begin{scope}[shift={(-.5,0)}]
	\draw[dashed] (-12,5) -- (-10,3) ;
	\draw[] (-8,5) -- (-10,3) ;
	\draw[dashed] (-12,1) -- (-10,3) ;
	\draw[] (-8,1) -- (-10,3) ;
	\draw[fill=gray!10] (-10,3) circle (1.2cm) ;
	\node at (-9.9,3) {$\amp^{\ct}$\,};
	\node at (-12.6,5) {$\phi$};
	\node at (-12.6,1) {$\phi$};
	\end{scope}
	
	\draw[dotted] (-7,5.5) -- (-7,.5) ;
	\node at (-6.3,5.) {$\sigma_i$};
	\node at (-6.3,1.) {$\sigma_j$};
	\node at (-7.8,5.) {$\sigma_i$};
	\node at (-7.8,1.) {$\sigma_j$};
	
	\begin{scope}[shift={(.5,0)}]
	\draw[] (-6,5) -- (-4,3) ;
	\draw[dashed] (-2,5) -- (-4,3) ;
	\draw[] (-6,1) -- (-4,3) ;
	\draw[dashed] (-2,1) -- (-4,3) ;
	\draw[fill=gray!10] (-4,3) circle (1.2cm) ;
	\node at (-1.4,5) {$\phi$};
	\node at (-1.4,1) {$\phi$};
	\node at (-3.9,3) {$\amp^{\ct}$\,};
	\end{scope}

	\end{scope}


	\begin{scope}[shift={(14,-7)}]

	\begin{scope}[shift={(-.5,0)}]
	\draw[dashed] (-12,5) -- (-10,3) ;
	\draw[] (-8,5) -- (-10,3) ;
	\draw[dashed] (-12,1) -- (-10,3) ;
	\draw[] (-8,1) -- (-10,3) ;
	\draw[fill=gray!10] (-10,3) circle (1.2cm) ;
	\node at (-10,3) {$\amp^\gr$\,};
	\node at (-12.6,5) {$\phi$};
	\node at (-12.6,1) {$\phi$};
	\end{scope}
	
	\draw[dotted] (-7,5.5) -- (-7,.5) ;
	\node at (-6.2,5.5) {$V_i^-$};
	\node at (-6.3,4.5) {$\sigma_i$};
	\node at (-6.2,1.5) {$V_i^+$};
	\node at (-6.3,0.5) {$\sigma_i$};
	\node at (-7.7,5.5) {$V_i^+$};
	\node at (-7.8,4.5) {$\sigma_i$};
	\node at (-7.7,1.5) {$V_i^-$};
	\node at (-7.8,0.5) {$\sigma_i$};
	
	\begin{scope}[shift={(.5,0)}]
	\draw[] (-6,5) -- (-4,3) ;
	\draw[dashed] (-2,5) -- (-4,3) ;
	\draw[] (-6,1) -- (-4,3) ;
	\draw[dashed] (-2,1) -- (-4,3) ;
	\draw[fill=gray!10] (-4,3) circle (1.2cm) ;
	\node at (-1.4,5) {$\phi$};
	\node at (-1.4,1) {$\phi$};
	\node at (-3.9,3) {$\amp^{\nm,\ct}$\,};
	\end{scope}

	\end{scope}
		
\end{tikzpicture}
\caption{\emph{
Cut diagrams contributing to the anomalous dimension of the leading four-scalar EFT coefficient. 
Eqs.~(\ref{eq:betac2GR}), (\ref{eq:betac2NM}), (\ref{eq:betaa2CT}), and (\ref{eq:betac2GRNM}) correspond, respectively, to the top-left, top-right, bottom-left, and bottom-right cuts. $\Phi = \phi, \sigma_i, \chi_i, V_i, h$ in the top-left cut.}}
\label{fig:cutsc2}
\end{figure}
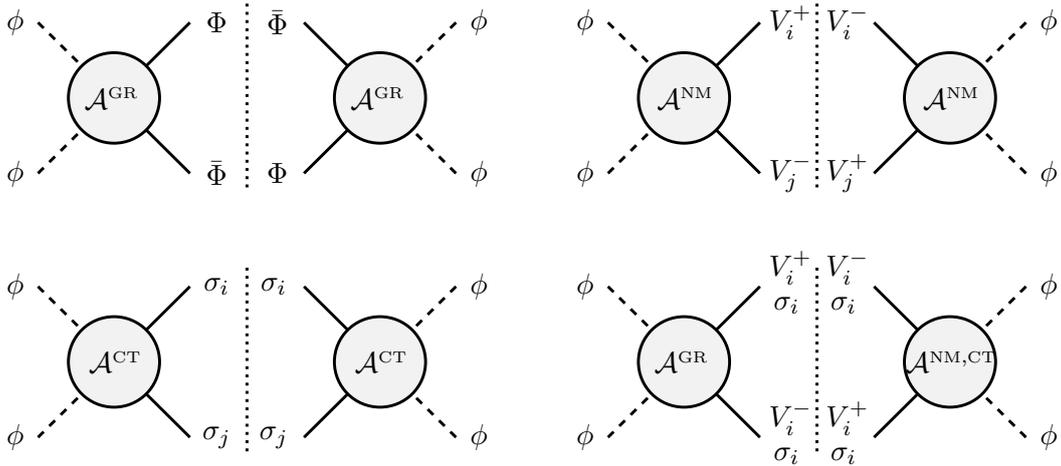

The non-minimal one-scalar-two-vector amplitude, \Eq{eq:phiV2}, via the four-point amplitude \Eq{eq:phi2V2}, contribute to the running as
\beq
a_2^\nm = - \frac{1}{8\pi^2} 
\sum_{i,j,k,l}^{N_V} c_{\phi V^2}^{ik} c_{\phi V^2}^{kj} c_{\phi V^2}^{il}c_{\phi V^2}^{lj} 
\,,
\label{eq:betac2NM}
\eeq
where we explicitly write all flavor sums. Note that the vectors involved in the cut necessarily contribute as well via the corresponding GR contribution in \Eq{eq:betac2GR}.

The contact amplitude between two shift-symmetric scalars and two distinct scalars of generic flavors, \Eq{eq:phi2S2}, leads to the following anomalous dimension 
\beq
a_2^\ct = - \frac{1}{16\pi^2} \sum_{i,j}^{N_\sigma} (c_{\phi^2\sigma^2}^{ij})^2 \,,
\label{eq:betaa2CT}
\eeq
arising from scalar cuts of the form $\amp^\ct_L \times \amp^\ct_R$. The scalars also contribute to the running as given in \Eq{eq:betac2GR}.

Since the helicities of $V_i$ and $\sigma_i$ involved in the four-point non-minimal amplitudes are the same as those in the GR amplitudes, the only difference being that the latter are flavor diagonal, mixed diagrams where on one side of the cut we have $\amp^\gr$ and on the other either $\amp^\nm$ or $\amp^\ct$ also contribute to the anomalous dimension,
\beq
a_2^{\gr \times \nm} + a_2^{\gr \times \ct}= + \frac{1}{6 \pi^2} \frac{1}{\mpl^2} \bigg( 
\sum_{i,k}^{N_V} (c_{\phi V^2}^{ik})^2 - \frac{1}{8} \sum_i^{N_\sigma} c_{\phi^2\sigma^2}^{ii} 
\bigg) \,, \label{eq:betac2GRNM}
\eeq

A summary of the relevant cuts (with all the channels being equivalent due to crossing symmetry) is depicted in \Fig{fig:cutsc2}.
While both the pure GR and pure non-minimal/contact contributions to the $\beta$-function are negative, leading to a larger $c_2$ toward lower energies, this is not the case for the mixed gravitational-non-minimal contribution in  \Eq{eq:betac2GRNM}.
This introduces the possibility that $c_2$ could run negative in the presence of couplings between the (shift-symmetric) scalar and massless vectors of the form $\phi V \tilde{V}$.
Indeed, let us focus on the full contribution to $a_2$ when the only non-gravitational interactions of $\phi$ are of this type, and, for simplicity, consider the case in which these are $V$-flavor diagonal and universal, i.e., $c_{\phi V^2}^{ij} = 0$ for $i \neq j$ and $c_{\phi V^2}^{ii} = c_{\phi V^2}$ $\forall i$. 
We find
\beq
a_2 = - \frac{1}{4\pi^2} \bigg( \frac{203}{80} \frac{1}{\mpl^4} + N_V \bigg( \frac{1}{15} \frac{1}{\mpl^4} - \frac{2}{3} \frac{c_{\phi V^2}^2}{\mpl^2} + \frac{1}{2} c_{\phi V^2}^4 \bigg) \bigg) \, .
\label{eq:betac2neg}
\eeq
The term scaling with the number of vectors is the smallest 
for $c_{\phi V^2}^2 \mpl^2 = 2/3$. 
If $N_V > 16$, such a contribution from the vectors surpasses that of the graviton and scalar, resulting in a positive $\beta$-function and, consequently, a negative running of the leading EFT coefficient of the four-scalar amplitude.

\subsection{Photon} \label{sec:photonbeta}

Similarly to the scalar case, we split the contributions to the $\beta$-function of the leading EFT coefficient of the four-photon MHV amplitude, $\gamma_2 = d\alpha_2/d\log \mu$, into purely GR ($1/\mpl$), non-minimal three-point couplings ($c_{F\chi^2}, c_{\sigma F^2}$), contact terms ($c_{\sigma^2 F^2}$), and mixed GR-non-minimal contributions.

Pure GR contributions to the running have been computed in \cite{Baratella:2021guc,Arkani-Hamed:2021ajd}. These arise from the same double cuts as in the scalar case, the relevant four-point amplitudes now involving photons instead of scalars, \Eq{eq:AGRtreegam} and \EqsRange{eq:gam2chi2}{eq:gam2h2}.
Their sum is given by
\beq
\gamma_2^\gr = 
- \frac{1}{4\pi^2} \frac{1}{\mpl^4} \left(\frac{137}{30} + \frac{N_\sigma}{60} + \frac{N_\chi}{20} + \frac{N_V}{5} \right) \,,
\label{eq:betaalpha2GR}
\eeq
where the first term in brackets is the contribution from cut gravitons and photons, and $N_\sigma, N_\chi, N_V$ are the number of scalars, fermions, and distinct vectors involved in the cut.

The non-minimal dipole and dilaton/axion three-point interactions, \Eqs{eq:dipole}{eq:dilaton}, via the four-point amplitudes \EqsRange{eq:g2chi2}{eq:gp2Vp2}, contribute to the running as
\bea
\gamma_2^\nm = - \frac{1}{12\pi^2} \bigg(
&& \!\!\!\!\!
\frac{1}{4} \sum_{i,j,k,l}^{N_\chi} c_{F\chi^2}^{ik} c_{F\chi^2}^{kj\,*} c_{F\chi^2}^{il\,*} c_{F\chi^2}^{lj}
\label{eq:betaalpha2NM} \\
&& \!\!\!\!\!
+ \sum_{i,j=0}^{N_V} \sum_{k,l}^{N_\sigma} \Big( c_{\sigma F^2}^{i;k} c_{\sigma F^2}^{j;k\,*} c_{\sigma F^2}^{i;l\,*} c_{\sigma F^2}^{j;l} + \frac{1}{4} c_{\sigma F^2}^{i;k} c_{\sigma F^2}^{j;k} c_{\sigma F^2}^{i;l\,*} c_{\sigma F^2}^{j;l\,*} + \frac{3}{4} c_{\sigma F^2}^{k} c_{\sigma F^2}^{ij;k\,*} c_{\sigma F^2}^{l\,*} c_{\sigma F^2}^{ij;l} 
\nonumber \\
&& \!\!\!\!\! \quad \qquad \qquad + \frac{3}{4} \big( c_{\sigma F^2}^{i;k}  c_{\sigma F^2}^{j;k} - c_{\sigma F^2}^{k} c_{\sigma F^2}^{ij;k} \big) 
\big( c_{\sigma F^2}^{i;l\,*}  c_{\sigma F^2}^{j;l\,*} - c_{\sigma F^2}^{l\,*} c_{\sigma F^2}^{ij;l\,*} \big)  \! \Big) \! \bigg) 
\,, \nonumber
\eea
where the first line is the contribution from cut fermions and, in the second line, the first term receives contributions from scalars and vectors in the cut, the second from cut scalars, and the rest from vector cuts. Note that the vector-flavor sums include the photon.

Similar to the scalar case, the helicities of $\sigma_i$, $\chi_i$, and $V_i$ in the four-point non-minimal amplitudes \EqsRange{eq:g2chi2}{eq:g2sigma2} are the same as those in the GR amplitudes, \EqsRange{eq:gam2chi2}{eq:gam2sigma2}. Therefore, mixed cuts of the form $\amp^\gr_L \times \amp^\nm_R$ (and $\amp^\nm_L \times \amp^\gr_R$ also) contribute to the anomalous dimension. In addition, a cut in which both sides involve the amplitude in \Eq{eq:gp2h} contributes at the same order in $\mpl$. The sum of all of these is given by
\beq
\gamma_2^{\gr \times \nm} = + \frac{1}{12 \pi^2} \frac{1}{\mpl^2} \bigg( \frac{1}{2} \sum_{i,j}^{N_\chi} |c_{F\chi^2}^{ij}|^2 + 2 \sum_{i}^{N_V} \sum_{k}^{N_\sigma} |c_{\sigma F^2}^{i;k}|^2
+ 6 \sum_k^{N_\sigma} |c_{\sigma F^2}^{k}|^2
\bigg) \,, 
\label{eq:betaalpha2GRNM}
\eeq
where note that we do not include here the photon in the vector-flavor sum.

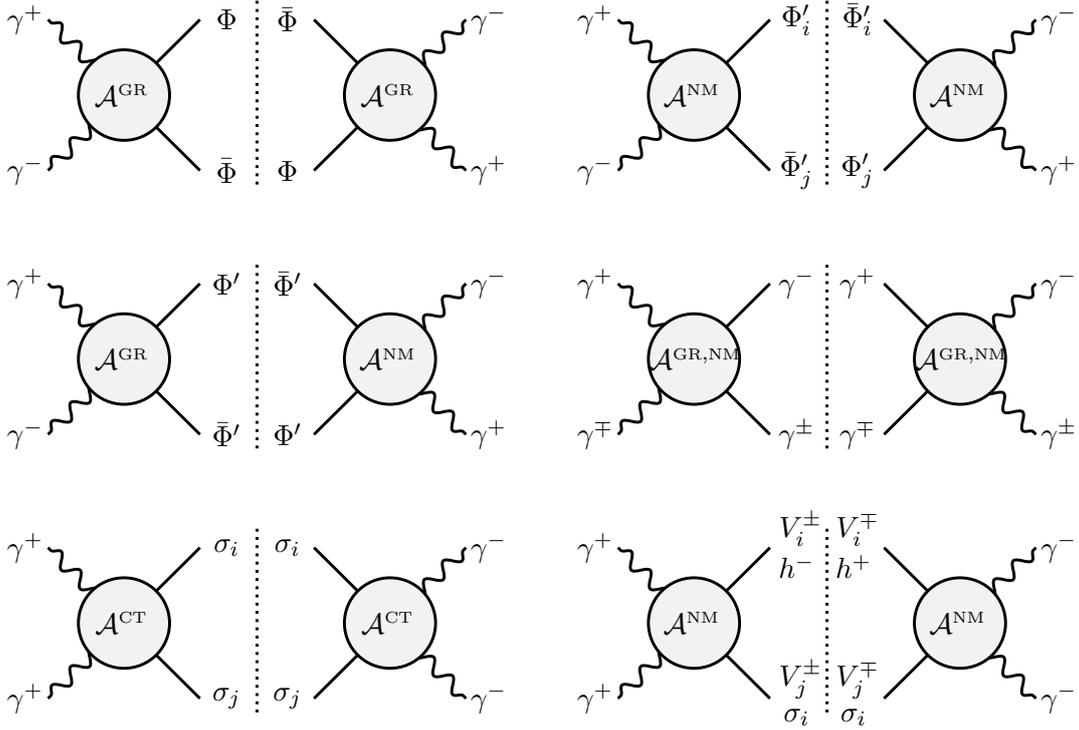
\begin{figure}[t]
\centering
\begin{tikzpicture}[line width=1.1 pt, scale=.50, baseline=(current bounding box.center)]
	
	\begin{scope}[shift={(-1,0)}]
	
	\begin{scope}[shift={(-.5,0)}]
	\draw[vector] (-12,5) -- (-10,3) ;
	\draw[] (-8,5) -- (-10,3) ;
	\draw[vector] (-12,1) -- (-10,3) ;
	\draw[] (-8,1) -- (-10,3) ;
	\draw[fill=gray!10] (-10,3) circle (1.2cm) ;
	\node at (-10,3) {$\amp^\gr$\,};
	\node at (-12.6,5) {$\gamma^+$};
	\node at (-12.6,1) {$\gamma^-$};
	\end{scope}
	
	\draw[dotted] (-7,5.5) -- (-7,.5) ;
	\node at (-6.2,5.) {$\bar{\Phi}$};
	\node at (-6.2,1.) {$\Phi$};
	\node at (-7.8,5.) {$\Phi$};
	\node at (-7.8,1.) {$\bar{\Phi}$};
	
	\begin{scope}[shift={(.5,0)}]
	\draw[] (-6,5) -- (-4,3) ;
	\draw[vector] (-2,5) -- (-4,3) ;
	\draw[] (-6,1) -- (-4,3) ;
	\draw[vector] (-2,1) -- (-4,3) ;
	\draw[fill=gray!10] (-4,3) circle (1.2cm) ;
	\node at (-1.4,5) {$\gamma^-$};
	\node at (-1.4,1) {$\gamma^+$};
	\node at (-4,3) {$\amp^\gr$\,};
	\end{scope}

	\end{scope}


	\begin{scope}[shift={(14,0)}]

	\begin{scope}[shift={(-.5,0)}]
	\draw[vector] (-12,5) -- (-10,3) ;
	\draw[] (-8,5) -- (-10,3) ;
	\draw[vector] (-12,1) -- (-10,3) ;
	\draw[] (-8,1) -- (-10,3) ;
	\draw[fill=gray!10] (-10,3) circle (1.2cm) ;
	\node at (-10,3) {$\amp^\nm$\,};
	\node at (-12.6,5) {$\gamma^+$};
	\node at (-12.6,1) {$\gamma^-$};
	\end{scope}

	\draw[dotted] (-7,5.5) -- (-7,.5) ;
	\node at (-6.2,5.) {$\bar{\Phi}'_i$};
	\node at (-6.2,1.) {$\Phi'_j$};
	\node at (-7.8,5.) {$\Phi'_i$};
	\node at (-7.8,1.) {$\bar{\Phi}'_j$};
		
	\begin{scope}[shift={(.5,0)}]
	\draw[] (-6,5) -- (-4,3) ;
	\draw[vector] (-2,5) -- (-4,3) ;
	\draw[] (-6,1) -- (-4,3) ;
	\draw[vector] (-2,1) -- (-4,3) ;
	\draw[fill=gray!10] (-4,3) circle (1.2cm) ;
	\node at (-1.4,5) {$\gamma^-$};
	\node at (-1.4,1) {$\gamma^+$};
	\node at (-4,3) {$\amp^\nm$\,};
	\end{scope}

	\end{scope}


	\begin{scope}[shift={(-1,-7)}]
	
	\begin{scope}[shift={(-.5,0)}]
	\draw[vector] (-12,5) -- (-10,3) ;
	\draw[] (-8,5) -- (-10,3) ;
	\draw[vector] (-12,1) -- (-10,3) ;
	\draw[] (-8,1) -- (-10,3) ;
	\draw[fill=gray!10] (-10,3) circle (1.2cm) ;
	\node at (-10,3) {$\amp^\gr$\,};
	\node at (-12.6,5) {$\gamma^+$};
	\node at (-12.6,1) {$\gamma^-$};
	\end{scope}
	
	\draw[dotted] (-7,5.5) -- (-7,.5) ;
	\node at (-6.2,5.) {$\bar{\Phi}'$};
	\node at (-6.2,1.) {$\Phi'$};
	\node at (-7.8,5.) {$\Phi'$};
	\node at (-7.8,1.) {$\bar{\Phi}'$};
	
	\begin{scope}[shift={(.5,0)}]
	\draw[] (-6,5) -- (-4,3) ;
	\draw[vector] (-2,5) -- (-4,3) ;
	\draw[] (-6,1) -- (-4,3) ;
	\draw[vector] (-2,1) -- (-4,3) ;
	\draw[fill=gray!10] (-4,3) circle (1.2cm) ;
	\node at (-1.4,5) {$\gamma^-$};
	\node at (-1.4,1) {$\gamma^+$};
	\node at (-4,3) {$\amp^\nm$\,};
	\end{scope}

	\end{scope}


	\begin{scope}[shift={(14,-7)}]

	\begin{scope}[shift={(-.5,0)}]
	\draw[vector] (-12,5) -- (-10,3) ;
	\draw[] (-8,5) -- (-10,3) ;
	\draw[vector] (-12,1) -- (-10,3) ;
	\draw[] (-8,1) -- (-10,3) ;
	\draw[fill=gray!10] (-10,3) circle (1.2cm) ;
	\node at (-9.9,3) {$\amp^{\gr,\nm}$\,};
	\node at (-12.6,5) {$\gamma^+$};
	\node at (-12.6,1) {$\gamma^\mp$};
	\end{scope}
	
	\draw[dotted] (-7,5.5) -- (-7,.5) ;
	\node at (-6.2,5.) {$\gamma^+$};
	\node at (-6.2,1.) {$\gamma^\mp$};
	\node at (-7.8,5.) {$\gamma^-$};
	\node at (-7.8,1.) {$\gamma^\pm$};
	
	\begin{scope}[shift={(.5,0)}]
	\draw[] (-6,5) -- (-4,3) ;
	\draw[vector] (-2,5) -- (-4,3) ;
	\draw[] (-6,1) -- (-4,3) ;
	\draw[vector] (-2,1) -- (-4,3) ;
	\draw[fill=gray!10] (-4,3) circle (1.2cm) ;
	\node at (-1.4,5) {$\gamma^-$};
	\node at (-1.4,1) {$\gamma^\pm$};
	\node at (-3.9,3) {$\amp^{\gr,\nm}$\,};
	\end{scope}

	\end{scope}


	\begin{scope}[shift={(-1,-14)}]

	\begin{scope}[shift={(-.5,0)}]
	\draw[vector] (-12,5) -- (-10,3) ;
	\draw[] (-8,5) -- (-10,3) ;
	\draw[vector] (-12,1) -- (-10,3) ;
	\draw[] (-8,1) -- (-10,3) ;
	\draw[fill=gray!10] (-10,3) circle (1.2cm) ;
	\node at (-10,3) {$\amp^\ct$\,};
	\node at (-12.6,5) {$\gamma^+$};
	\node at (-12.6,1) {$\gamma^+$};
	\end{scope}
	
	\draw[dotted] (-7,5.5) -- (-7,.5) ;
	\node at (-6.2,5.) {$\sigma_i$};
	\node at (-6.2,1.) {$\sigma_j$};
	\node at (-7.8,5.) {$\sigma_i$};
	\node at (-7.8,1.) {$\sigma_j$};
	
	\begin{scope}[shift={(.5,0)}]
	\draw[] (-6,5) -- (-4,3) ;
	\draw[vector] (-2,5) -- (-4,3) ;
	\draw[] (-6,1) -- (-4,3) ;
	\draw[vector] (-2,1) -- (-4,3) ;
	\draw[fill=gray!10] (-4,3) circle (1.2cm) ;
	\node at (-1.4,5) {$\gamma^-$};
	\node at (-1.4,1) {$\gamma^-$};
	\node at (-4,3) {$\amp^\ct$\,};
	\end{scope}

	\end{scope}


	\begin{scope}[shift={(14,-14)}]

	\begin{scope}[shift={(-.5,0)}]
	\draw[vector] (-12,5) -- (-10,3) ;
	\draw[] (-8,5) -- (-10,3) ;
	\draw[vector] (-12,1) -- (-10,3) ;
	\draw[] (-8,1) -- (-10,3) ;
	\draw[fill=gray!10] (-10,3) circle (1.2cm) ;
	\node at (-10,3) {$\amp^\nm$\,};
	\node at (-12.6,5) {$\gamma^+$};
	\node at (-12.6,1) {$\gamma^+$};
	\end{scope}
	
	\draw[dotted] (-7,5.5) -- (-7,.5) ;
	\node at (-6.2,5.5) {$V^\mp_i$};
	\node at (-6.3,4.5) {$h^+$};
	\node at (-6.2,1.5) {$V^\mp_j$};
	\node at (-6.3,0.5) {$\sigma_i$};
	\node at (-7.7,5.5) {$V^\pm_i$};
	\node at (-7.8,4.5) {$h^-$};
	\node at (-7.7,1.5) {$V^\pm_j$};
	\node at (-7.8,0.5) {$\sigma_i$};
	
	\begin{scope}[shift={(.5,0)}]
	\draw[] (-6,5) -- (-4,3) ;
	\draw[vector] (-2,5) -- (-4,3) ;
	\draw[] (-6,1) -- (-4,3) ;
	\draw[vector] (-2,1) -- (-4,3) ;
	\draw[fill=gray!10] (-4,3) circle (1.2cm) ;
	\node at (-1.4,5) {$\gamma^-$};
	\node at (-1.4,1) {$\gamma^-$};
	\node at (-4,3) {$\amp^\nm$\,};
	\end{scope}

	\end{scope}
	
\end{tikzpicture}
\caption{\emph{
Cut diagrams contributing to the anomalous dimension of the leading (MHV) four-photon EFT coefficient.
$\Phi = \sigma_i, \chi_i, V_i, h$ in the top-left cut and $\Phi' = \sigma, \chi, V$ in the top-right and middle-left cuts. 
The top-right cut also includes contributions with $\Phi_i, \bar{\Phi}'_j = \gamma^+, V_j^-$ and $\Phi_i, \bar{\Phi}'_j = V_i^+, \gamma^-$, and the bottom-right cut contributions with $V_i^\pm = \gamma^-$.
}}
\label{fig:cutsalpha2}
\end{figure}

Finally, the contact term in \Eq{eq:gp2sigma2} gives rise to the following anomalous dimension 
\beq
\gamma_2^\ct = - \frac{1}{16\pi^2} \sum_{k,l}^{N_\sigma} |c_{\sigma^2 F^2}^{kl}|^2 \, .
\label{eq:betaalpha2CT}
\eeq
While the contact amplitude can interfere with the GR amplitude \Eq{eq:gam2sigma2}, this does not give rise to the MHV four-photon amplitude.

With all the relevant cuts computed, see \Fig{fig:cutsalpha2} for a summary (the $s$ and $t$ channels are equivalent due to crossing symmetry), let us discuss the possibility of a negative running of $\alpha_2$. The only positive contributions to the $\beta$-function are the mixed gravitational-non-minimal contributions, \Eq{eq:betaalpha2GRNM}. Therefore, we focus on the full contribution to $\gamma_2$ when the only non-gravitational interactions of the photon are of either $\bar{\chi} F \chi$ or $\sigma F^2$ type.
When dipoles are non-vanishing, we can simply consider the scenario in which these are flavor diagonal in pairs and of universal strength $c_{F\chi^2}$,%
\footnote{In our notation, this can be implemented as $c_{F\chi^2}^{i=2k-1,j=2l} = c_{F\chi^2} \delta_{k,l}$, where $k, l = 1, \dots, N_\chi/2$ and recall $c_{F\chi^2}^{ij} = - c_{F\chi^2}^{ij}$.}
where one finds
\beq
\gamma_2 = - \frac{1}{4\pi^2} \bigg( \frac{137}{30} \frac{1}{\mpl^4} + \frac{N_\chi}{2} \bigg( \frac{1}{10} \frac{1}{\mpl^4} - \frac{1}{3} \frac{|c_{F\chi^2}|^2}{\mpl^2} + \frac{1}{6} |c_{F\chi^2}|^4 \bigg) \bigg) \, .
\label{eq:betaalpha2negdip}
\eeq
The term scaling with the number of fermion pairs $(N_\chi/2)$ is the smallest 
for $|c_{F\chi^2}|^2 \mpl^2 = 1$. If $N_\chi/2 > 68$, the contribution from the fermions surpasses that of the graviton and photon, resulting in a positive $\beta$-function for the leading EFT coefficient of the four-photon (MHV) amplitude, i.e., $\alpha_2$ running negative toward low energies.

When dilaton/axion couplings are non-vanishing, several scenarios are interesting to consider. Let us start with a theory with no additional vectors, thus where the scalars are non-minimally coupled to photons only, with a coupling that we take to be universal, i.e., $c_{\sigma F^2}^k = c_{\sigma F^2}$ $\forall k$. The corresponding $\beta$-function is given by
\beq
\gamma_2 = - \frac{1}{4\pi^2} \bigg( \frac{137}{30} \frac{1}{\mpl^4} + N_\sigma \bigg( \frac{1}{60} \frac{1}{\mpl^4} - 2 \frac{|c_{\sigma F^2}|^2}{\mpl^2} + N_\sigma \frac{2}{3} |c_{\sigma F^2}|^4 \bigg) \bigg) \, .
\label{eq:betaalpha2posda}
\eeq
The smallest value of the term scaling with $N_\sigma$ is obtained for $N_\sigma =1$, but it is not negative enough 
to yield a net positive contribution. 
One can next consider a set of scalars, each non-minimally coupled to two photons and to the photon and one vector, i.e., $c_{\sigma F^2}^k, c_{\sigma F^2}^{k;k} \neq 0$. In this case, although the positive contribution to the anomalous dimension from \Eq{eq:betaalpha2GRNM} is enhanced, the negative contribution from \Eq{eq:betaalpha2NM} is as well, ultimately preventing a net positive $\beta$-function \cite{Arkani-Hamed:2021ajd}. The scenario that minimizes the contribution from purely non-minimal couplings is the one in which both a dilaton and an axion couple to the photon and one vector with the same strength $c_{\phi F^2}$, effectively forming a complex scalar. With $N_V$ vectors and dilaton/axion pairs and universal couplings,%
\footnote{In our notation, 
$c_{\sigma F^2}^{i;k} = c_{\phi F^2} \delta_{i,(k+1)/2}$ for odd $k$ (dilaton) and $c_{\sigma F^2}^{i;k} = i c_{\phi F^2} \delta_{i,k/2}$ for even $k$ (axion), with $c_{\phi F^2}$ real.}
one finds
\beq
\gamma_2 = - \frac{1}{4\pi^2} \bigg( \frac{137}{30} \frac{1}{\mpl^4} + N_V \bigg( \frac{7}{30} \frac{1}{\mpl^4} - \frac{4}{3} \frac{|c_{\phi F^2}|^2}{\mpl^2} + \frac{4}{3} |c_{\phi F^2}|^4 \bigg) \bigg) \, .
\label{eq:betaalpha2negda}
\eeq
Similar to the dipole case, the term scaling with the number of vectors is the smallest 
for $|c_{\phi F^2}|^2 \mpl^2 = 1/2$, and the $\beta$-function is positive if $N_V > 45$.
\Eqs{eq:betaalpha2negdip}{eq:betaalpha2negda} match the results derived in \cite{Arkani-Hamed:2021ajd}.

\subsection{Graviton} \label{sec:gravitonbeta}

Regarding the running of the leading EFT coefficient of the four-graviton MHV amplitude, $\beta_4 = dg_4/d\log \mu$, we first recall that there are no pure GR contributions at one loop. Therefore, a non-vanishing anomalous dimension requires non-minimal three-point or contact interactions. We split the different contributions into those involving only non-minimal three-point couplings ($c_{C V^2}, c_{\sigma V^2}$), those where both minimal ($1/\mpl$) and non-minimal ($c_{\sigma C^2}, c_{C V^2}$) interactions are required, and those involving amplitudes with contact terms ($c_{\sigma^2 C^2}$).

The non-minimal three-point interactions \Eqs{eq:phiC2}{eq:sigmaV2}, via the four-point amplitude \Eq{eq:hp2Vp2}, contribute to the running as
\beq
\beta_4^\nm = - \frac{1}{16\pi^2} \sum_{i,j}^{N_V} \sum_{k,l}^{N_\sigma} \Big(
c_{\sigma C^2}^k c_{\sigma V^2}^{ij;k} c_{\sigma C^2}^{l\,*} c_{\sigma V^2}^{ij;l\,*} +
c_{\sigma C^2}^k c_{\sigma V^2}^{ij;k\,*} c_{\sigma C^2}^{l\,*} c_{\sigma V^2}^{ij;l}
\Big) \,.
\eeq
Diagrams that involve both minimal and non-minimal three-point couplings are of two types. On the one hand, there are contributions from cuts of the form $\amp^\gr \times \amp^\nm$, where the relevant gravitational four-point amplitudes are given in \Eqs{eq:h2sigma2}{eq:h2V2}, and (\ref{eq:AGRtreeh}), while those with non-minimal couplings are \Eqs{eq:h2SiSj}{eq:h2ViVj}, and (\ref{eq:ANMtreeh}).
On the other hand, there are also contributions from cuts of the form $\amp^\nm \times \amp^\nm$, where $\amp^\nm$ involves both a minimal and a non-minimal interaction, \Eqs{eq:h3sigma}{eq:h2Vp2}.
The resulting anomalous dimension is given by
\beq
\beta_4^{\gr \times \nm} = - \frac{1}{20 \pi^2} \bigg( \frac{1}{4} \sum_{i,j}^{N_V} |c_{CV^2}^{ij}|^2 - \frac{26}{3} \frac{1}{\mpl^2} 
\sum_k^{N_\sigma} |c_{\sigma C^2}^{k}|^2 
\bigg) \,.
\label{eq:betag4GRNM}
\eeq

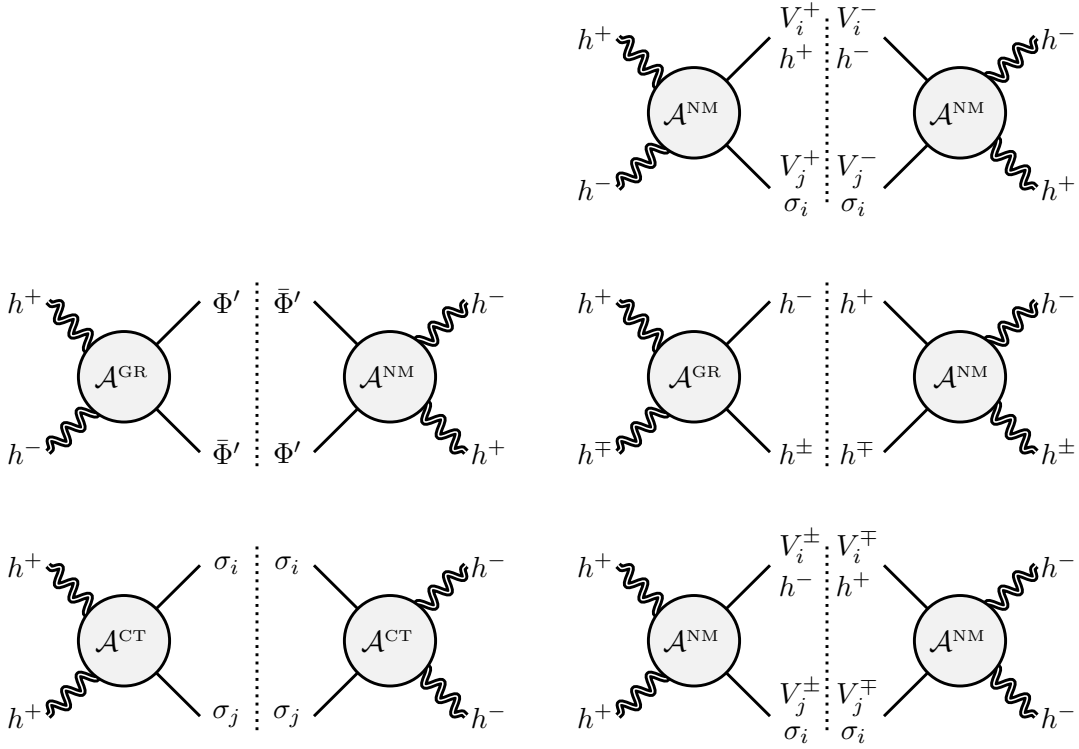
\begin{figure}[t]
\centering
\begin{tikzpicture}[line width=1.1 pt, scale=.50, baseline=(current bounding box.center)]
	
	\begin{scope}[shift={(14,0)}]

	\begin{scope}[shift={(-.5,0)}]
	\draw[graviton] (-12,5) -- (-10,3) ;
	\draw[] (-8,5) -- (-10,3) ;
	\draw[graviton] (-12,1) -- (-10,3) ;
	\draw[] (-8,1) -- (-10,3) ;
	\draw[fill=gray!10] (-10,3) circle (1.2cm) ;
	\node at (-10,3) {$\amp^\nm$\,};
	\node at (-12.6,5) {$h^+$};
	\node at (-12.6,1) {$h^-$};
	\end{scope}

	\draw[dotted] (-7,5.5) -- (-7,.5) ;
	\node at (-6.2,5.5) {$V^-_i$};
	\node at (-6.3,4.5) {$h^-$};
	\node at (-6.2,1.5) {$V^-_j$};
	\node at (-6.3,0.5) {$\sigma_i$};
	\node at (-7.7,5.5) {$V^+_i$};
	\node at (-7.8,4.5) {$h^+$};
	\node at (-7.7,1.5) {$V^+_j$};
	\node at (-7.8,0.5) {$\sigma_i$};

	\begin{scope}[shift={(.5,0)}]
	\draw[] (-6,5) -- (-4,3) ;
	\draw[graviton] (-2,5) -- (-4,3) ;
	\draw[] (-6,1) -- (-4,3) ;
	\draw[graviton] (-2,1) -- (-4,3) ;
	\draw[fill=gray!10] (-4,3) circle (1.2cm) ;
	\node at (-1.4,5) {$h^-$};
	\node at (-1.4,1) {$h^+$};
	\node at (-4,3) {$\amp^\nm$\,};
	\end{scope}

	\end{scope}


	\begin{scope}[shift={(-1,-7)}]
	
	\begin{scope}[shift={(-.5,0)}]
	\draw[graviton] (-12,5) -- (-10,3) ;
	\draw[] (-8,5) -- (-10,3) ;
	\draw[graviton] (-12,1) -- (-10,3) ;
	\draw[] (-8,1) -- (-10,3) ;
	\draw[fill=gray!10] (-10,3) circle (1.2cm) ;
	\node at (-10,3) {$\amp^\gr$\,};
	\node at (-12.6,5) {$h^+$};
	\node at (-12.6,1) {$h^-$};
	\end{scope}
	
	\draw[dotted] (-7,5.5) -- (-7,.5) ;
	\node at (-6.2,5.) {$\bar{\Phi}'$};
	\node at (-6.2,1.) {$\Phi'$};
	\node at (-7.8,5.) {$\Phi'$};
	\node at (-7.8,1.) {$\bar{\Phi}'$};
	
	\begin{scope}[shift={(.5,0)}]
	\draw[] (-6,5) -- (-4,3) ;
	\draw[graviton] (-2,5) -- (-4,3) ;
	\draw[] (-6,1) -- (-4,3) ;
	\draw[graviton] (-2,1) -- (-4,3) ;
	\draw[fill=gray!10] (-4,3) circle (1.2cm) ;
	\node at (-1.4,5) {$h^-$};
	\node at (-1.4,1) {$h^+$};
	\node at (-4,3) {$\amp^\nm$\,};
	\end{scope}

	\end{scope}


	\begin{scope}[shift={(14,-7)}]

	\begin{scope}[shift={(-.5,0)}]
	\draw[graviton] (-12,5) -- (-10,3) ;
	\draw[] (-8,5) -- (-10,3) ;
	\draw[graviton] (-12,1) -- (-10,3) ;
	\draw[] (-8,1) -- (-10,3) ;
	\draw[fill=gray!10] (-10,3) circle (1.2cm) ;
	\node at (-9.9,3) {$\amp^\gr$\,};
	\node at (-12.6,5) {$h^+$};
	\node at (-12.6,1) {$h^\mp$};
	\end{scope}
	
	\draw[dotted] (-7,5.5) -- (-7,.5) ;
	\node at (-6.2,5.) {$h^+$};
	\node at (-6.2,1.) {$h^\mp$};
	\node at (-7.8,5.) {$h^-$};
	\node at (-7.8,1.) {$h^\pm$};
	
	\begin{scope}[shift={(.5,0)}]
	\draw[] (-6,5) -- (-4,3) ;
	\draw[graviton] (-2,5) -- (-4,3) ;
	\draw[] (-6,1) -- (-4,3) ;
	\draw[graviton] (-2,1) -- (-4,3) ;
	\draw[fill=gray!10] (-4,3) circle (1.2cm) ;
	\node at (-1.4,5) {$h^-$};
	\node at (-1.4,1) {$h^\pm$};
	\node at (-3.9,3) {$\amp^\nm$\,};
	\end{scope}

	\end{scope}


	\begin{scope}[shift={(-1,-14)}]

	\begin{scope}[shift={(-.5,0)}]
	\draw[graviton] (-12,5) -- (-10,3) ;
	\draw[] (-8,5) -- (-10,3) ;
	\draw[graviton] (-12,1) -- (-10,3) ;
	\draw[] (-8,1) -- (-10,3) ;
	\draw[fill=gray!10] (-10,3) circle (1.2cm) ;
	\node at (-10,3) {$\amp^\ct$\,};
	\node at (-12.6,5) {$h^+$};
	\node at (-12.6,1) {$h^+$};
	\end{scope}
	
	\draw[dotted] (-7,5.5) -- (-7,.5) ;
	\node at (-6.2,5.) {$\sigma_i$};
	\node at (-6.2,1.) {$\sigma_j$};
	\node at (-7.8,5.) {$\sigma_i$};
	\node at (-7.8,1.) {$\sigma_j$};
	
	\begin{scope}[shift={(.5,0)}]
	\draw[] (-6,5) -- (-4,3) ;
	\draw[graviton] (-2,5) -- (-4,3) ;
	\draw[] (-6,1) -- (-4,3) ;
	\draw[graviton] (-2,1) -- (-4,3) ;
	\draw[fill=gray!10] (-4,3) circle (1.2cm) ;
	\node at (-1.4,5) {$h^-$};
	\node at (-1.4,1) {$h^-$};
	\node at (-4,3) {$\amp^\ct$\,};
	\end{scope}

	\end{scope}


	\begin{scope}[shift={(14,-14)}]

	\begin{scope}[shift={(-.5,0)}]
	\draw[graviton] (-12,5) -- (-10,3) ;
	\draw[] (-8,5) -- (-10,3) ;
	\draw[graviton] (-12,1) -- (-10,3) ;
	\draw[] (-8,1) -- (-10,3) ;
	\draw[fill=gray!10] (-10,3) circle (1.2cm) ;
	\node at (-10,3) {$\amp^\nm$\,};
	\node at (-12.6,5) {$h^+$};
	\node at (-12.6,1) {$h^+$};
	\end{scope}
	
	\draw[dotted] (-7,5.5) -- (-7,.5) ;
	\node at (-6.2,5.5) {$V^\mp_i$};
	\node at (-6.3,4.5) {$h^+$};
	\node at (-6.2,1.5) {$V^\mp_j$};
	\node at (-6.3,0.5) {$\sigma_i$};
	\node at (-7.7,5.5) {$V^\pm_i$};
	\node at (-7.8,4.5) {$h^-$};
	\node at (-7.7,1.5) {$V^\pm_j$};
	\node at (-7.8,0.5) {$\sigma_i$};
	
	\begin{scope}[shift={(.5,0)}]
	\draw[] (-6,5) -- (-4,3) ;
	\draw[graviton] (-2,5) -- (-4,3) ;
	\draw[] (-6,1) -- (-4,3) ;
	\draw[graviton] (-2,1) -- (-4,3) ;
	\draw[fill=gray!10] (-4,3) circle (1.2cm) ;
	\node at (-1.4,5) {$h^-$};
	\node at (-1.4,1) {$h^-$};
	\node at (-4,3) {$\amp^\nm$\,};
	\end{scope}

	\end{scope}
	
\end{tikzpicture}
\caption{\emph{
Cuts that contribute to the anomalous dimension of the leading (MHV) four-graviton EFT coefficient.
$\Phi' = \sigma, V$ in the middle-left.
}}
\label{fig:cutsbeta4}
\end{figure}

Finally, the contact amplitude \Eq{eq:hp2sigma2}
gives rise to the following anomalous dimension
\beq
\beta_4^\ct = - \frac{1}{16\pi^2} \sum_{k,l}^{N_\sigma} |c_{\sigma^2 C^2}^{kl}|^2 \, .
\label{eq:betag4CT}
\eeq 

A summary of the relevant cuts is shown in \Fig{fig:cutsbeta4}.
In contrast to the scalar and photon cases, the scenario in which $g_4$ decreases toward the IR is straightforward to identify: a theory of scalars non-minimally coupled to gravity. Considering $N_\sigma$ such scalars, all with the same interaction strength, we find
\beq
\beta_4 = + \frac{13}{30 \pi^2} \frac{1}{\mpl^2} N_\sigma |c_{\sigma C^2}|^2 \, .
\label{eq:betag4pos}
\eeq

\subsection{Positivity in the decoupling limit} \label{sec:positivity}

The fact that a positive $\beta$-function for the leading coefficients of the scalar, photon, and graviton EFTs vanishes in the decoupling limit $\mpl \to \infty$, see Eqs.~(\ref{eq:betac2neg}), (\ref{eq:betaalpha2negda}), and (\ref{eq:betag4pos}) respectively, should not come as a surprise, as this is no accident. As we will argue below, the requirement of dynamical gravitons is tied to their universal exchange in the $t$ channel of elastic processes.

As discussed in the beginning of \Sec{sec:betas}, one-loop anomalous dimensions $\gamma_i = dc_i/d\log\mu$ are fixed by unitarity as the sum over double cuts, in the $s$, $t$, and $u$ channels for a four-point amplitude,
\bea
\frac{\gamma_i }{c_i} \amp_i(1_\Psi,2_{\Omega},3_{\bar\Psi},4_{\bar\Omega}) = && \label{eq:cuts} \\
= - \tfrac{1}{4\pi^3} \sum_{\Phi,\Phi'} \sigma_{\Phi,\Phi'} \bigg( \!\!&&\!\! 
\int d{\rm LIPS} \, \amp(1_\Psi,2_{\Omega},-\ell_{\bar{\Phi}},-\ell'_{\bar{\Phi'}}) \amp(\ell'_{\Phi'},\ell_{\Phi},3_{\bar\Psi},4_{\bar\Omega}) \nonumber \\
\!\!&&\!\! 
+ (-1)^{F_t} \int d{\rm LIPS} \, \amp(1_\Psi,3_{\bar\Psi},-\ell_{\bar{\Phi}},-\ell'_{\bar{\Phi'}}) \amp(\ell'_{\Phi'},\ell_{\Phi},2_{\Omega},4_{\bar\Omega}) \nonumber \\
\!\!&&\!\! 
+ (-1)^{F_u} \int d{\rm LIPS} \, \amp(1_\Psi,4_{\bar\Omega},-\ell_{\bar{\Phi}},-\ell'_{\bar{\Phi'}}) \amp(\ell'_{\Phi'},\ell_{\Phi},2_{\Omega},3_{\bar\Psi}) \bigg) \,, \nonumber
\eea
where $F_{t (u)}$ is the number of fermionic permutations required to move $3_{\bar\Psi} (4_{\bar\Omega})$ next to $1_\Psi$. We are focusing on an elastic amplitude as this is the case for all the amplitudes of interest in this work.%
\footnote{Under crossing,
\beq
\amp(1_\Psi,2_{\Omega},-4_{\bar\Omega},-3_{\bar\Psi}) \equiv 
\sigma^{-1}_{\Psi,\Omega} \, \amp(1_\Psi,2_{\Omega} \to 3_{\Psi},4_{\Omega}) \,.
\label{eq:crossing}
\eeq
}
It is important to notice that \Eq{eq:cuts} holds only in the absence of IR divergences; otherwise, the $d$LIPS integrals would be ill-defined.
Moreover, when there are no IR divergences, the contributions to the two-cuts from triangles and boxes -- which take the form of logarithmic terms, $\log(-s_{ij}/s_{ik})$ (see \App{sec:bubble}) -- cancel in the sum over channels \cite{Baratella:2020lzz}. This cancellation explains why, unlike in \Eq{eq:bs}, the operator $\mathcal{R}$ does not appear in \Eq{eq:cuts}.

Once the validity of \Eq{eq:cuts} is established, we can gain information on the sign of the anomalous dimension $\gamma_i$ by considering its forward limit, $t = (p_1+p_3)^2 = (p_2 + p_4)^2 \to 0$.
Using crossing relations, such as \Eq{eq:crossing}, and taking into consideration our conventions (see footnote \ref{fn:conventions}), we arrive at
\bea
\frac{\gamma_i }{c_i} \amp_i(1_\Psi,2_{\Omega} \to 1_{\Psi},2_{\Omega}) 
= && \label{eq:cutsforward} \\
= - \tfrac{1}{4\pi^3} \sum_{\Phi,\Phi'}  \int d{\rm LIPS}
\!\!&&\!\!  \, \bigg( |\amp(1_\Psi,2_{\Omega} \to \ell_{\Phi},\ell'_{\Phi'})|^2 
+ |\amp(1_\Psi,-2_{\bar\Omega} \to \ell_{\Phi},\ell'_{\Phi'})|^2 \nonumber \\
&& \quad + \sigma_{\Psi,\Omega} \sigma_{\Phi,\Phi'} 
\, \amp(1_\Psi,-1_{\bar\Psi},-\ell_{\bar{\Phi}},-\ell'_{\bar{\Phi'}}) \amp(\ell'_{\Phi'},\ell_{\Phi},2_{\Omega},-2_{\bar\Omega}) \bigg) \, .
\nonumber
\eea
Aside from the $t$-channel cut in the last line, the right-hand side is proportional to a sum of (exclusive) cross sections, given that $\int d{\rm LIPS} \, |\amp(1_\Psi,2_{\Omega} \to \ell_{\Phi},\ell'_{\Phi'})|^2 = \alpha \, s \, \sigma_{\Psi \Omega \to \Phi \Phi'}(s) $ and $\int d{\rm LIPS} \, |\amp(1_\Psi,-2_{\bar \Omega} \to \ell_{\Phi},\ell'_{\Phi'})|^2 = \alpha \, u \, \sigma_{\Psi \bar \Omega \to \Phi \Phi'}(u)$, with $\alpha > 0$ and $u = -s$ in the forward limit.
Note that while the first term is positive for standard physical kinematics, $s > 0$, the second is evaluated in $u$-channel kinematics and is therefore positive when $u = -s > 0$. 

\Eq{eq:cutsforward} shows that sufficient conditions for a negative anomalous dimension are that, in the forward limit, the amplitude $\amp_i$ scales with even powers of $s$ and the $t$-channel cut vanishes.
The first requirement is satisfied by all EFT amplitudes of interest in this work. 
We will show below that the second condition holds as long as gravitons are non-dynamical, and we will clarify how the inclusion of GR amplitudes leads to a non-vanishing $t$-cut in the limit $t \to 0$.

Before doing so, a few comments on \Eq{eq:cutsforward} are in order.
In theories where the only minimal interactions are gravitational, there are no collinear divergences (see \App{sec:irdivs}). This implies that the $1/\epsilon^2$ terms in the one-loop amplitude -- associated with soft-collinear divergences -- cancel. In any case, these do not contribute to any cut.
As for soft divergences, since they are associated with $(1/\epsilon) \log(-s_{ij})$ terms in the amplitude, their cancellation must take place in each channel separately, implying that each double cut is IR finite.
This becomes evident after taking the forward limit: if there are no IR divergences, the cross sections must be free of singularities throughout the ($d$LIPS) phase space and therefore regular. Moreover, triangle and box contributions must vanish within each individual cut, since tree-level IR-finite cross sections do not have branch cuts. 

Finally, while the positivity implied by \Eq{eq:cutsforward} for a forward-vanishing $t$-cut is well known, particularly in the context of bounds derived from forward-limit dispersion relations (see, e.g., \cite{Distler:2006if,Bellazzini:2019xts,Bellazzini:2020cot,Trott:2020ebl,Arkani-Hamed:2020blm}), here the focus is on the operational conditions required for this to hold, along the lines of \cite{Trott:2020ebl,Baratella:2021guc,Liao:2025npz}. In other words, our emphasis is on the prerequisites for the forward limit to be well defined, taking into account the possibility that triangles and boxes contribute to the two-cuts due to the presence of non-trivial three-point amplitudes.

\subsubsection{Partial-wave decomposition} \label{sec:pw}

A partial-wave analysis is often the most convenient approach for understanding the dependence of amplitudes, or their phase-space integrals, on the Mandelstam variables. The partial-wave expansion of a four-point amplitude can be written as (see, e.g., \cite{Hebbar:2020ukp,Baratella:2020dvw}):
\beq
\amp(1_{h_1},2_{h_2},3_{h_3},4_{h_4}) = e^{i\phi(h_{12}-h_{43})} \sum_J n_J \, d^{J}_{h_{12},h_{43}}(\theta) \, 
a_J^{\{ h_1,h_2,h_4,h_3 \}}(s) \,,
\label{eq:pwexp}
\eeq
where the polar coordinates $(\theta, \phi)$ parametrize the direction of the outgoing particle pair $\{-3_{-h_3},-4_{-h_4}\}$ with respect to the $z$-axis defined by the direction of particles $1$ and $2$ in the center-of-mass frame.%
\footnote{Recall \Eq{eq:crossing} and note that we have omitted the factor $i^{F[h_3,h_4]}$ in \Eq{eq:pwexp} since it equals one in all cases of interest in this work. In addition, the chosen frame is reached by taking
\beq
\begin{pmatrix}
|3\rangle \\
|4\rangle  
\end{pmatrix} = R 
\begin{pmatrix}
|1\rangle \\
|2\rangle  
\end{pmatrix} \, , \quad 
\begin{pmatrix}
|3] \\
|4]  
\end{pmatrix} = -R^{*} 
\begin{pmatrix}
|1] \\
|2]  
\end{pmatrix} \, , \qquad
R = 
\begin{pmatrix}
c_{\theta/2} & -s_{\theta/2} e^{-i\phi} \\
s_{\theta/2} e^{i\phi} & c_{\theta/2}
\end{pmatrix} \,,
\nonumber
\eeq
where $c_{\theta/2} \equiv \cos(\theta/2)$ and $s_{\theta/2} \equiv \sin(\theta/2)$.}
The sum runs over the allowed angular momenta $J$ of the pair $\{1,2\}$ (i.e., in the $s$ channel), $n_J = 16\pi (2 J + 1)$, $h_{ij} = h_i - h_j$, and $d^J_{h,h'}(\theta)$ are the Wigner $d$-functions.
We will also be using the helicity-stripped $d$-functions \cite{Caron-Huot:2022ugt},
\beq
\tilde{d}^J_{h,h'} (c_\theta) = \left( \frac{1- c_\theta}{2} \right)^{\frac{h'-h}{2}} \left( \frac{1+ c_\theta}{2} \right)^{-\frac{h+h'}{2}} d^J_{h,h'}(\theta) \,,
\label{eq:hstripWd}
\eeq
where $c_\theta \equiv \cos\theta$ and $-t/s = (1- c_\theta)/2$, $-u/s = (1+ c_\theta)/2$. 
The Legendre polynomials correspond to $P_J(c_\theta) = \tilde{d}^J_{0,0} (c_\theta)$.
The dynamics is encoded in the partial-wave coefficients, $\bar{a}_J(s) \equiv 16\pi a_J(s)$,
which can be obtained from the amplitude via the relation
\beq
\bar{a}_J^{\{ h_1,h_2,h_4,h_3 \}}(s) = \frac{1}{2}
\int_0^\pi d\theta \, s_\theta \, d^J_{h_{12},h_{43}}(\theta) \,\amp(1_{h_1},2_{h_2},3_{h_3},4_{h_4})|_{\phi = 0} \,.
\label{eq:aJ}
\eeq

The partial-wave decomposition -- which is not always well-defined, elastic amplitudes mediated by graviton exchange being an important example -- can significantly simplify the computation of phase-space integrals, such as those in \Eqs{eq:bs}{eq:cuts}. For instance, the $s$-channel double cut is given by \cite{Baratella:2020dvw}
\bea
\int d{\rm LIPS} \, \amp _L(1_{h_1},2_{h_2},-\ell_{\bar{\Phi}},-\ell'_{\bar{\Phi'}}) \amp_R(\ell'_{\Phi'},\ell_{\Phi},3_{h_3},4_{h_4}) = && \label{eq:bspw} \\
= \frac{\pi}{2} e^{i\phi(h_{12}-h_{43})} 16\pi \sum_J n_J \!\!&&\!\! d^J_{h_{12},h_{43}} (\theta) \, a^L_J(s) \, a^R_J(s) \, ,
\nonumber
\eea
where it is understood that $a^L_J = a_J^{\{ h_1,h_2,h_{\Phi'},h_{\Phi} \}}$ and similarly for $a^R_J$.

To illustrate the utility of this approach, let us work out some of its implications in a simple example: the EFT of U(1) Nambu-Goldstone boson with self-interactions only. We focus on the renormalization of the four-scalar amplitude at one loop, thus only four-point contact terms are relevant. An independent basis for these reads
\beq
\amp^\eft(1_\phi,2_\phi,3_\phi,4_\phi) = \sum_{p,q} c_{p,q} \, (s^2 + t^2 + u^2)^p \, (stu)^q \,, 
\label{eq:AEFTtreephigen}
\eeq
with $p,q \geqslant 0$. Note that \Eq{eq:AEFTtreephi} contains the leading terms, in an expansion in momenta, within this sum, with the redefined coefficients $ c_{p,q} \equiv c_{2p + 3q}$. The Wilson coefficients scale with powers of the cutoff as $c_{p,q} \sim 1/\coff^{2(2p+3q)}$, with multiple terms sharing the same scaling dimension from order $1/\coff^8$ onward.

Importantly, the angular momentum in each term of \Eq{eq:AEFTtreephigen} is bounded from above by $J \leqslant 2p+2q$. This is evident upon recalling that the maximum power of $t$ in a Legendre polynomial is determined by its spin, $P_J \sim t^J$ at large $t$.%
\footnote{In more detail, $P_J(1+2t/s) = \sum_{k=0}^J \binom{n}{k} \binom{n+k}{k} (t/s)^k$, where we have used $c_\theta = 1 +2t/s$.}
Moreover, crossing symmetry implies that only even $J$ appear.
Then, applying \Eq{eq:bspw} to this case, and with no explicit knowledge of the partial-wave coefficients needed,%
\footnote{Whenever needed, the partial-wave coefficients can be obtained via \Eq{eq:aJ}. For instance, for the terms with Wilson coefficients $c_{n,0}$,
\beq
a^J_{n,0} = s^{2n} \frac{2^n n!(J-1)!!}{16\pi J!!} \sum_{k=0}^n \frac{(-1)^{n+k+J/2} (n-k)!}{k! (2n-2k-J)!! (2n-2k+J+1)!!} \,.
\nonumber
\eeq}
readily implies that the $s$-channel cut, or equivalently the $s$-channel bubble coefficient as there are no triangles nor boxes, depends on the Mandelstam variables as
\beq
b_s(1_\phi,2_\phi,3_\phi,4_\phi)
\sim c_{p,q} \, c_{p',q'} \, s^{2(p+p')+3(q+q')-2\min[p+q,p'+q']} \, f_{p,p';q,q'}(s,t) \,,
\label{eq:bsgen}
\eeq
where $f_{p,p';q,q'}(s,t) \sim t^{2\min[p+q,p'+q']}$ at large $t$. This follows from the fact that the sum over angular momenta in \Eq{eq:bspw} runs up to $J_{\max} = 2\min[p+q,p'+q']$, while dimensional analysis fixes the remaining $s$ dependence. Note that even if the amplitudes involved in the double cut do not vanish in the $s \to 0$ limit, the bubble coefficient does. \Eq{eq:bsgen} implies as well that the leading divergence in $s$ associated with the $t^{2n}$ term in the amplitude is given by $\partial^{2n}\amp(s,t)/\partial t^{2n}|_{t = 0} \sim s^{2(n+1)} \log(-s)$ \cite{Bellazzini:2021oaj}.\\

In the following, we aim to provide a simple understanding of why the $t$-cut in \Eq{eq:cuts} vanishes in the limit $t \to 0$, thereby explaining why purely non-minimal and contact interactions contribute negatively to the anomalous dimensions of scalars, photons, and gravitons, and why the source of positivity (negative running) lies in minimal gravitational interactions. Accordingly, we will analyze the partial-wave expansion in the $t$ channel rather than the $s$ channel, which is obtained from \Eq{eq:pwexp} simply via the exchange $2 \leftrightarrow 3$, with the polar coordinates parametrizing the direction of the outgoing pair $\{-4_{-h_4}, -2_{-h_2} \}$.
In any case, recall that $s$- and $t$-channel cuts are related by crossing symmetry.
In what follows, amplitudes are evaluated at physical momenta.

\subsubsection{Scalar} \label{sec:pwscalar}

The partial-wave decomposition of the amplitudes Eqs.~(\ref{eq:phi2S2}), (\ref{eq:phi2V2}), and (\ref{eq:phi2Phi2}) ($\Phi = \sigma, \, V^\pm$) is given by
\bea
\amp^\ct(1_{\phi},3_{\phi},2_{\sigma},4_{\sigma}) = c_{\phi^2 \sigma^2} t \,, 
&&
\bar{a}_J = c_{\phi^2 \sigma^2} t \, \delta_{J,0} \,, 
\label{eq:phi2S2pw} \\
\amp^\nm(1_{\phi},3_{\phi},2_{V^+},4_{V^-}) = - c_{\phi V^2}^2 t \,,
&&
\bar{a}_{J\geqslant2} = \frac{-2 c_{\phi V^2}^2 t}{\sqrt{(J+2) (J+1) J (J-1)}} \,,
\label{eq:phi2V2pw} \\
\amp^\gr(1_{\phi},3_{\phi},2_{\sigma},4_{\sigma}) = 
-\frac{s (s+t)}{\mpl^2 t} \,,
&&
\bar{a}_J = \frac{t}{6 \mpl^2} \left( \delta_{J,0} - \frac{1}{5} \delta_{J,2} \right) \,, 
\label{eq:phi2S2GRpw} \\
\amp^\gr(1_{\phi},3_{\phi},2_{V^+},4_{V^-}) = 
-\frac{s (s+t)}{\mpl^2 t} \,,
&&
\bar{a}_{J\geqslant2} = \frac{t}{5 \sqrt{6} \mpl^2} \delta_{J,2} 
\label{eq:phi2V2GRpw} \,,
\eea
where, for the purposes of this section, we focus on a single flavor, without loss of generality.
We note that the factorization poles of $\amp^\nm$ do not lead to any singularity for real momenta, and instead the amplitude simply scales with $t$, much like $\amp^\ct$. This is contrast with $\amp^\gr$, which exhibit the well-known universal Coulomb singularity ($t/s \to 0$ limit) characteristic of gravitational interactions and source of soft gravitational divergences.
Nevertheless, since $-s(s+t)/t = t s_{\bar \theta}^2/4$ with $\bar \theta$ the scattering angle in the center-of-mass frame of particles 1 and 3, the $t$-channel partial-wave coefficients of $\amp^\gr$ are well-defined (unlike in the $s$-channel, where they would not be).

With the partial-wave coefficients at hand, the calculation of the $t$-channel double cut using \Eq{eq:bspw} is straightforward. However, let us recall that our aim is to gain a simple understanding of its dependence on $t$.
For instance, it is immediately clear that, from the contact amplitude alone, $b_t^\ct \sim c_{\phi^2 \sigma^2}^2 t^2$, since only the spin-0 partial wave contributes. As a result, the $t$-cut vanishes in the forward limit and the corresponding anomalous dimension must be negative, as found in \Eq{eq:betaa2CT}.

The situation is less straightforward for the purely non-minimal contribution, since all (even) partial waves contribute in \Eq{eq:phi2V2pw},
\bea
\frac{2}{\pi} \! \int d{\rm LIPS} \, \amp_L^\nm(1_\phi,3_\phi,-\ell_{V^+},-\ell'_{V^-}) 
\!\!&&\!\!
\amp_R^\nm(\ell'_{V^+},\ell_{V^-},2_\phi,4_\phi) = \label{eq:tcutnm} \\
&& = c_{\phi V^2}^4 t^2 \sum_{J \, \rm{even}} \frac{2(2J+1)P_J(c_{\bar \theta})}{(J+2) (J+1) J (J-1)} \nonumber \\
&& = c_{\phi V^2}^4 t \left( t - 2 s \log\Big(\frac{-s}{t}\Big) + 2 (s+t) \log\Big(\frac{s+t}{t}\Big)  \right) \,,
\nonumber
\eea
where the last equality has been derived using standard $d$LIPS integration \cite{Baratella:2020lzz,Baratella:2021guc}.
Besides the bubble coefficient, $b_t^\nm = c_{\phi V^2}^4 t^2$, \Eq{eq:tcutnm} includes contributions from triangles and boxes. These vanish in the limit $t \to 0$, as expected from the discussion following \Eq{eq:cutsforward}.
Therefore, to understand why the $t$-cut vanishes with $t$, it suffices to focus on the behavior of the bubble coefficient.
This can be done by explicitly examining the $d$LIPS integrand in \Eq{eq:tcutnm}, 
\beq
\amp_L^\nm \amp_R^\nm = c_{\phi V^2}^4 t^2 \frac{\an{1\ell'}\an{2\ell} \sq{\ell 1} \sq{\ell' 2}}{\an{1\ell} \an{2\ell'}\sq{\ell 2} \sq{\ell' 1}} \propto t^2 \frac{\an{2\ell} \sq{\ell' 2}}{\an{2\ell'}\sq{\ell 2}} \xrightarrow{d{\rm LIPS}} t^2 \, .
\label{eq:tcutnm2}
\eeq
In the first step, we used that the phase-space integral can be parametrized as a rotation of the loop momentum spinors, $|\ell\rangle = \cos \eta |1\rangle - \sin \eta \, e^{i\varphi} |3\rangle$ and  $|\ell'\rangle = \sin \eta \, e^{-i\varphi} |1\rangle + \cos \eta |3\rangle$, along with the corresponding rotation of the square spinors given by complex conjugation (which satisfy the constraint $p_{\ell} + p_{\ell'} = p_1+p_3$).
One can then observe that no $1/t$ pole can arise from $\an{2\ell} \sq{\ell' 2}/\an{2\ell'}\sq{\ell 2}$, as $|\ell] (|\ell'\rangle)$ must be evaluated, after $d$LIPS integration, at either $|1] (|1\rangle)$ or $|3] (|3\rangle)$, with no further dependence on $t$ ($\an{13}$, $[31]$) given that $p_1$ and $p_3$ are on the same side of the cut.%
\footnote{The fact that the tree-level amplitudes on either side of the cut do not contain $t \to 0$ singularities, and in fact vanish for $t \to 0$, implies that the phase-space integral behaves in the same way. This result can be argued to follow in generality from the Coleman-Norton theorem.} 
In fact, when considering only the bubble contribution, any potential pole cancels out.
This becomes most transparent when the phase-space integral is deformed via a BCFW shift of the cut legs ($|\ell] \to |\ell] + z |\ell']$, $|\ell'\rangle \to |\ell'\rangle - z |\ell\rangle$), and the pole at infinity is picked \cite{Arkani-Hamed:2008owk}.
One finds ${\rm Res}[ \amp_L^\nm(z) \amp_R^\nm(z)/z,\infty ] \propto t^2$, which is independent of $\theta$ and hence vanishes in the forward limit, also after the $d$LIPS integration.
Since the $t$-channel double cut vanishes in the forward limit, the purely non-minimal contribution to the anomalous dimension must be negative, as found in \Eq{eq:betac2NM}.

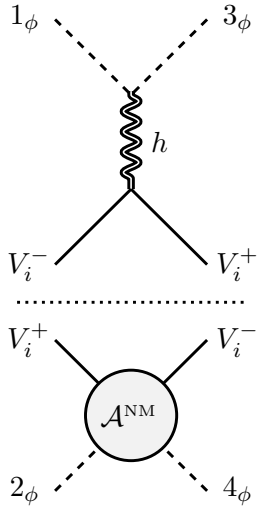
\begin{figure}[t]
\centering
\begin{tikzpicture}[line width=1.1 pt, scale=.50, baseline=(current bounding box.center)]
	
	\begin{scope}[shift={(6.5,0)}]

	\draw[dashed] (-12,13.5) -- (-10,11.5) ;
	\draw[dashed] (-8,13.5) -- (-10,11.5) ;
	\draw[graviton] (-10,11.6) -- (-10,9) ;
	\node at (-9.25,10.25) {$h$} ;
	\draw[] (-12,7) -- (-10,9) ;
	\draw[] (-8,7) -- (-10,9) ;
	\node at (-7.2,13.5) {$3_\phi$};
	\node at (-12.8,13.5) {$1_\phi$};

	\draw[dotted] (-13,6) -- (-7,6) ;
	\node at (-7.2,7) {$V_i^+$} ;
	\node at (-12.7,7) {$V_i^-$} ;
	\node at (-7.2,5.) {$V_i^-$} ;
	\node at (-12.7,5.) {$V_i^+$} ;

	\draw[] (-12,5) -- (-10,3) ;
	\draw[] (-8,5) -- (-10,3) ;
	\draw[dashed] (-12,1) -- (-10,3) ;
	\draw[dashed] (-8,1) -- (-10,3) ;
	\draw[fill=gray!10] (-10,3) circle (1.2cm) ;
	\node at (-10,3) {$\amp^\nm$\,} ;
	\node at (-7.2,1) {$4_\phi$} ;
	\node at (-12.8,1) {$2_\phi$} ;

	\end{scope}
			
\end{tikzpicture}
\caption{\emph{
Cut diagram involving the $t$-channel exchange of a minimally coupled graviton, responsible for the negative running of the leading four-scalar EFT coefficient.}}
\label{fig:cutforward}
\end{figure}

Finally, it is much simpler to understand why $t$-cuts involving an amplitude mediated by a minimally coupled graviton do not vanish in the forward limit, hence the corresponding contribution to the anomalous dimension need not be negative. Specifically, conservation of angular momentum implies that only the spin-2 partial wave contributes (and, in certain cases, the spin-0 as well), see \Eqs{eq:phi2S2GRpw}{eq:phi2V2GRpw}. For instance, in the case of $\amp^\gr \times \amp^\nm$ (with $c_{\bar \theta} = 1+2s/t$), 
\beq
b_t^{\gr \times \nm} = 32\pi \, n_2 \, a_{2}^\nm a_{2}^\gr P_2(c_{\bar \theta})= - \frac{c_{\phi V^2}^2}{3\mpl^2} \left(t^2 + 6 s (s+t)\right) \,,
\label{eq:bubbletcuts}
\eeq
which is non-vanishing and negative in the $t \to 0$ limit.%
\footnote{Proceeding as in \Eq{eq:tcutnm2}, one can arrive at the same conclusion. Specifically,
\beq
\amp_L^\nm \amp_R^\gr = \frac{c_{\phi V^2}^2}{\mpl^2} \frac{\an{1\ell'}\an{2\ell}^2 \sq{\ell 1} \sq{\ell' 2}^2}{\an{1\ell} \sq{\ell' 1}} \propto \an{2\ell}^2 \sq{\ell' 2}^2 \xrightarrow{d{\rm LIPS}} s^2 + O(st) \, ,
\nonumber
\eeq
where the right-hand side is shown as an expansion around small $t$.} Therefore, we conclude that the spin-2 $1/t$ pole of GR is the source of the negative running. This is illustrated in \Fig{fig:cutforward}.
We should point out that the mixed gravitational-non-minimal contribution in \Eq{eq:bubbletcuts} is associated with the interference term $2\,{\rm Re}[\amp^\gr (\amp^\nm)^*]$ of the cross section $\sigma(\phi \phi \to V^+ V^-)$ in \Eq{eq:cutsforward}. As such, it needs not be positive in general. Therefore, the correct statement is that the total bubble coefficient, from gravitational and non-minimal amplitudes, is not guaranteed to be positive because ${\rm lim}_{t \to 0}\,b_t^{\gr \times \nm} \neq 0$.

\subsubsection{Photon} \label{sec:pwphoton}

As in the scalar case, a partial-wave analysis is illuminating for understanding the behaviour of the $t$-cut in the forward limit. We focus here on the case where only $c^{i;k}_{\sigma F^2}$ and $c^{kl}_{\sigma^2 F^2}$ are non-zero and real  for a single vector $i$ and scalar $k, l$ flavor. The analysis for other couplings relevant to the renormalization of the leading (MHV) four-photon EFT coefficient proceeds analogously and leads to the same qualitative conclusions. 
The partial-wave decomposition of the amplitudes Eqs.~(\ref{eq:g2V2}), (\ref{eq:g2sigma2}), (\ref{eq:gam2V2}) and (\ref{eq:gam2sigma2}) is given by
\bea
\amp^\nm(1_{\gamma^+},3_{\gamma^-},2_{V^+},4_{V^-}) = 
- c_{\sigma F^2}^2 s
&&
\bar{a}_{J\geqslant2} = \frac{6 c_{\sigma F^2}^2 t}{(J+2) (J+1) J (J-1)} \,, 
\label{eq:g2V2pw} \\
\amp^\nm(1_{\gamma^+},3_{\gamma^-},2_\sigma,4_\sigma) = - c_{\sigma F^2}^2 t \,,
&&
\bar{a}_{J\geqslant2  \, \rm{even}} = \frac{-2 c_{\sigma F^2}^2 t}{\sqrt{(J+2) (J+1) J (J-1)}} \,, 
\label{eq:g2sigma2pw} \\
\amp^\gr(1_{\gamma^+},3_{\gamma^-},2_{V^+},4_{V^-}) = 
-\frac{s^2}{\mpl^2 t} \,,
&&
\bar{a}_{J\geqslant2} = -\frac{t}{5\mpl^2} \delta_{J,2} \,, 
\label{eq:gam2V2pw} \\
\amp^\gr(1_{\gamma^+},3_{\gamma^-},2_\sigma,4_\sigma) = 
-\frac{s (s+t)}{\mpl^2 t}
&&
\bar{a}_{J\geqslant2} = \frac{t}{5 \sqrt{6} \mpl^2} \delta_{J,2} \,. 
\label{eq:gam2sigma2pw}
\eea
Note that we show neither $\amp^\ct(1_{\gamma^+},2_{\gamma^+},3_{\sigma_i},4_{\sigma_j})$ nor $\amp^\nm(1_{\gamma^+},2_{\gamma^+},3_{V^+},4_{V^+})$, even though they involve the couplings $c_{\sigma^2 F^2}$ and $c_{\sigma F^2}$ (see \Eqs{eq:gp2sigma2}{eq:gp2Vp2}), since they trivially do not contribute to the $t$-channel cut.
For the contact amplitude, this automatically implies that its contribution to the anomalous dimension must be negative, as shown in \Eq{eq:betaalpha2CT}.

Similarly to the scalar case, the factorization poles of $\amp^\nm$ do not lead to singularities for real momenta, and their angular momentum decomposition involves an infinite number of partial waves. In contrast, $\amp^\gr$ display a Coulomb singularity and only the spin-2 partial wave contributes. 
Consequently, the partial-wave decomposition is not entirely transparent when trying to understand the $t$-dependence of the $t$-cut in the non-minimal case. 
Through standard $d$LIPS integration, we find
\bea
\frac{2}{\pi} \! 
\int d{\rm LIPS}  \amp_L^\nm(1_{\gamma^+},3_{\gamma^-},-\ell_{V^+},-\ell'_{V^-}) 
\!\!&&\!\! 
\amp_R^\nm(\ell'_{V^+},\ell_{V^-},2_{\gamma^-},4_{\gamma^+}) = \label{eq:gamtcutnmV} \\
 = \frac{c_{\sigma F^2}^4 t}{6(s+t)^2} \!\!&&\!\! \left( (s+t)(11s^2+7st+2t^2) - 6 s^3 \log\Big(\frac{-s}{t}\Big) \right) \,,
\nonumber
\eea
for the vector cut, and
\bea
\frac{1}{\pi} \! \int d{\rm LIPS} \amp_L^\nm(1_{\gamma^+},3_{\gamma^-},-\ell_\sigma,-\ell'_\sigma) 
\!\!&&\!\!  
\amp_R^\nm(\ell'_\sigma,\ell_\sigma,2_{\gamma^-},4_{\gamma^+}) = \label{eq:gamtcutnms} \\
&& = \frac{c_{\sigma F^2}^4 t^2}{2(s+t)^2} \left( (s+t)(s-t) - 2 s t \log\Big(\frac{-s}{t}\Big) \right) \,,
\nonumber
\eea
for the scalar cut.
Both bubble coefficients in \Eqs{eq:gamtcutnmV}{eq:gamtcutnms}, as well as the contributions from triangles and boxes, vanish in the limit $t \to 0$. To better grasp why this is the case for the former, we examine the corresponding phase-space integrands, similarly to the scalar case. The vector cut integrand in \Eq{eq:gamtcutnmV} is given by
\beq
\amp_L^\nm \amp_R^\nm|_{V^+V^-} = -c_{\sigma F^2}^4 \frac{\an{3\ell'}^2\an{2\ell}^2 \sq{\ell 1} \sq{\ell' 4}}{\an{1\ell} \an{4\ell'}} \propto t \frac{\an{2\ell}^2 \sq{\ell' 4}}{\an{4\ell'}} \xrightarrow{d{\rm LIPS}} t s + O(t^2) \, ,
\label{eq:tcutnm2V}
\eeq
where the right-hand shows the scaling of the leading term in an expansion around small $t$. 
To arrive at this result, we have explicitly evaluated the spinor contractions $\an{3\ell'}^2 \sq{\ell 1}/\an{1\ell}$ with the loop momenta parametrized in terms of $p_1$ and $p_3$. From the resulting expression, it is clear that the overall scaling with $t$ will remain after the phase-space integration, since $\ell$ and $\ell'$ must be proportional to $p_1$ or $p_3$, with the proportionally factor independent of $t$ (one should further note that $|\ell\rangle^2 [\ell'|/|\ell'\rangle$ must carry the helicity weights of $|3\rangle^2 [1|^2$).
The same analysis for the scalar cut in \Eq{eq:gamtcutnms} leads to
\beq
\amp_L^\nm \amp_R^\nm|_{\sigma \sigma} = -c_{\sigma F^2}^4 t^2 \frac{\an{3\ell} \an{3\ell'} \sq{\ell 4} \sq{\ell' 4}}{\an{1\ell} \an{1\ell'} \sq{\ell 2} \sq{\ell' 2}} \propto t^2 \frac{\sq{\ell 4} \sq{\ell' 4}}{\sq{\ell 2} \sq{\ell' 2}} \xrightarrow{d{\rm LIPS}} t^2 + O(t^3/s) \, .
\label{eq:tcutnm2s}
\eeq
The discussion above shows that the complex poles of non-minimal amplitudes, while they give rise to $1/u$ (higher-order) poles in bubble coefficients as well as to contributions from triangle and box topologies, do not introduce any $1/t$ singularities and their double cuts in fact vanish as $t \to 0$. As in the scalar case, an appropriate BCFW shift of the phase-space integral confirms these results.

This behaviour is instead not satisfied by amplitudes generated by the exchange of a minimally-coupled graviton, which feature a universal $1/t$ pole and lead to non-zero bubble coefficients in the $t \to 0$ limit. This is easily understood from angular momentum selection rules, specifically from the fact only the spin-2 partial wave contributes to the double cuts.
Indeed, for the mixed $\amp^\gr \times \amp^\nm$ contributions, we find, for the vector cut
\beq
b_t^{\gr \times \nm}|_{V^+V^-} = 32\pi \, n_2 \, (a_{2}^\nm a_{2}^\gr)|_{V^+V^-} \, d^2_{2,2}(\bar \theta) = - \frac{c_{\sigma F^2}^2}{2\mpl^2} (s+t)^2 \,,
\eeq
and for the scalar cut
\beq
b_t^{\gr \times \nm}|_{\sigma \sigma} = 32\pi \, n_2 \, (a_{2}^\nm a_{2}^\gr)|_{\sigma \sigma} \, d^2_{2,2}(\bar \theta) = - \frac{c_{\sigma F^2}^2}{6\mpl^2} (s+t)^2 \,,
\eeq
both of which are non-vanishing and negative.%
\footnote{Recall that leading EFT coefficient of the four-photon MHV amplitude is $\alpha_2 \an{23}^2 \sq{14}^2 = \alpha_2 (s+t)^2$.} 

\subsubsection{Graviton} \label{sec:pwgraviton}

We focus here on the case where only the non-minimal scalar-graviton three-point coupling $c_{\sigma C^2}$ is non-vanishing, since this is the only one leading to a negative running of the leading (MHV) four-graviton EFT coefficient, \Eq{eq:betag4GRNM}. 
This contribution arises in combination with a gravitational amplitude and should therefore vanish in the limit $\mpl \to \infty$. 
Additionally, it is not necessarily positive even if the $t$-channel double cuts $\amp^\gr \times \amp^\nm$ vanish in the $t \to 0$ limit, since these are associated with interference terms of the cross sections $\sigma(h^+ h^{\mp} \to h^+ h^{\pm})$ and $\sigma(h^+ h^- \to \sigma \sigma)$. 
Moreover, in the case of cut gravitons, the gravitational four-graviton amplitude, \Eq{eq:AGRtreeh}, which lacks a well-defined partial-wave decomposition, introduces IR divergences that render \Eq{eq:cuts} invalid.
For these reasons, in the following we simply present the results for the individual bubble coefficients.

For the $t$-channel bubble coefficient, we find
\bea
b_t|_{h^+h^-} &=& -\frac{c_{\sigma C^2}^2}{\mpl^6} \frac{t^2 (147\, t^5 + 872\, t^4 s + 2143\, t^3 s^2 + 2777\, t^2 s^3 + 1968\, t s^4 + 669\, s^5)}{30(s+t)^3} \,, \quad \\
b_t|_{\sigma\sigma} &=& \frac{c_{\sigma C^2}^2}{\mpl^6} \frac{t^2 (t-s) (t^4 + 9\, t^3 s + 46\, t^2 s^2 + 9\, t s^3 + s^4)}{30(s+t)^3} \,, \quad \\
b_t|_{h^{\pm}\sigma} &=& 2 \frac{c_{\sigma C^2}^2}{\mpl^6} \frac{t^2 (t-s) (6\, t^4 - 31\, t^3 s - 44\, t^2 s^2 - 31\, t s^3 + 6\, s^4)}{30(s+t)^3} \,, \quad
\eea
while the $s$-channel bubble coefficient is obtained from the $t$-channel one by crossing $s \leftrightarrow t$.
Note that we find a negative $b_s|_{h^+h^-} = -\frac{49}{10}\an{23}^4 \sq{14}^4$ in the forward limit.
In the $u$ channel, we obtain
\bea
b_u|_{h^+h^+} &=& 0 \,, \\
b_u|_{h^{-}\sigma} &=& \frac{c_{\sigma C^2}^2}{\mpl^6} (s+t)^4 \,.  
\eea
As expected, the sum over all channels yields a local answer, $b_s + b_t + b_u = -\frac{52}{12} \an{23}^4 \sq{14}^4$. After multiplying by the $-1/8\pi^2$ loop factor and factoring out a $1/\mpl^4$ normalization, this matches the result given in \Eq{eq:betag4GRNM}.

\section{Gravitational dispersion relations} \label{sec:dispersions}

It is interesting to contrast the result that the leading coefficients of the scalar, photon, and graviton EFTs can run negative at low energies with what can be inferred from dispersion relations, which rely only on the general principles of unitarity, locality, and causality. In particular, it is well known that at tree level and in the absence of gravitational interactions these principles are incompatible with negative coefficients.

We consider dispersion relations in the complex $s$ plane at fixed $t < 0$, of the form
\beq
\frac{1}{i\pi}\oint_C ds \frac{A(s,t)}{\left(s+\frac{t}{2}\right)^{2n+1}} = B_n(\bar{s},t)|_{\ir} - B_n(\bar{s},t)|_{\uv} = 0 \, ,
\label{eq:dispersion}
\eeq
with $A(s,t)$ a scattering amplitude with helicity weights subtracted and $s$-$u$ symmetrized,
\bea
A^{4\phi}(s,t) &\equiv& \amp(1_\phi,2_\phi,3_\phi,4_\phi) \,, 
\label{eq:A4phi} \\
A^{4\gamma}(s,t) &\equiv& \frac{1}{2} \left( \frac{\amp(1_{\gamma^+},2_{\gamma^-},3_{\gamma^-},4_{\gamma^+})}{\an{23}^2 \sq{14}^2} + \frac{\amp(1_{\gamma^+},2_{\gamma^+},3_{\gamma^-},4_{\gamma^-})}{\an{34}^2 \sq{12}^2} \right) \,,
\label{eq:A4pho} \\
A^{4h}(s,t) &\equiv& \frac{1}{2} \left( \frac{\amp(1_{h^+},2_{h^-},3_{h^-},4_{h^+})}{\an{23}^4 \sq{14}^4} + \frac{\amp(1_{h^+},2_{h^+},3_{h^-},4_{h^-})}{\an{34}^4 \sq{12}^4} \right) \,,
\label{eq:A4h}
\eea
for shift-symmetric scalars, photons, and gravitons, respectively. 

We assume a unitary $S$-matrix and that the amplitudes are analytic in the upper-half plane, satisfy $s \leftrightarrow u$ crossing symmetry and hermitian analyticity, $A(s^*,t) = A^*(s,t)$, and are bounded in the Regge limit as $\lim_{|s| \to \infty}\amp(s,t)/s^2 \to 0$ after suitably smearing over the momentum transfer $p = \sqrt{-t}$ \cite{Haring:2022cyf,Haring:2024wyz}; see also \Sec{sec:smear}. This implies $A_{4\phi} \lesssim s^2$, $A_{4\gamma} \lesssim s^0$, and $A_{4h} \lesssim s^{-2}$ for $-|s|/t \gg 1$. 

\begin{figure}[t]
\centering
    \begin{tikzpicture}
    	\draw[black!60!white,->] (0,-0.5) -- (0,3.2);
	\draw[black!60!white,->] (-2.2,0) -- (3,0);
	\node (a) at (2.5,2.75) {$s$};
	\draw (a.north west) -- (a.south west) -- (a.south east);	
	
	\draw (0,-0.115) node[cross out, draw, thick, inner sep=1.5pt] {};
	\draw[decoration={snake, amplitude=1}, decorate, thick, blue] (3,-0.115) -- (0,-0.115) ;
	\draw[decoration={snake, amplitude=1}, decorate, very thick, red] (3,-0.115) -- (2.0,-0.115) ;	

	\draw (0.4,0) node[cross out, draw, inner sep=1.5pt] {};
		
	\draw (0.8,0.115) node[cross out, draw, thick, inner sep=1.5pt] {};
	\draw[decoration={snake, amplitude=1}, decorate, thick, blue] (-1.2,0.115) -- (0.8,0.115) ;
	\draw[decoration={snake, amplitude=1}, decorate, very thick, red] (-2.2,0.115) -- (-1.2,0.115) ;	
	
	\draw[thick, postaction={decorate}, decoration={markings, mark= at position 0.5 with {\arrow{<}}}] (3.0,+0.2) arc (0:+180:2.6) ;
	\draw[thick, postaction={decorate}, decoration={markings, mark= at position 0.5 with {\arrow{>}}}] (1.4,+0.2) arc (0:+180:1.0) ;
	
	\draw (0,0) node[xshift=-0.2cm, yshift=-0.4cm] {\scriptsize $0$};
	\draw (0.8,0) node[xshift=+0.2cm, yshift=-0.4cm] {\scriptsize $-t$};
	\draw (0.4,0) node[xshift=0cm, yshift=-0.4cm] {\scriptsize $-\frac{t}{2}$};
	\draw (2.15,0) node[xshift=0cm, yshift=0.6cm] {$\ell_1$};
	\draw (-1.45,0) node[xshift=0cm, yshift=0.6cm] {$\ell_2$};
	\draw (1.0,1.35) node[xshift=0cm, yshift=0cm] {$C_{\bar{s}}$};
	\draw (-1.8,2.35) node[xshift=0cm, yshift=0cm] {$C_{\infty}$};
	
	\draw[thick, postaction={decorate}, decoration={markings, mark= at position 0.5 with {\arrow{<}}}] (1.4,+0.2) -- (3,+0.2);
        \draw[thick, postaction={decorate}, decoration={markings, mark= at position 0.5 with {\arrow{<}}}] (-2.2,+0.2) -- (-0.6,+0.2);
        			
\end{tikzpicture}
\caption{\emph{
Contour of integration $C$ in the complex $s$ plane associated with the dispersion relations \Eq{eq:dispersion}, along with the non-analyticities of the integrand.}}
\label{fig:splane}
\end{figure}
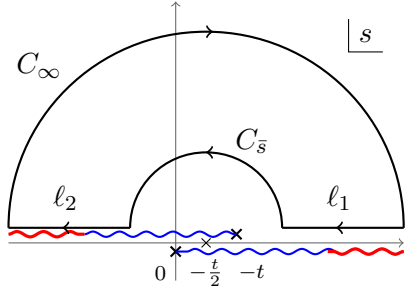

The closed contour of integration, $C$, is depicted in \Fig{fig:splane}, along with the discontinuities of $A(s,t)$, which consist of single poles and the $s$- and $u$-channel branch cuts. Note that the latter cover the entire real axis, overlap for $s \in [0,-t]$, and, for $s \gtrsim M^2$ and $s \lesssim -(M^2+t)$ (colored in red) lie outside the EFT range of predictivity. In \Eq{eq:dispersion}, we have split the dispersion relation into a low- and a high-energy representation, $B_n(\bar{s},t)|_{\ir}$ and $B_n(\bar{s},t)|_{\uv}$ respectively, which must cancel each other. The IR representation is given by the contour integral over the low energy half-arc $C_{\bar{s}}$, centered at $s = -t/2$ and with radius $\bar{s}+t/2$, where $\bar{s} < M^2$, 
\beq
B_n(\bar{s},t)|_{\ir} = \frac{1}{i \pi} \int_{C_{\bar{s}}} ds \frac{A(s,t)}{\left(s+\frac{t}{2}\right)^{2n+1}} \, .
\label{eq:irrep}
\eeq
Provided that the integral over the semicircle at infinity, $C_\infty$, vanishes, the UV representation is given by the integral over the lines $\ell_1$ and $\ell_2$, which can be combined into
\beq
B_n(\bar{s},t)|_{\uv} = \frac{2}{\pi} \int_{\bar{s}}^{\infty} ds \frac{\textrm{Im}[A(s,t)]}{\left(s+\frac{t}{2}\right)^{2n+1}} \, .
\label{eq:uvrep}
\eeq
Since we wish to be agnostic about UV physics, we can use the partial-wave decomposition, \Eq{eq:pwexp} (with stripped helicity weights), to write the UV representation as
\bea
B_{n \geqslant 1}^{4\phi}(\bar{s},t)|_{\uv}
&=& \frac{2}{\pi}  \sum_J n_J \int_{\bar{s}}^{\infty} ds \frac{\rho_J^{\phi\phi}(s) \,\tilde{d}^J_{0,0}\left(1+\frac{2t}{s}\right)}{\left(s+\frac{t}{2}\right)^{2n+1}} \,,
\label{eq:B4phi} \\
B_{n \geqslant 0}^{4\gamma}(\bar{s},t)|_{\uv}
&=& \frac{1}{\pi}  \sum_J n_J \int_{\bar{s}}^{\infty} ds \frac{\rho_J^{\gamma^+\gamma^+}(s) \, \tilde{d}^J_{0,0}\left(1+\frac{2t}{s}\right) + \rho_J^{\gamma^+\gamma^-}(s) \, \tilde{d}^J_{2,2}\left(1+\frac{2t}{s}\right)}{s^2 \left(s+\frac{t}{2}\right)^{2n+1}} \,, 
\label{eq:B4pho} \\
B_{n \geqslant -1}^{4h}(\bar{s},t)|_{\uv}
&=& \frac{1}{\pi}  \sum_J n_J \int_{\bar{s}}^{\infty} ds \frac{\rho_J^{h^+h^+}(s) \, \tilde{d}^J_{0,0}\left(1+\frac{2t}{s}\right) + \rho_J^{h^+h^-}(s) \, \tilde{d}^J_{4,4}\left(1+\frac{2t}{s}\right)}{s^4 \left(s+\frac{t}{2}\right)^{2n+1}} \,, 
\label{eq:B4h}
\eea
for scalars, photons, and gravitons, respectively. We have defined the spectral densities $\rho_J^{\phi\phi}(s) \equiv \textrm{Im}[a_J^{\phi\phi\phi\phi}(s)]$, $\rho_J^{\gamma^+\gamma^\pm}(s) \equiv \textrm{Im}[a_J^{\gamma^+\gamma^\pm \gamma^\mp \gamma^-}(s)]$, and $\rho_J^{h^+h^\pm}(s) \equiv \textrm{Im}[a_J^{h^+h^\pm h^\mp h^-}(s)]$. These are constrained by unitarity,
\beq
0 \leqslant \rho_J^{\phi\phi}(s), \, \rho_J^{\gamma^+\gamma^+}(s), \, \rho_J^{h^+h^+}(s) \leqslant 2 \,, \quad 0 \leqslant \rho_J^{\gamma^+\gamma^-}(s), \, \rho_J^{h^+h^-}(s) \leqslant 1 \, ,
\label{eq:densities}
\eeq
and satisfy that  $\rho_J^{\phi\phi}, \rho_J^{\gamma^+\gamma^+}, \rho_J^{h^+h^+} = 0$ for odd $J$, and $\rho_{J<2}^{\gamma^+\gamma^-}, \rho_{J<4}^{h^+h^-} = 0$.
In \EqsRange{eq:B4phi}{eq:B4h} we have also specified the number of subtractions ($n$) necessary for the vanishing of the contour at infinity given the assumed Regge behaviour.

Since our main interest lies in the leading operators of the scalar, photon, and graviton EFTs, we will focus in the following  on those dispersion relations that are sensitive at tree level to the corresponding Wilson coefficients, i.e., $n = 1$ for scalars and $n = 0$ for photons and gravitons.

\subsection{Background and Setup} \label{sec:previous}

Standard dispersion relations in the pure (shift-symmetric) scalar and photon EFTs, without gravity in particular, admit a well-defined forward limit that leads to a sum rule relating the corresponding leading coefficients to (an integral of) a cross section, implying their positivity, $c_2, \, \alpha_2 > 0$ \cite{Pham:1985cr,Adams:2006sv,Cheung:2014ega,Bellazzini:2016xrt,Bellazzini:2019xts,Arkani-Hamed:2020blm,Bellazzini:2020cot,Henriksson:2021ymi}. Likewise, EFT coefficients associated with terms that in the forward limit are proportional to an even power of $s$ can be shown to be positive as well, e.g., $c_4, \, \alpha_4 > 0$. 

At tree level, the inclusion of gravity precludes taking the forward limit in dispersion relations sensitive to the graviton $1/t$ pole, namely $n = 1$ for scalars and $n = 0$ for photons. 
For instance, in scalar gravitational scattering, the low energy representation reads
\beq
B_{1}^{4\phi}(\bar{s},t)|_{\ir} = -\frac{1}{\mpl^2 t} + 2 c_2 - c_3 t + O(t/\coff^2) \, 
\label{eq:B4phi1tree}
\eeq
where $O(t/\coff^2)$ encodes the contributions from the higher-derivative EFT terms in \Eq{eq:AEFTtreephi} that contain a term scaling as $s^2$. Tree-level IR representations such as \Eq{eq:B4phi1tree} can be improved so that they only depend on a finite number of EFT coefficients (e.g., $c_2$ and $c_3$) \cite{Sinha:2020win,Caron-Huot:2020cmc,Caron-Huot:2021rmr}.
Finite-$t$ partial improvements, which can be consistently implemented also in the presence of loop effects, have been developed in \cite{Beadle:2024hqg}.

By enforcing the positivity of the $n=1$ UV representation in \Eq{eq:B4phi}, the dispersion relation associated with \Eq{eq:B4phi1tree} has been shown to lead to the bound \cite{Caron-Huot:2021rmr,Chang:2025cxc}
\beq
c_2 \geqslant \frac{\zeta}{\coff^2 \mpl^2} \frac{1}{\epsilon} \,, \quad \epsilon < 0 \,.
\label{eq:c2treebound}
\eeq
This result is derived by smearing the sum rule over the momentum transfer $p = \sqrt{-t}$, which suppresses the graviton pole contribution relative to the EFT contributions -- rendering it finite rather than divergent as in $p = 0$ limit  -- while ensuring that the UV representation is positive \cite{Caron-Huot:2021rmr,Caron-Huot:2022ugt} (and the assumed Regge behaviour holds). Specifically, 
\beq
\int_0^q dp^{1-2\epsilon} \, \psi(p) \, B_{1}^{4\phi}(\bar{s},-p^2)|_{\ir} = \int_0^q dp^{1-2\epsilon} \, \psi(p) \, B_{1}^{4\phi}(\bar{s},-p^2)|_{\uv} \geqslant 0 \,,
\label{eq:smearedsr}
\eeq
where $\psi(p)$ is the smearing functional and $q$ the maximum transfer momentum considered.
This is, however, not possible to achieve in $d = 4$ for all impact parameters $b = 2J/\sqrt{s}$ in \Eq{eq:B4phi}, since positivity is compromised at large $b$, which explains why the lower bound scales as $1/\epsilon = (4-d)/2$ from the smearing of the graviton pole \cite{Chang:2025cxc}. 
In addition, let us note that the bound in \Eq{eq:c2treebound} depends on a numerical coefficient $\zeta$, which is determined by the details of the improvement and smearing procedures of the dispersion relation. In particular, if one is allowed to consider $q^2 \simeq \bar{s} \simeq M^2$, then $\zeta = O(10)$, while the bound typically degrades as $\zeta \sim (M/q)^2$ if $q^2 \lesssim \bar{s} \lesssim M^2$.
Tree-level bounds analogous to \Eq{eq:c2treebound} for photon scattering, $\alpha_2 \gtrsim (\coff \mpl)^{-2}/\epsilon$, were derived in \cite{Henriksson:2022oeu}. 

Gravitational loops introduce several important effects, adding to the generic subtleties that arise at loop level. Specifically, once loop effects are included, all EFT couplings enter any given dispersion relation \cite{Bellazzini:2020cot}. This complicates the construction of improved sum rules that are (mostly) sensitive to a finite number of coefficients -- which is important in order to derive robust, optimal bounds -- and typically requires the introduction of assumptions on the degree of convergence of the EFT.
One-loop gravitational corrections, in particular, introduce low-energy contributions that are singular in the forward limit (in $d \leqslant 6$), thereby precluding the use of forward limits and requiring either finite-$t$ improvements or the use of crossing-symmetric dispersion relations \cite{Beadle:2024hqg,Chang:2025cxc,Beadle:2025cdx}.

Importantly, these effects bring no qualitative changes to the tree-level bound on $c_2$ in \Eq{eq:c2treebound}.%
\footnote{This statement requires clarification. In $d = 4 -2\epsilon$ dimensions, it holds provided one preserves the gravitational decoupling limit $(M/4\pi\mpl)^2 \to 0$ even when approaching the four-dimensional limit $\epsilon \to 0$; in other words, the gravitational loop expansion parameter $(M/4\pi\mpl)^2/\epsilon$ is kept finite and small. While this prevents taking $d = 4$, \cite{Bellazzini:2025bay} has shown that, by working with suitably constructed IR-safe amplitudes, a sensible result in four dimensions is possible. In practice, IR divergences are replaced by finite logarithms controlled by the resolution with which soft gravitons can be measured; see Apps.~\ref{sec:irdivs}, \ref{sec:IRsafe} for further details.}
The optimal bound at one loop, deteriorates quantitatively, mainly because, in order to maintain EFT predictivity, the couplings (coefficients) must be evaluated at smaller energies and momentum transfers, namely $q^2 \lesssim \bar{s} \lesssim M^2$ in \Eq{eq:smearedsr}. 

More interestingly, gravitational loop effects generically extend the allowed negativity of $c_2$ to all higher-derivative EFT coefficients. 
For instance, while the positivity of $c_4$ in \Eq{eq:AEFTtreephi} remains at tree level even when $\mpl$ is finite, we have checked (see \Sec{sec:infrared} for details) that at one loop the bound parametrically becomes
\beq
c_4(\bar{s}) \geqslant 
- \frac{O(1)}{16 \pi^2 q^2 \bar{s} \mpl^4} \frac{1}{\epsilon^2} + \dots \,,
\label{eq:c4loopbound}
\eeq
where the ellipsis contain a number of a priori subleading corrections, of $O(q^2/\coff^2)$ and $O(\bar{s}^2/\coff^4)$ relative to the l.h.s.~from higher-derivative EFT terms and their $\beta$-functions, and of $O(\epsilon)$ relative to the r.h.s., as well as contributions of $O(c_2/(4\pi\mpl)^2\bar{s})$ and $O(1/(4\pi\mpl^2 \bar{s})^2)$.
In addition, note that we have explicitly included the dependence on $\bar{s}$ of $c_4$ from its RGE, $c_4(\bar{s}) = c_4 + ( ac_2^2 + bc_3/\mpl^2 ) \log(\bar{s}/\mu^2)$ with $a > 0$ and $b < 0$.

The situation of $c_4$ in scalar scattering bares some resemblance to the one of the leading EFT coefficient in graviton scattering, $g_4$.
At tree level, the positivity bound $g_4 > 0$ follows from the $n = 0$ sum rule in \Eq{eq:B4h} in the forward limit, which can be consistently taken given that the graviton pole does not contribute to the dispersion relation \cite{Bellazzini:2015cra,Caron-Huot:2022ugt}. In addition, there exists a tree-level lower bound on $g_4$ dependent on the non-minimal scalar-graviton three-point coupling \cite{Dong:2024omo}
\beq
g_4 \geqslant \frac{|c_{\sigma C^2}|^4 \coff^2}{O(100) \mpl^2} \epsilon \, .
\label{eq:g4treebound}
\eeq

In the following sections we will parametrically reproduce the tree- and one-loop bounds on the leading EFT amplitude in scalar scattering, and extend the tree-level bounds on the leading EFT coefficients for photons and gravitons to one loop. As we have discussed in \Sec{sec:betas}, the RGE of these coefficients can be negative if certain non-minimal three-point amplitudes are present, yet only if dynamical gravitons are involved. We will show that the tree-level gravitational contribution in \Eq{eq:c2treebound} is sufficient to preserve the positivity of the IR representation even after including such a negative running of $c_2$ or $\alpha_2$, as long as the number of light non-minimal coupled particles is bounded. In the case of gravitons, given that $g_4$ can run negative, we will see that one-loop gravitational corrections can restore the apparent inconsistency with the tree-level positivity bound.

As emphasized above, in this work we focus on parametric, rather than optimal, constraints. Accordingly, we will restrict our dispersion relations to the regime $-t \lesssim \bar{s} \ll \coff^2$, by which we mean that $i)$ all higher-$t$ contributions to the sum rules are assumed to be sufficiently suppressed -- in \Eq{eq:B4phi1tree}, this corresponds, in the notation of \Eq{eq:AEFTtreephigen}, to the conditions $c_{p,0} t^{2(p-1)}, -c_{p,1} t^{2p+1} \ll c_2$, and $ii)$ the perturbativity condition $\beta_n \bar{s}^n \ll c_m(\bar{s}) \bar{s}^m$ is satisfied \cite{Bellazzini:2020cot} for $n > m$ -- with $m = 2$ for scalars, 
such that loop corrections to the sum rules from higher-derivative operators are sufficiently suppressed, even for strongly coupled EFTs at the cutoff.%
\footnote{In addition, if the EFT is assumed to be weakly coupled, i.e., $\beta_n \bar{s}^n \ll c_m(\bar{s}) \bar{s}^m$ holds for $\bar{s} \simeq M^2$, the condition $\bar{s} \ll \coff^2$ can be relaxed (while keeping $-t \lesssim \coff^2$).}
These conditions are enough to ensure the parametric validity of our bounds near saturation.

Finally, we assume there exists a well-defined gravitational loop expansion in $d = 4 - 2\epsilon$ dimensions, i.e., 
\beq
\frac{\bar{s}}{(4\pi\mpl)^2} \frac{1}{\epsilon} \ll 1 \,.
\label{eq:gravexp}
\eeq
Note that, a priori, this condition precludes our analysis from applying in four dimensions. An extension along the lines of \cite{Bellazzini:2025bay}, which we present in \App{sec:IRsafe}, is required. Nonetheless, anticipating that the $1/\epsilon$ divergence is ultimately replaced by a physical, finite logarithm, the same parametric bounds should hold, which is all we aim to establish here.

Before turning to the details of our analysis, let us note that our bounds are similar in spirit to those derived in \cite{Caron-Huot:2024lbf}, in that the negative contributions to the IR representation of the dispersion relation cannot be attributed to modifications of the Regge behaviour. This is in contrast to, e.g., \cite{Tokuda:2020mlf,Herrero-Valea:2020wxz,deRham:2022gfe,Caron-Huot:2024tsk}, where negative low-energy contributions arise from light particles $m^2 \ll \bar{s}$ that are nevertheless heavier than the momentum transfer considered, $m^2 \gg -t$; in those cases, such contributions cancel between IR and UV.

\subsection{Smearing, IR regularization, and UV positivity} \label{sec:smear}

Positivity bounds crucially rely on showing that the UV representation is positive. This is non-trivial because the (helicity-stripped) Wigner $d$-functions in \EqsRange{eq:B4phi}{eq:B4h} oscillate around zero for all $s$ and $J$ at fixed $t \neq 0$. The proposal of \cite{Caron-Huot:2021rmr,Caron-Huot:2022ugt} to deal with this problem is to integrate the dispersion relation over momentum transfers $p = \sqrt{-t}$, weighted by a smearing functional $\psi(p)$ chosen such that
\beq
\int_0^q dp^{1-2\epsilon} \, \psi(p) \, B_n(\bar{s},-p^2)|_{\uv} \geqslant 0 \,.
\label{eq:smeareduv}
\eeq
Because of unitarity, in the form of \Eq{eq:densities}, ensuring that the high-energy representation is positive is tantamount -- whether for scalars, photons, or gravitons -- to identifying the functionals for which
\beq
\Psi_J(s) \equiv n_J^{(4-2\epsilon)} s^{\epsilon} \int_0^q dp^{1-2\epsilon} \frac{\psi(p) \,\tilde{d}^{J,-\epsilon}_{h,h}\left(1-2p^2/s\right)}{\left(1-p^2/2s\right)^{2n+1}} \geqslant 0 \,,
\label{eq:positive}
\eeq
since then the smeared high-energy representation in \Eq{eq:smeareduv} is manifestly positive,
\beq
\frac{2}{\pi} \sum_J \int_{\bar{s}}^{\infty} ds \frac{\rho_J(s) \Psi_J(s)}{s^{2n+1+m}} \geqslant 0 \,,
\eeq
with $m = 0, 2, 4$ for scalars, photons, and gravitons, respectively.

Because we work in $d = 4 - 2\epsilon$ throughout, \Eq{eq:positive} involves a $d$-dimensional extension of the partial-wave decomposition in \Eq{eq:pwexp}, without helicity weights.%
\footnote{This extension is not crucial as far as the positivity of the UV representation is concerned; the same conclusions can be reached with $\epsilon = 0$. We choose to work here in $d = 4 - 2 \epsilon$ not only for overall consistency, but also as a check that keeping $\epsilon \neq 0$ is innocuous.}
The extended Wigner $d$-functions are given in terms of Jacobi polynomials,
\beq
\tilde{d}^{J,\alpha}_{h,h'} (c_\theta) = \sqrt{\tfrac{\Gamma(1+\alpha)^2 \Gamma(1+J) \Gamma(1+J-h) \Gamma(1+J+h+2\alpha)}{\Gamma(1+J+2\alpha) \Gamma(1+J-h'+\alpha) \Gamma(1+J+h'+\alpha)}} \,
 P_{J-h}^{(h-h'+\alpha,h+h'+\alpha)}(c_\theta) \,, \,\,\, \alpha = \frac{d-4}{2} \,,
\label{eq:hstripWdd}
\eeq
and reduce to the standard $d$-functions for $\alpha = 0$ and to the usual Gegenbauer polynomials (up to a proportionality factor \cite{Vilenkin:1991aaa}) for $h = h' = 0$.%
\footnote{The extended Wigner $d$-functions can be derived by acting with raising and lowering operators on the $h = h' = 0$ Jacobi polynomials (basis of the scalar partial-wave expansion in $d$ dimensions) \cite{Vilenkin:1991aaa, Buric:2023ykg}. Their normalization is fixed such that
\beq
\int_0^\pi d\theta d^{J,(d-4)/2}_{h,h'} d^{J',(d-4)/2}_{h,h'} (\sin\theta)^{d-3} = \frac{\delta_{JJ'} \Omega_{d-2}}{n_J^{(d)} \Omega_{d-3}} \,.
\nonumber
\eeq}
The normalization of the partial-wave decomposition,
\beq
n_J^{(d)} = \frac{4 (4\pi)^{d-2} n_d(J)}{\Omega_{d-2}} = \frac{4(4\pi)^{d/2-1}(2J+d-3) \Gamma(J+d-3)}{\Gamma(1+J) \Gamma(d/2-1)} \,,
\eeq
together with the factor $s^{(d-4)/2}$ also appearing in \Eq{eq:positive}, is such that the unitarity of the $S$-matrix in $d$ dimensions remains as in \Eq{eq:densities} \cite{Hebbar:2020ukp,Caron-Huot:2021rmr}. Here, $n_d(J)$ is the number of states with angular momentum $J$ in $d$ dimensions, and $\Omega_{d-2}$ is the solid angle of the $(d-2)$-sphere.

It has been discussed in \cite{Caron-Huot:2021rmr,Caron-Huot:2022ugt,Haring:2022cyf} that, for \Eq{eq:positive} to hold, the smearing functional must satisfy
\bea
&& \psi(p) \overset{p \to 0}{\sim} a_1 p \, , \quad a_1 > 0 \,, \\
\label{eq:condpsi1}
&& \psi(p) \overset{p \to q}{\sim} \left( 1- \frac{p}{q} \right)^2 \,,
\label{eq:condpsi2}
\eea
as limiting behaviours around $p = 0$ and $p = q$. 
Both conditions can be shown to arise from positivity in the eikonal regime, i.e., in the limit of large $J$ and $s$ ($J\,, -s/t \to \infty$) with impact parameter $b = 2J/\sqrt{s}$ fixed. 

Let us analyze this limit in detail, given its particular connection with gravitational $1/\epsilon$ IR divergences in $d = 4-2\epsilon$ dimensions. We follow closely the analysis of \cite{Henriksson:2022oeu} and \cite{Chang:2025cxc}.
In the eikonal limit,
\beq
\lim_{s \to \infty} \Psi_{\frac{b\sqrt{s}}{2}}(s) = 8 (2\pi)^{1-\epsilon} b\sqrt{s} \int_0^q dp \, \psi(p) (bp)^{-\epsilon} J_{-\epsilon} (bp) \,,
\label{eq:fourierpsi}
\eeq
irrespective of whether the scattered states are scalars, photons, or gravitons as long as the scattering is elastic \cite{Bellazzini:2022wzv}, since the equal-spin Wigner $d$-functions become a Bessel function. 
In addition, note that the eikonal limit of $\Psi_J(s)$ corresponds to the Fourier transform of $\psi(p)$. Since this must be positive for all impact parameters, and therefore also its integrals over $b$, \Eq{eq:condpsi1} follows \cite{Henriksson:2022oeu}.

In this regard, it is important to point out that by working in $d = 4 - 2 \epsilon$ dimensions, the smearing of the IR representation in those cases where the graviton $1/p^2$ pole is present, see, e.g., \Eqs{eq:B4phi1tree}{eq:smearedsr}, is regularized even if $\psi(p) \sim p$ toward $p = 0$ ($a_1 = 1$),
\beq
\frac{1}{\mpl^2} \int_0^q dp^{1-2\epsilon} \frac{\psi(p)}{p^2} = - \frac{1}{2 \mpl^2 \epsilon} + O(\epsilon^0) \, ,
\eeq
where we note that the leading term is insensitive to the details of the smearing function at $O(p^{r > 1})$.
We will show below that gravitational loops are regularized in a similar fashion \cite{Chang:2025cxc}.
Besides, in contrast to regularizing the graviton pole in $d = 4$ with an IR cutoff on the range of momentum transfers, $p > 1/m_{\ir}$ \cite{Caron-Huot:2021rmr,Caron-Huot:2022ugt,Henriksson:2022oeu}, working in dimensional regularization has the advantage of preserving positivity at large impact parameters.

Indeed, for $\psi(p)$ a generic polynomial in $p$, the leading behaviour of \Eq{eq:fourierpsi} at large impact parameters is oscillatory. Let us consider one such polynomial term,
\beq
\lim_{s \to \infty} \Psi_{\frac{b\sqrt{s}}{2}}(s) |_{\psi(p) = p^r} \sim \frac{1}{(bq)^r} \frac{2^r \Gamma(\frac{1+r}{2}-\epsilon)}{\Gamma(\frac{1-r}{2})} 
- \frac{\cos(bq+\frac{\pi\epsilon}{2})-\sin(bq+\frac{\pi\epsilon}{2})}{\sqrt{\pi}(bq)^{1/2+\epsilon}} 
+ \dots \,,
\label{eq:fourierpr}
\eeq
where subleading terms, encoded in the ellipsis, are of order $1/(bq)^{(3/2, \, 5/2, \,\dots)+\epsilon}$ and oscillatory as well. 
This illustrates the need for the funcional to be sufficiently suppressed around the maximum momentum transfer involved in the smearing, in particular given that the non-oscillatory term vanishes for odd $r$, and $r \geqslant 1$ as explained above. The condition in \Eq{eq:condpsi2} successfully achieves this, which we implement by restricting the set of functionals to those of the form \cite{Henriksson:2022oeu}
\beq
\psi(p) = q \sum_{r=1}^{7} a_r \left(\frac{p}{q}\right)^r \left(1 + \frac{r-5}{2} \left(\frac{p}{q}\right)^{3-r} - \frac{r-3}{2} \left(\frac{p}{q}\right)^{5-r} \right) \,.
\label{eq:psi}
\eeq
In this way, the expansion of \Eq{eq:fourierpsi} at large impact parameter limit is controlled by a term $\sim - a_2/(bq)^2$, which is positive as long as $a_2 < 0$, while the oscillating terms scale as $1/(bq)^{5/2-\epsilon}$. Note that the fact that we have kept $\epsilon \neq 0$ does not affect this conclusion.
In addition, for our purposes it is sufficient to consider polynomials up to $r = 7$.

Certainly, to fully ensure the positivity of the UV representation, regimes other than the eikonal must be considered: finite $J$ and $s$, finite $J$ and large $s$, and large $J$ and finite $s$. We relegate the discussion of how enforcing UV positivity in these regions constrains the smearing functional to \App{sec:details}, and refer the reader to \cite{Henriksson:2022oeu,Dong:2024omo} for further details.

\subsection{One-loop infrared representation and bounds} \label{sec:infrared}

In the following, we evaluate the IR representation, \Eq{eq:irrep}, of the dispersion relations that are sensitive at tree level to the leading EFT coefficients for scalars ($n = 1$), photons ($n = 0$), and gravitons ($n = 0$). We include minimal gravitational couplings, as well as the non-minimal three-point couplings that induce a negative running of $c_2$, $\alpha_2$, and $g_4$. The corresponding one-loop amplitudes can be found in \App{sec:loop}. We then present our parametric (non-optimal) smeared bounds. Before doing so, a few general remarks, applicable to all the amplitudes under consideration, are in order. 

Since our goal is to derive a lower bound on the aforementioned coefficients, we require that their smeared low-energy contribution is positive. We do not require that the remaining contributions be minimized, as we are not aiming at optimal bounds.

For ease of exposition, the IR representation is given as an expansion in $-t/\bar{s}$, and in the numerical analysis leading to a suitable smearing functional we keep a hierarchy $p^2/\bar{s} \ll 1$. 
While this is not required in general, it may be useful if the EFT is weakly coupled at the cutoff, since loop effects from higher-derivative contact interactions are then subleading even when evaluated at $\bar{s} = M^2$, while the hierarchy $p^2/\coff^2 \ll 1$ must still be maintained. Note that the positivity of the UV representation holds irrespective of whether $\bar{s} \ll \coff^2$ or $\bar{s} \simeq \coff^2$.

Finally, we identify the running EFT coefficients as
\bea
c_2(\bar{s}) &=& c_2 + \frac{a_2}{2} \log \Big(\frac{\bar{s}}{\mu^2}\Big) \, , \\ 
\alpha_2(\bar{s}) &=& \alpha_2 + \frac{\gamma_2}{2} \log \Big(\frac{\bar{s}}{\mu^2}\Big) \, , \\
g_4(\bar{s}) &=& g_4 + \frac{\beta_4}{2} \log \Big(\frac{\bar{s}}{\mu^2}\Big) \, ,
\eea
where the $1/\epsilon$ UV divergence along with any finite (polynomial) contribution of the one-loop amplitude is cancelled by the corresponding counterterm.

\subsubsection{Scalar} \label{sec:IRscalar}

The low-energy representation of the spin-2 subtracted dispersion relation for the 4-point (shift-symmetric) scalar amplitude reads, at tree level,
\beq
B_{1}^{4\phi}(\bar{s},t)|_{\ir}^\tree = -\frac{1}{\mpl^2 t} + 2 c_2 - c_3 t + O(t/\coff^2) \, .
\eeq
For the one-loop contribution from gravitons and vectors, including GR and $\phi V \tilde{V}$ interactions, we find
\bea
B_{1}^{4\phi}(\bar{s},t)|_{\ir}^\loop &=&
\frac{1}{8 \pi^2 \mpl^4} \bigg( - \frac{2 \bar{s} (-t/\mu^2)^{-\epsilon}}{\epsilon t} + \log_{-\bar{s}/t} \Big( \frac{1}{\epsilon} - \log_{-t} + \log_{-\bar{s}/t} \Big) + \frac{7}{6} \pi^2 \bigg) \nonumber \\
&& + \, a_2 \log_{\bar{s}} - \frac{b_t|_{t = 0}}{16\pi^2} \log_{-\bar{s}/t} + O(-t/\bar{s})+O(\bar{s}^2/\coff^4) \, .
\label{eq:sumrulephi}
\eea
To avoid unnecessary clutter, we used the shorthand $\log(-\bar{s}/t)$, $\log(-t/\mu^2)$, and $\log(\bar{s}/\mu^2)$, $a_2$ is given in \Eq{eq:betac2neg}, and $b_t|_{t = 0}$ is the $t$-channel bubble coefficient in the forward limit, $b_t|_{t = 0} = -(283 + N_V (12 - 120 c_{\phi V^2}^2 \mpl^2))/60\mpl^4$.
We kept the factor $(-t)^{-\epsilon}/t$ unexpanded in $\epsilon$, since, upon smearing it, all terms of its would-be expansion 
contribute equally. 
We also recall that the $1/\epsilon$ pole is associated with the soft gravitational divergences (see \App{sec:irdivs}).

After smearing the sum rule with the functional \Eq{eq:psi}, we arrive at the following lower bound on the running coefficient of the leading EFT scalar amplitude,
\beq
c_2(\bar{s}) \geqslant \frac{\zeta}{2 q^2 \mpl^2} \frac{1}{\epsilon} \left( 1 + \frac{2\bar{s}}{\epsilon(4\pi\mpl)^2} \right) + \dots \,, \qquad \epsilon < 0
\label{eq:c2loopbound}
\eeq
where the ellipsis contain subleading corrections, from the sum rule \Eq{eq:sumrulephi} of $O(q^2/\coff^2)$, $O(\bar{s}^2/\coff^4)$ and 
$O(1)$ yet not logarithmically enhanced relative to the l.h.s., and of $O(\epsilon)$ and $O(\epsilon q^2/\bar{s})$, times a $\log q^2$ or $\log\bar{s}$, 
relative to the r.h.s.. 
The numerical factor we obtain from the smearing procedure is $\zeta \approx 90$.%
\footnote{Specifically, we find numerically that the parameters $a_1 = 1$, $a_2 \approx -8.845$, $a_4 \approx -0.152$, $a_6 \approx 62.04$, and $a_7 \approx -21.57$, for $q^2/\bar{s} \approx 0.1$, satisfy all the necessary requirements. We have also checked that, for smaller values of $q^2/\bar{s}$, $\zeta$ remains of $O(100)$.}
As shown in \cite{Chang:2025cxc}, the allowed negativity of $c_2$ due to tree-level gravity is modified at one (graviton) loop, with the correction being proportional to the loop-expansion parameter $(\sqrt{\bar{s}}/4\pi\mpl)^2/\epsilon$. 
Note that the one-loop divergence, of $O(1/\epsilon^2)$, arises solely from the first term in \Eq{eq:sumrulephi}, since this is the only one with a $1/t$ pole.
We consider such a correction to be subleading, following the discussion around \Eq{eq:gravexp} and in line with 
\cite{Bellazzini:2025bay}, where a physical interpretation of this bound in $d = 4$ is derived. As shown in \App{sec:IRsafe}, sum rules constructed from IR-finite amplitudes lead to a bound in which the $1/\epsilon$ terms are replaced by finite logarithms.

In the situation where $c_2$ is allowed to evolve down for sufficiently long, such that its boundary condition (or matching contribution), $c_2(\coff^2)$, becomes irrelevant, \Eq{eq:c2loopbound} can be approximated by $a_2 \log(\coff^2/\bar{s}) \lesssim - \zeta/q^2\mpl^2\epsilon$.
In the large-$N_V$ limit and maximizing the $\beta$-function by setting $c_{\phi V^2}^2 = 2/3\mpl^2$, one then obtains the bound
\beq
N_V \lesssim O(1.4 \times 10^4) \frac{(4\pi\mpl)^2}{\coff^2} \, ,
\label{eq:NVphibound}
\eeq
where we have taken $-\epsilon \log(\coff^2/\bar{s}) \sim 1$ and $q^2/\coff^2 = 1/100$.

\subsubsection{Photon} \label{sec:IRphoton}

The low-energy representation of the spin-2 subtracted dispersion relation for the 4-point MHV photon amplitude reads, at tree level,%
\footnote{We note that a 4-photon MHV amplitude mediated by two insertions of a non-minimal photon-graviton $\amp(1_{\gamma^+},2_{\gamma^+},3_{h^+})$ contributes to the sum rule as $\alpha_3$ does.}
\beq
B_{0}^{4\gamma}(\bar{s},t)|_{\ir}^\tree =  -\frac{1}{\mpl^2 t} + \alpha_2 - \frac{\alpha_3 t}{2} + O(t/\coff^2) \, .
\eeq
The one-loop contribution from gravitons and scalar-vector pairs with GR and $\sigma F^2$ interactions -- a similar analysis holds for the one-loop of fermions coupled via $\bar{\chi} F \chi$ -- reads
\bea
B_{0}^{4\gamma}(\bar{s},t)|_{\ir}^\loop &=&
\frac{1}{8 \pi^2 \mpl^4} \bigg( - \frac{2 \bar{s} (-t/\mu^2)^{-\epsilon}}{\epsilon t} + \log_{-\bar{s}/t} \Big( \frac{1}{\epsilon} - \log_{-t} + \log_{-\bar{s}/t} \Big) + \frac{7}{6} \pi^2 \bigg) \nonumber \\
&& + \, \frac{\gamma_2}{2} \log_{\bar{s}} - \frac{b_t|_{t = 0}}{16\pi^2} \log_{-\bar{s}/t} + O(-t/\bar{s})+O(\bar{s}^2/\coff^4) \,,
\label{eq:sumrulepho}
\eea
where $\gamma_2$ is given in \Eq{eq:betaalpha2negda}, and $b_t|_{t = 0}$ is the $t$-channel bubble coefficient in the forward limit, $b_t|_{t = 0} = -(7 + N_V (7 - 40 c_{\phi F^2}^2 \mpl^2))/30\mpl^4$.
Note that both the tree-level and one-loop graviton contributions to the IR representation are the same as in the scalar case.

After smearing the dispersion relation, a bound on the running coefficient of the leading EFT photon amplitude is derived,
\beq
\alpha_2(\bar{s}) \geqslant \frac{\zeta}{q^2 \mpl^2} \frac{1}{\epsilon} \left( 1 + \frac{2\bar{s}}{\epsilon(4\pi\mpl)^2} \right) + \dots \,, \qquad \epsilon < 0
\label{eq:alpha2loopbound}
\eeq
where the ellipsis contain the same type of subleading corrections as in the scalar case.
The numerical factor from our smearing procedure is $\zeta \approx 70$.
Analogously to the scalar case, the one-loop modification of the allowed tree-level negativity of $\alpha_2$ is proportional to the gravitational loop-expansion parameter.

Assuming a sufficiently large hierarchy between the cutoff and the scale where the coefficient $\alpha_2$ is measured, such that this is dominated by its low-energy running, we derive the following bound on the number $N_V$ of vectors and dilaton/axion pairs,
\beq
N_V \lesssim O(3.5 \times 10^4) \frac{(4\pi\mpl)^2}{\coff^2} \, ,
\label{eq:NVphobound}
\eeq
in the large-$N_V$ limit and maximizing the $\beta$-function by setting $c_{\phi F^2}^2 = 1/2\mpl^2$. As in the scalar case, we have taken $-\epsilon \log(\coff^2/\bar{s}) \sim 1$ and $q^2/\coff^2 = 1/100$.

\subsubsection{Graviton} \label{sec:IRgraviton}

The low-energy representation of the spin-4 subtracted dispersion relation for the 4-point MHV graviton amplitude reads, at tree level,
\beq
\mpl^4 \, B_{0}^{4h}(\bar{s},t)|_{\ir}^\tree = g_4 - \left( \frac{g_3^2}{\mpl^2} + \frac{g_5}{2} \right) t + O(t/\coff^2) \, .
\eeq
For the one-loop contribution from gravitons and scalars, including GR and $\sigma C^2$ couplings, we find
\bea
\mpl^4 \, B_{0}^{4h}(\bar{s},t)|_{\ir}^\loop &=&
\frac{1}{4 \pi^2} \bigg( \frac{(-t/\mu^2)^{-\epsilon}}{\epsilon \bar{s} t} - \frac{3}{4 \bar{s}^2} \Big( \frac{1}{\epsilon} - \log_{-t} + \frac{2}{3} \log_{-\bar{s}/t} - \frac{402+N_\sigma}{360} \Big) \bigg) \nonumber \\
&& + \, \frac{\beta_4}{2} \log_{\bar{s}} + O(-t/\bar{s})+O(\bar{s}^2/\coff^4) \,,
\label{eq:sumruleh}
\eea
where $\beta_4$ is given in \Eq{eq:betag4pos} (in this case $b_t|_{t = 0} = 0$). Note that $O(c_{\sigma C^2}^4)$ contributions (see also \Eq{eq:g4treebound}) are relatively suppressed by $c_{\sigma C^2}^2 \bar{s}^2/\mpl^2$ and are therefore encoded in the subleading $O(\bar{s}^2/\coff^4)$ term.
Importantly, in contrast with the scalar and photon cases, there is no gravitational term at tree level, and the one-loop contribution is positive. 

After smearing, we obtain the following bound on the running coefficient of the leading EFT graviton amplitude,
\beq
g_4(\bar{s}) \geqslant - \frac{\zeta}{8 \pi^2 q^2 \bar{s}} \frac{1}{\epsilon^2} - \frac{\zeta' N_\sigma}{1920 \pi^2 \bar{s}^2}
+ \dots \,,
\label{eq:g4loopbound}
\eeq
where the ellipsis denote subleading corrections, from the sum rule \Eq{eq:sumruleh} of $O(q^2/\coff^2)$, $O(\bar{s}^2/\coff^4)$ relative to the l.h.s., and of $O(\epsilon)$, $O(\epsilon q^2/\bar{s})$, multiplied by $\log q^2$, $\log\bar{s}$, relative to the r.h.s.. We explicitly display the one such contribution that scales with the number of scalars.
The numerical factors we obtain from our smearing functional are $\zeta \approx 70 $ and $\zeta' = 0.75$.

We thus find that the one-loop contribution from GR controls the leading allowed negativity of $g_4$.
For finite $\epsilon$, this is of the same order of magnitude as a (positive) one-loop contribution from a single minimally coupled heavy particle of mass $M \sim (q^2 \bar{s})^{1/4}$ \cite{Caron-Huot:2022ugt}.

\Eq{eq:g4loopbound} restores consistency with the fact that $g_4(\bar{s})$ can decrease toward low energies, provided that
\beq
N_\sigma c_{\sigma C^2}^2 \lesssim O(600) \frac{(4\pi\mpl)^2}{\coff^4} \, ,
\label{eq:Nsigmahbound}
\eeq
where we have taken $-\epsilon \log(\coff^2/\bar{s}) \sim 1$ and $q^2/\bar{s} \simeq \bar{s}/M^2 = 1/10$ (such that $q^2/\coff^2 = 1/100$ as in the previous cases), and neglected the finite term.
The same parametric bound follows from demanding the absence of time advances in a gravitational background \cite{Serra:2022pzl}.

\section{Conclusions} \label{sec:conclusions}

In this work we have studied in generality the gravitationally induced one-loop running of the most relevant amplitudes in the shift-symmetric scalar, photon, and graviton EFTs.
At energies sufficiently below the cutoff, renormalization group evolution is expected to control the associated Wilson coefficients, rendering direct matching contributions irrelevant.

We have found that in specific EFTs with a moderately large number of new particles non-minimally coupled to the scalar, photon, or graviton, running can drive the aforementioned coefficients to smaller values toward the infrared, yet only when the corresponding positive $\beta$-functions are suppressed by the Planck scale.
We have traced the origin of this result to the $t$-channel exchange of gravitons in one-loop elastic processes, see \Fig{fig:cutforward}. In the gravitational decoupling limit, any contribution to the $t$-channel branch-cut discontinuity vanishes in the forward limit, rendering the total contribution to the running manifestly positive. This is no longer the case once dynamical gravitons are included.

We have examined the consistency of these findings with what can be inferred solely from the assumptions of unitarity, locality, and causality. Quite nicely, we have shown that the graviton $1/t$ pole is also responsible for the positive contributions to the dispersion relations necessary to preserve positivity in the ultraviolet.
In the scalar and photon cases, we conclude that the magnitude of the negative running is consistent with the amount of negativity allowed by tree-level graviton exchange in GR, provided that the number of non-minimally coupled states is bounded by the species scale, $N \lesssim (4\pi\mpl/\coff)^2$.
In the graviton case, the positive contributions required to counterbalance the negative running arise only at one loop, as predicted by the universal soft divergences associated with minimal gravitational coupling. They are of the correct size as long as the number and interaction strength of the non-minimally coupled scalars satisfy $N c_{\sigma C^2}^2 \lesssim (4\pi\mpl/\coff^2)^2$.

While the bounds on the number of species that we have obtained are not new \cite{Veneziano:2001ah,Dvali:2007hz}, the specific dispersion relations used to derive them are. Admittedly, our bounds are not as general as those derived in \cite{Caron-Huot:2024lbf}. However, we have shown that a large number of light, non-minimally coupled particles underlies the saturation of the maximal gravitationally allowed negativity of the leading coefficients in the scalar, photon, and graviton EFTs.

Our results constitute an example of the fact that loop effects can drive EFT coefficients outside of the region allowed by tree-level positivity constraints \cite{Bellazzini:2020cot,Arkani-Hamed:2020blm,Bellazzini:2021oaj,Chala:2021wpj}, of relevance in the context of the swampland program (e.g., with implications for the weak gravity conjecture \cite{Arkani-Hamed:2021ajd}).
While in this work we have focussed on gravitational theories, understood as those in which the only minimal coupling is that of GR, we are aware of other sources of negative running in the photon and graviton EFTs that do not involved minimally coupled gravitons but instead arise from other minimal interactions. These will be discussed in a separate publication \cite{Fernandez2026}.


\section*{Acknowledgments}

We thank Brando Bellazzini, Stefano De Angelis, Adam Falkowski, Agustin Sabio Vera, and Francesco Riva for useful conversations. The work of J.S.~has been supported by the following grants: RYC-2020-028992-I, PID2022-142545NB-C22 and CNS2023-145069, funded by MICIU/AEI/10.13039/501100011033 and by the European Union ESF/EU, ERDF/EU, and NextGenerationEU/PRTR, CSIC-20223AT023, and ``IFT Centro de Excelencia Severo Ochoa CEX2020-001007-S''. 
J.F.~is supported by a Ph.D.~contract ``contrato predoctoral para formaci\'on de doctores'' (PRE2021-099269) associated with the aforementioned Severo Ochoa grant CEX2020-001007-S-21-4. 
M.R.~is supported by NSF Grant PHY-2310429, Simons Investigator Award No.~824870, DOE HEP QuantISED award \#100495, the Gordon and Betty Moore Foundation Grant GBMF7946, and the U.S.~Department of Energy (DOE), Office of Science, National Quantum Information Science Research Centers, Superconducting Quantum Materials and Systems Center (SQMS) under contract No.~DEAC02-07CH11359.


\appendix

\section{SUGRA}\label{sec:sugra}

Introduction of massless gravitini, $\psi_i$, is particularly constrained by consistent factorization of four-point amplitudes, leading to supersymmetric theories \cite{McGady:2013sga} (see also \cite{Bellazzini:2025shd,Gherghetta:2025tlx}). In this appendix, we summarize the relevant results for the anomalous dimensions of the leading EFT correction to the MHV four-photon and four-graviton amplitudes in such theories. 

The relevant minimal gravitational three-point amplitudes are
\bea
\amp^\gr(1_{h^+},2_{\sigma_i},3_{\sigma_i}) &=& -\frac{1}{\mpl}\frac{\sq{12}^2 \sq{13}^2}{\sq{23}^2} \,, 
\label{eq:h1sigma2GR}\\
\amp^\gr(1_{h^+},2_{\chi_i^+},3_{\chi_i^-}) &=& \frac{1}{\mpl}\frac{\sq{12}^3 \sq{13}}{\sq{23}^2} \,, 
\label{eq:h1chi2GR}\\
\amp^\gr(1_{h^+},2_{V_i^+},3_{V_i^-}) &=& \frac{1}{\mpl}\frac{\sq{12}^4}{\sq{23}^2} \,, 
\label{eq:h1V2GR}\\
\amp^\gr(1_{h^+},2_{\psi_i^+},3_{\psi_i^-}) &=& -\frac{1}{\mpl}\frac{\sq{12}^5}{\sq{13} \sq{23}^2} \,, 
\label{eq:h1psi2GR}\\
\amp^\gr(1_{h^+},2_{h^+},3_{h^-}) &=& -\frac{1}{\mpl}\frac{\sq{12}^6}{\sq{13}^2 \sq{23}^2} \,.
\label{eq:h3GR}
\eea
Additionally, minimal amplitudes involving gravitini may be obtained using Supersymmetric Ward Identities (SWI). For instance, in $\mathcal{N}=1$ SUGRA, we find a three-point amplitude involving the photon, the gravitino, and a photino $\lambda$,
\beq
\amp^{\mathcal{N}=1}(1_{\gamma^+},2_{\psi^+},3_{\lambda^-}) = \frac{1}{\mpl}\frac{\sq{12}^3}{\sq{13}} \,.
\eeq

With the four-point amplitudes,
\bea
\amp^{\mathcal{N}=1}(1_{\gamma^+},2_{\gamma^-},3_{\lambda^+},4_{\lambda^-}) &=& - \frac{1}{\mpl^2} [1|3|2\rangle \sq{13} \an{24} \left( \frac{1}{s} + \frac{1}{u} \right)\,, \\
\amp^{\mathcal{N}=1}(1_{\gamma^+},2_{\gamma^-},3_{\psi^+},4_{\psi^-}) &=&  \frac{1}{\mpl^2} [3|2|4\rangle \sq{13}^2 \an{24}^2 \frac{1}{st} \,, 
\eea
and following the procedure of \Sec{sec:betas}, we obtain the $\beta$-function for $\alpha_2$ in $\mathcal{N}=1$ SUGRA \cite{vanNieuwenhuizen:1976bg},
\beq
\gamma_2 = 
- \frac{25}{26\pi^2} \frac{1}{\mpl^4} < 0 \,.
\label{eq:betaalpha2N1SUGRA}
\eeq
While this is larger than that in \Eq{eq:betaalpha2GR}, it is not possible to reverse the sign with minimal couplings. 

The inclusion of non-minimal couplings for chiral multiplets was considered in \cite{Arkani-Hamed:2021ajd}, where the $\beta$-function was shown to remain negative.
Specifically, adding $N_{\sigma\chi}$ chiral multiplets, formed by a complex scalar $\sigma$ and a fermion $\chi$, with the dipole- and dilaton-like interactions given in \Eqs{eq:dipole}{eq:dilaton},%
\footnote{In the notation of \Sec{sec:photon}, $\chi_1 = \chi$, $\chi_2 = \lambda$, and the scalar is decomposed into a dilaton and an axion, $\sigma_{1,2}$. In this way, $c_{F\chi^2}^{12} = g_{\sigma\chi}$ and $c_{\sigma F^2}^{k} = i^{k-1} g_{\sigma\chi}/\sqrt{2}$.}
\bea
\amp^\nm(1_{\gamma^+},2_{\gamma^+},3_{\sigma}) &=& \frac{g_{\sigma\chi}}{\mpl}\sq{12}^2\,, \\
\amp^\nm(1_{\gamma^+},3_{\chi^+},3_{\lambda^+}) &=& \frac{g_{\sigma\chi}}{\mpl}\sq{12}\sq{13}\,, 
\eea
leads to
\beq
\gamma_2 = 
- \frac{1}{4\pi^2} \frac{1}{\mpl^4} \bigg( \frac{25}{6} + N_{\sigma\chi} \bigg(\frac{1}{12} -2g_{\sigma\chi}^2+\frac{3}{4}N_{\sigma\chi}g_{\sigma\chi}^4 \bigg) \bigg) < 0 \,.
\label{eq:betaalpha2N1SUGRANM}
\eeq
This result differs from that in \cite{Arkani-Hamed:2021ajd}; we have checked that the four-photino elastic amplitude exhibits the appropriate running according to SWIs and \Eq{eq:betaalpha2N1SUGRANM}.

In extended SUGRA, the four-graviphoton $\alpha_2$ coupling does not run, since the four-graviton amplitude is not renormalized by minimal couplings. The $\mathcal{N}=2$ case was studied in \cite{Arkani-Hamed:2021ajd}, which showed how SUGRA contributions to the $\beta$-function cancel. For the Maxwell multiplet and the hypermultiplet, the inclusion of non-minimal interactions like those of \Sec{sec:photon}, with appropriately tuned couplings as dictated by supersymmetry, is necessary to cancel the contributions from minimal interactions. The supergravity multiplet yields a vanishing anomalous dimension when the coupling of the graviphoton to the gravitini is included,
\beq
\amp^{\mathcal{N}=2}(1_{\gamma^-},2_{\psi_1^+},3_{\psi_2^+}) = -\frac{1}{\mpl}\frac{\sq{23}^4}{\sq{12}\sq{13}} \,.
\eeq
This gives rise to the four-point amplitude
\beq
\amp^{\mathcal{N}=2}(1_{\gamma^+},2_{\gamma^-},3_{\psi_i^+},4_{\psi_i^-}) = -\frac{1}{\mpl^2} [3|2|4\rangle \sq{13}^2 \an{24}^2 \frac{1}{su} \,,
\eeq
whose corresponding double cut cancels the first term in \Eq{eq:betaalpha2GR}.

Finally, let us comment on the effects of a non-minimal three-point interaction between a graviton, a gravitino, and a fermion. In $\mathcal{N}=1$,
\beq
\amp(1_{h^+},2_{\psi^+},3_{\chi_i^+}) = -\frac{c_{\sigma C^2}^i}{\mpl^2} \sq{12}^3 \sq{13}\,,
\label{eq:hpsichi}
\eeq
corresponds to the supersymmetric version of \Eq{eq:phiC2}, as derived from SWIs. Since a $\sigma C^2$ interaction contributes positively to the leading EFT coefficient of the MHV four-graviton amplitude, see \Eq{eq:betag4pos},%
\footnote{Note that in this case the scalar is complex, thus the number of degrees of freedom doubles.}
it is interesting to consider the contributions associated with \Eq{eq:hpsichi} to the running of $\beta_4$.

The four-point amplitudes that determine the relevant double cuts are
\bea
\amp(1_{h^+},2_{h^-},3_{\chi_i^+},4_{\chi_j^-}) &=&  [1|3|2\rangle^3 \sq{13} \an{24} \left( -\frac{\delta_{ij}}{\mpl^2}\frac{1}{stu} + \frac{c_{\sigma C^2}^ic_{\sigma C^2}^{j\,*}}{\mpl^4}\frac{1}{t} \right)\,, \\
\amp(1_{h^+},2_{h^-},3_{\psi^+},4_{\psi^-}) &=&  [1|3|2\rangle \sq{13}^3 \an{24}^3 \left( \frac{1}{\mpl^2}\frac{1}{stu} - \sum_i^{N_{\sigma\chi}}\frac{|c_{\sigma C^2}^i|^2}{\mpl^4}\frac{1}{t} \right)\,, \\
\amp(1_{h^+},2_{h^-},3_{\psi^+},4_{\chi_i^+}) &=& \frac{c_{\sigma C^2}^i}{\mpl^3} \langle2|1|3|2\rangle^2 \sq{13}^3 \sq{14} \frac{1}{stu} \,, 
\eea
where we have considered $N_{\sigma\chi}$ chiral multiplets. Taking the same coupling $c_{\sigma C^2}$ for all of them,  \Eq{eq:betag4pos} is modified as
\beq
\beta_4 = - \frac{1}{4 \pi^2} \frac{1}{\mpl^2} N_{\sigma\chi} |c_{\sigma C^2}|^2 \left( - \frac{52}{15} + \frac{7}{30} \right)=\frac{97}{120 \pi^2} \frac{1}{\mpl^2} N_{\sigma\chi} |c_{\sigma C^2}|^2\, .
\eeq
We observe that, while gravitini give rise to a negative contribution, this is insufficient to reverse the sign of $\beta_4$.

\section{Bubbles, triangles and boxes}\label{sec:bubble}

Via the Passarino-Veltman decomposition, any massless four-point one-loop amplitude can be expressed as a linear combination of bubble, triangle, and box integrals,
\beq
\amp_\loop = \sum_{i} b_i I_2^{(i)} + \sum_{i} c_i I_3^{(i)} + \sum_{j} d_j I_4^{(j)} + R \,,
\label{eq:pv}
\eeq
where $i = s, t, u$, $j = st, su, tu$, $R$ is a rational term, and the basis integrals, 
\beq
I_n = (-1)^n \tilde{\mu}^{4 - d} \int\frac{d^d\ell}{(2\pi)^d} \frac{-i}{\ell^2 (\ell - p_1)^2 \cdots (\ell+p_n)^2} \,,
\eeq
where $d = 4 - 2 \epsilon$ and $\tilde{\mu}^2 = e^{\gamma_{E}} \mu^2/4\pi$.
These integrals evaluate to
\bea
I_2(s) &=& r_2(\epsilon) \tilde{\mu}^{2\epsilon} \frac{(-s)^{-\epsilon}}{\epsilon} \,, \\
I_3(s) &=& -r_3(\epsilon) \tilde{\mu}^{2\epsilon} \frac{(-s)^{-\epsilon}}{s \, \epsilon^2} \,, \\
I_4(s,t) &=& r_4(\epsilon) \frac{1}{st} \left(2\tilde{\mu}^{2\epsilon} \frac{(-s)^{-\epsilon} + (-t)^{-\epsilon}}{\epsilon^2} - \log^2(s/t) - \pi^2\right) \,, 
\eea
for $s, t < 0$ and $r_3(\epsilon) = r_4(\epsilon) = r_2(\epsilon) (1-2\epsilon)$, with
\beq
r_2(\epsilon) = \frac{\Gamma(1-\epsilon)^2 \Gamma(1+\epsilon)}{(4\pi)^{2-\epsilon} \Gamma(2-2\epsilon)} \,.
\eeq
In an expansion in $\epsilon$, the expressions read
\bea
I_2(s) &=& \frac{1}{16\pi^2} \left(\frac{1}{\epsilon} - \log\Big(\frac{-s}{\mu^2}\Big)  \right) \,, \\
I_3(s) &=& -\frac{1}{32\pi^2} \frac{1}{s} \left(\frac{2}{\epsilon^2} - \frac{2}{\epsilon} \log\Big(\frac{-s}{\mu^2}\Big) + \log^2\Big(\frac{-s}{\mu^2}\Big) \right) \,, \\
I_4(s,t) &=& \frac{1}{8\pi^2} \frac{1}{st} \left(\frac{2}{\epsilon^2} - \frac{1}{\epsilon} \left( \log\Big(\frac{-s}{\mu^2}\Big) + \log\Big(\frac{-t}{\mu^2}\Big) \right) + \log\Big(\frac{-s}{\mu^2}\Big) \log\Big(\frac{-t}{\mu^2}\Big) - \frac{\pi^2}{2} \right) \,. \quad 
\eea
We recall that only the bubble integral is UV divergent (regulated by $\epsilon > 0$), while the triangle and box integrals are IR divergent (and require $\epsilon < 0$).
Consequently, anomalous dimensions are determined by the rational coefficients of the bubble integrals, which can be extracted from their double cuts, obtained by putting the corresponding propagators on shell.
Since triangle and box integrals also have non-vanishing double cuts,
\bea
{\rm Cut}_2^{(s)}[I_2(s)] &=& - \frac{1}{8\pi^2} \,, \label{eq:cutbubble} \\
{\rm Cut}_2^{(s)}[I_3(s)] &=& \frac{1}{8\pi^2} \frac{1}{s} \left(\frac{1}{\epsilon} - \log\Big(\frac{-s}{\mu^2}\Big) \right) \,, \label{eq:cuttriangle} \\
{\rm Cut}_2^{(s)}[I_4(s,t)] &=& - \frac{1}{4\pi^2} \frac{1}{st} \left(\frac{1}{\epsilon} - \log\Big(\frac{-t}{\mu^2}\Big) \right) \,. \label{eq:cutbox}
\eea
their contribution needs to be appropriately subtracted.

\section{Infrared divergences}\label{sec:irdivs}

Complementing the discussion in \Sec{sec:betas}, in this appendix we review how minimal gravitational interactions lead to soft but not collinear divergences, and extend the analysis to the case of a scalar-photon three-point interaction $\phi F^2$, as an illustration of the generic IR behaviour associated with non-minimal couplings. Instead of analyzing loop amplitudes, we focus on the real emission of collinear and soft particles at tree level. As is well known \cite{Bloch:1937pw,Kinoshita:1962ur,Lee:1964is,Weinberg:1965nx}, IR divergences arise when integrating the corresponding cross section over the collinear or soft phase space and cancel those appearing in virtual (one-loop) corrections, rendering physical observables (with experimentally resolvable particles) finite.

In the strict collinear limit of two particles, $a$ and $b$, their momenta $p_a$ and $p_b$ become parallel, such that $s_{ab} = (p_a + p_b)^2 = 0$. One can parametrize the momenta as $p_a = x p_c$, $p_b = (1-x) p_c$, where the total collinear momentum $p_c = p_a + p_b$ is distributed according to the momentum fraction $x$. 
Tree-level amplitudes factorize in the collinear limit,
\beq
\amp^\tree_n(\dots, a_{h_a},b_{h_b},\dots) \overset{1 || 2}{\longrightarrow} {\textrm{Split}}_{h_c}^\tree(x,a_{h_a},b_{h_b}) \amp^\tree_{n-1}(\dots,c_{-h_{c}},\dots) \,,
\label{eg:colfact}
\eeq
where the sum over helicities $h_c$ is understood, and where the splitting functions are given by ${\textrm{Split}}_{h_c}^\tree(x,a_{h_a},b_{h_b}) = \lim_{p^2 \to 0} \amp^\tree(a_{h_a},b_{h_b},-c_{h_c})/p^2$.
In practice (see, e.g., \cite{Badger:2023eqz}), it is convenient to approach the collinear configuration as the $z \to 0$ limit of the three-point amplitude with momenta parametrized as $|a\rangle = \cos \eta |c\rangle - z \sin \eta |r\rangle$ and  $|b\rangle = \sin \eta |c\rangle + z \cos \eta |r\rangle$, along with the same rotation on the square spinors, where $p_c = p_a + p_b + O(z^2)$ and $r$ a null reference momentum not parallel to $p_c$.

In this way one obtains the splitting functions in GR (which are not corrected at loop level \cite{Bern:1998sv}),
\bea
{\textrm{Split}}_{h^-}^\gr(x,a_{h^+},b_{h^-}) &=& -\frac{1}{\mpl}  \frac{x^3}{(1-x)} \frac{\an{ab}}{\sq{ba}} \,, \\
{\textrm{Split}}_{h^+}^\gr(x,a_{h^-},b_{h^-}) &=& -\frac{1}{\mpl} \frac{1}{x(1-x)} \frac{\an{ab}}{\sq{ba}} \,,
\eea
with the other splittings given by ${\textrm{Split}}_{h_c}^\gr(x,a_{-h_a},b_{-h_b}) = {\textrm{Split}}_{-h_c}^\gr(x,a_{h_a},b_{h_b})|_{\an{ab} \leftrightarrow \sq{ba}}$ and ${\textrm{Split}}_{h^-}^\gr(x,1_{h^-},2_{h^-}) = 0$.

To show that the collinear configuration does not lead to IR divergences, one can consider the differential cross section for the splitting of a graviton into two collinear gravitons, $d\sigma_{n}(\dots,a_h,b_h,\dots) = f_{h h \to h} \, dP_{\rm col} \, d\sigma_{n-1}(\dots,c_h,\dots)$, where the phase-space measure of the two collinear particles in the collinear region, $s_{ab} < \Er^2$, 
is given by \cite{Giele:1991vf},
\beq
dP_{\rm col} = \frac{1}{(4\pi)^{2-\epsilon} \Gamma(1-\epsilon)} ds_{ab} \, dx \, [s_{ab}x(1-x)]^{-\epsilon} \,, 
\eeq
with $x \in (x_1,1-x_2)$ to avoid overlapping with the soft region at $x_1, \, x_2 \to 0$.
The collinear factor $f_{h h \to h}$ depends on the gravitons helicities and is determined by $|{\textrm{Split}}_{h_c}^\gr(x,a_{h_a},b_{h_b})|^2$. For instance, if one includes all possible splittings,
\beq
f_{h h \to h} = 2 \frac{1+x^8+(1-x)^8}{x^2(1-x)^2} \,.
\eeq 
The precise dependence on the momentum fraction is not relevant for our discussion. The important point is the absence of any collinear $1/s_{ab}$ pole, such that the integral over the collinear phase space is finite,
\beq
\int_0^{\Er^2} ds_{ab} f_{h h \to h} \propto \Er^2 \,.
\label{eq:colint}
\eeq

A similar analysis can be carried out for non-minimal three-point amplitudes.
For instance, considering the interaction of a scalar and two photons $\amp(1_{\phi},2_{\gamma^+},3_{\gamma^+}) =  c_{\phi F^2} \sq{23}^2$, one can derive, analogously to \Eq{eg:colfact}, the splitting functions
\bea
{\textrm{Split}}_{\phi}^\nm(x,a_{\gamma^-},b_{\gamma^-}) &=& c_{\phi F^2} \frac{\an{ab}}{\sq{ba}} \,, \\
{\textrm{Split}}_{\gamma^-}^\nm(x,a_{\phi},b_{\gamma^-}) &=& -c_{\phi F^2} x \frac{\an{ab}}{\sq{ba}} \,.
\eea
These exhibit the same behaviour with the collinear momenta as the splitting function in GR. 
Consequently, the collinear factors $f_{\gamma\gamma \to 0}$ and $f_{0\gamma \to \gamma}$ do not exhibit singularities with the invariant mass of the collinear pair, and the corresponding phase-space integral over $s_{ab}$ is finite as in \Eq{eq:colint}.

Concerning the behaviour of scattering processes when a particle $s$ goes soft, $p_s \to 0$, tree-level amplitudes in GR (and, generally, in minimally coupled gauge theories) display a universal factorizing behaviour when a graviton has very low energy \cite{Low:1958sn,Weinberg:1965nx,Cachazo:2014fwa},
\beq
\amp^\tree_{n+1}(s_h, 1_{h_1}, \dots, n_{h_n}) \overset{p_s \to 0}{\longrightarrow} \mathcal{S}(s_h,1,\dots,n) \amp^\tree_{n}(1_{h_1}, \dots, n_{h_n}) \,,
\label{eg:softfact}
\eeq
where the soft gravitational function $\mathcal{S}$ depends on the momentum and helicity of the soft graviton, and on the momenta of the hard scattered particles only. At leading soft order
\beq
\mathcal{S}^{(0)}(s_{h^+},1,\dots,n) = \frac{1}{\mpl} \sum_a^n \frac{\sq{sa}\an{xa}\an{ya}}{\an{sa}\an{xs}\an{ys}} \,,
\label{eq:softgr}
\eeq
where $x$ and $y$ are reference spinors associated with graviton polarization, and we note that all particles contribute because of the universality of minimal gravitational interactions \cite{Weinberg:1964kqu}.
The individual terms in the sum above can be derived via a BCFW shift of the soft graviton leg and one of the hard particles, e.g.~$n$, $|s\rangle \to |s\rangle + z |n\rangle$ and $|n] \to |n] - z |s]$. Factorization then yields \cite{Cachazo:2014fwa,Arkani-Hamed:2008owk}
\beq
\mathcal{S}^{(0)}(s_{h^+},a;n) = \frac{1}{\mpl}  \frac{\sq{sa}\an{na}^2}{\an{sa}\an{ns}^2} \,,
\eeq
whose sum (from $a = 1, \dots, n-1$) is equal to \Eq{eq:softgr}.

To show that the soft limit does lead to IR divergences, one can consider the cross section for the emission of a soft graviton, $d\sigma_{n+1}(s_h,1_{h_1},\dots,n_{h_n}) = f_\gr \, dP_{\rm soft} \, d\sigma_{n}(1_{h_1},\dots,n_{h_n})$, where the soft factor, obtained from \Eq{eq:softgr}, is given by \cite{Weinberg:1965nx}
\beq
f_\gr = \frac{1}{\mpl^2} \sum_{a,\, b = 1}^n \frac{s_{ab}^2}{s_{as}s_{bs}} \,,
\label{eq:fsoft}
\eeq
and the soft phase-space measure can be parametrized as \cite{Giele:1991vf},
\beq
dP_{\rm soft} = \frac{1}{(4\pi)^{2-\epsilon} \Gamma(1-\epsilon)} \frac{ds_{as} \, ds_{bs} \, (\eta_a s_{as} \eta_b s_{bs})^{-\epsilon}}{s_{ab}} \,,
\eeq
where $a$ and $b$ are the two hard particles in the process and the soft region is $\eta_a s_{as}, \eta_b s_{bs} < \Er \sqrt{\eta_a \eta_b s_{ab}}$ 
$\forall \, a,b$ (with $\eta_a = \mp1$ for $a$ in the initial/final state).

The gravitational soft factor, \Eq{eq:fsoft}, displays poles at $s_{as} = 0$ and $s_{bs} = 0$, implying that the integral over the soft phase space is IR divergent,
\beq
(\tilde{\mu}\mu)^{2\epsilon} s_{ab} \int_0^{\bar{s}_{as}} \frac{ds_{as}}{s_{as} (\eta_a s_{as})^\epsilon} \int_0^{\bar{s}_{bs}} \frac{ds_{bs}}{s_{bs} (\eta_b s_{bs})^\epsilon} = \frac{s_{ab}}{\epsilon^2} \left(\frac{\eta_a \eta_b s_{ab} \Er^2}{\tilde{\mu}^2\mu^2}\right)^{-\epsilon} \,.
\label{eq:softint}
\eeq
where $\bar{s}_{as} = \eta_a \Er \sqrt{\eta_a \eta_b s_{ab}}$ and likewise for $\bar{s}_{bs}$. 
Importantly, the double pole in $\epsilon$ cancels due to momentum conservation upon summing over all the terms in \Eq{eq:fsoft}, as in one-loop gravitational amplitudes \cite{Weinberg:1965nx}. For instance, in a four-particle ($n = 4$) process \cite{Addazi:2016ksu}
\beq
\int dP_{\rm soft} f_\gr = \frac{1}{4\pi^2\mpl^2\epsilon} \left( s \log\Big(\frac{s}{\mu^2}\Big) + t \log\Big(\frac{-t}{\mu^2}\Big) + u \log\Big(\frac{-u}{\mu^2}\Big) \right) \,,
\eeq
which has the expected form to cancel the virtual infrared divergences arising at one-loop (recalling that $ \log(s) = \textrm{Re}[\log(-s)]$).

A similar analysis carries over to non-minimal three-point interactions.
In this case, however, there is no soft factor that can be factorized as in \Eq{eg:softfact} -- i.e., leaving another tree-level amplitude without soft particle -- due to the inelasticity inherent in non-minimal three-point amplitudes.
Nevertheless, a procedure analogous to the GR case, using BCFW shifts, is still useful to understand the suppression in the soft limit associated with an insertion of a non-minimal coupling.
For instance, focussing on the interaction of a scalar and two photons, when one of the photons goes soft the amplitude scales as
\beq
\lim_{p_s \to 0}\amp^{\tree}_{\nm}(s_{\gamma^+}, 1_{h_1}, \dots, n_{h_n}) \sim \mathcal{S}^{(0)}_{\nm}(s_{\gamma^+},a;n) = c_{\phi F^2} \frac{\sq{sa}}{\an{sa}} \,,
\label{eq:softphoton}
\eeq
where $a$ is the momentum of the ($+$ helicity) photon or scalar involved in $\amp(1_{\phi},2_{\gamma^+},3_{\gamma^+}) =  c_{\phi F^2} \sq{23}^2$ where, in the case of the scalar, the rest of the amplitude must carry a $|a\rangle^2$ helicity weight.
In the limit where the scalar momentum is small, one finds
\beq
\lim_{p_s \to 0}\amp^{\tree}_{\nm}(s_{\phi}, 1_{h_1}, \dots, n_{h_n}) \sim \mathcal{S}^{(0)}_{\nm}(s_{\phi},a;n) = c_{\phi F^2} \frac{\an{ns}^2\sq{sa}}{\an{na}^2\an{sa}} \,,
\label{eq:softphi}
\eeq
where $n$ is the other BCFW-shifted momentum, while $a$ is the momentum of a photon in the non-minimal interaction. 

In both cases, it is already apparent that the integration over the phase space of the radiated soft particle does not introduce any IR divergences,
\bea
\int dP_{\rm soft} |\mathcal{S}^{(0)}_{\nm}(s_{\gamma^+},a;n)|^2 &=& \frac{\Er^2}{16 \pi^2} \,, \\
\int dP_{\rm soft} |\mathcal{S}^{(0)}_{\nm}(s_{\phi},a;n)|^2 &=& \frac{\Er^4}{48 \pi^2 s} \,.
\eea
Finally, it is worth noting that the real emission of a soft scalar, \Eq{eq:softphi}, is suppressed by its momentum, unlike the emission of a soft photon, \Eq{eq:softphoton}.
This is expected from the shift symmetry of $\phi F^2$ and underlies the additional $t$-suppression of the $t$-channel double cut in the scalar case, \Eq{eq:tcutnm2}, relative to the photon case, \Eq{eq:tcutnm2V}.

\section{One-loop amplitudes}\label{sec:loop}

In this appendix we collect the full one-loop amplitudes relevant for our analysis, in particular for the discussion of dispersion relations in \Sec{sec:dispersions}.
As in the main text, we organize them according to the couplings involved: gravitational ($1/\mpl$), non-minimal three-point ($c_{\phi V^2}, c_{\sigma F^2}, c_{\sigma C^2}$), and their combination.

We have computed the cut-constructible pieces of the amplitudes -- i.e., the coefficients of the box, triangle, and bubble scalar integrals -- via their quadruple, triple, and double cuts, following the methods of \cite{Forde:2007mi,Badger:2008cm,Arkani-Hamed:2008owk}. In those cases where the amplitude is free of infrared divergences, we have checked the consistency of the result against the contributions to the two-cuts from triangles and boxes; see \Eqs{eq:cutbubble}{eq:cutbox}. The rational terms have been determined by requiring that the one-loop amplitude does not exhibit a more singular behaviour in the limit $s, t, u \to 0$ than at tree level. This excludes finite terms associated with UV renormalization, which are absorbed into the corresponding EFT coefficient.

We cite the references with the original computations whenever we could find them.

\subsection*{Scalar}

The one-loop four-scalar amplitude with a graviton and $N_V$ minimally coupled vectors in the loop is given by \cite{Baratella:2021guc,Chang:2025cxc}
\bea
\amp_\loop^\gr(1_\phi,2_\phi,3_\phi,4_\phi) &=& \frac{1}{\mpl^4} \bigg( (s^4+t^4) I_4(s,t) + (s^4+u^4) I_4(s,u) + (t^4+u^4) I_4(t,u) \\
&& - 2s (3 s^2 - 4 tu) I_3(s) - 2t (3 t^2 - 4 su) I_3(t)  - 2u (3 u^2 - 4 st) I_3(u) \nonumber \\
&& + \frac{163 s^2 - 283 tu}{60} I_2(s) + \frac{163 t^2 - 283 su}{60} I_2(t) + \frac{163 u^2 - 283 st}{60} I_2(u) \nonumber \\
&& + \frac{N_V}{30} \Big( (s^2 - 6 tu) I_2(s) + (t^2 - 6 su) I_2(t) + (u^2 - 6 st) I_2(u) \Big) \bigg) \,. \nonumber
\eea
For the mixed gravitational-non-minimal contribution, we find
\beq
\amp_\loop^{\gr \times \nm}(1_\phi,2_\phi,3_\phi,4_\phi) = -\frac{N_Vc_{\phi V^2}^2}{3\mpl^2} \Big( (s^2 - 6 tu) I_2(s) + (t^2 - 6 su) I_2(t) + (u^2 - 6 st) I_2(u) \Big) \,,
\eeq
and for the amplitude with non-minimally coupled vectors running in the loop,
\bea
\amp_\loop^\nm(1_\phi,2_\phi,3_\phi,4_\phi) &=& N_V c_{\phi V^2}^4 \Big( s^2 t^2 I_4(s,t) + s^2 u^2 I_4(s,u) + t^2 u^2 I_4(t,u) \nonumber \\
&& \qquad \qquad - 2s^3 I_3(s) - 2t^3 I_3(t)  - 2u^3 I_3(u)  \nonumber \\
&& \qquad \qquad + s^2 I_2(s) + t^2 I_2(t) + u^2 I_2(u)  \Big) \,.
\eea
Note that while this amplitude receives contributions from triangles and boxes, it exhibits no IR divergences.

\subsection*{Photon}

For the one-loop four-photon MHV amplitude with a graviton and $N_V$  minimally coupled vectors and dilaton/axion pairs, we find
\bea
\frac{\amp_\loop^\gr(1_{\gamma^+},2_{\gamma^-},3_{\gamma^-},4_{\gamma^+})}{\an{23}^2 \sq{14}^2} &=& \frac{1}{\mpl^4} \bigg( \\
&& \hspace{-4cm} \frac{2 s^6+4 s^5 t+7 s^4 t^2+8 s^3 t^3+7 s^2 t^4+4 s t^5+2 t^6}{2 u^4} I_4(s,t) + u^2 I_4(s,u) + u^2 I_4(t,u) \nonumber \\
&& \hspace{-4cm} - \frac{8 s^4+23 s^3 t+32 s^2 t^2+23 s t^3+8 t^4}{u^4} \Big( s I_3(s) + t I_3(t) \Big) \nonumber \\
&& \hspace{-4cm} - \frac{267 s^3+511 s^2 t+311 s t^2+7 t^3}{30 u^3} I_2(s) - \frac{7 s^3+311 s^2 t+511 s t^2+267 t^3}{30 u^3} I_2(t) \nonumber \\
&& \hspace{-4cm} - \frac{1}{16\pi^2} \frac{st}{u^2} + \frac{7 N_V}{30} \Big( I_2(s) + I_2(t) \Big) \bigg) \,.\nonumber
\eea
The mixed gravitational-non-minimal amplitude at one loop is given solely by bubbles \cite{Arkani-Hamed:2021ajd}
\beq
\frac{\amp_\loop^{\gr \times \nm}(1_{\gamma^+},2_{\gamma^-},3_{\gamma^-},4_{\gamma^+})}{\an{23}^2 \sq{14}^2} = -\frac{4N_Vc_{\sigma F^2}^2}{3\mpl^2} \Big( I_2(s) + I_2(t) \Big) \,,
\eeq
while the pure non-minimal contribution we find takes the form
\bea
\frac{\amp_\loop^\nm(1_{\gamma^+},2_{\gamma^-},3_{\gamma^-},4_{\gamma^+})}{\an{23}^2 \sq{14}^2} &=& 4 N_V c_{\sigma F^2}^4 \bigg( \frac{s^2 t^2 (s^2 + t^2)}{2 u^4} I_4(s,t) + \frac{s t (s^2 + t^2)}{2 u^4} \Big( s I_3(s) + t I_3(t) \Big) \nonumber \\
&& \qquad \qquad + \frac{s (2s^2 + st + 5t^2)}{3 u^3} I_2(s) + \frac{t (2 t^2 + st + 5s^2)}{3 u^3} I_2(t) \nonumber \\
&& \qquad \qquad - \frac{1}{16\pi^2} \frac{st}{u^2} \bigg) \,.
\eea
As in the scalar case, the contributions from triangles and boxes to this amplitude are such that all the corresponding $1/\epsilon$ terms cancel.

\subsection*{Graviton}

The one-loop four-graviton MHV amplitude with a graviton and $N_\sigma$ scalars running in the loop is given by \cite{Dunbar:1994bn,Dunbar:1995ed}
\bea
\frac{\amp_\loop^\gr(1_{h^+},2_{h^-},3_{h^-},4_{h^+})}{\an{23}^4 \sq{14}^4} &=& \frac{1}{\mpl^4} \bigg( \\
&& \hspace{-3cm} \frac{s^8 + t^8 + u^8}{2 u^8} I_4(s,t) + I_4(s,u) + I_4(t,u) + \frac{s^8 + t^8 - u^8}{u^8} \Big( \frac{1}{t} I_3(s) + \frac{1}{s} I_3(t) \Big) \nonumber \\
&& \hspace{-3cm} +\frac{261 s^4+809 s^3 t+1126 s^2 t^2+809 s t^3+261 t^4}{30 u^7} \Big( (t-s) I_2(s) + (s-t) I_2(t) \Big) \nonumber \\
&& \hspace{-3cm} + \frac{1}{2880\pi^2} \frac{1682 s^4+5303 s^3 t+7422 s^2 t^2+5303 s t^3+1682 t^4}{u^6} \bigg) \nonumber \\
&& \hspace{-1.8cm} + \frac{N_\sigma}{\mpl^4} \bigg( \frac{s^4 t^4}{2 u^8} I_4(s,t) + \frac{s^4 t^4}{u^8} \Big( \frac{1}{t} I_3(s) + \frac{1}{s} I_3(t) \Big) \nonumber \\
&& \hspace{-0.5cm} + \frac{s^4+9 s^3 t+46 s^2 t^2+9 s t^3+t^4}{60 u^7} \Big( (t-s) I_2(s) + (s-t) I_2(t) \Big) \nonumber \\
&& \hspace{-0.5cm} + \frac{1}{5760\pi^2} \frac{2 s^4+23 s^3 t+222 s^2 t^2+23 s t^3+2 t^4}{u^6} \bigg) \,, \nonumber
\eea
For the one-loop contribution with two GR couplings and two non-minimal scalar-graviton three-point couplings, we obtain
\bea
\frac{\amp_\loop^{\gr \times \nm}(1_{h^+},2_{h^-},3_{h^-},4_{h^+})}{\an{23}^4 \sq{14}^4} &=& \frac{N_\sigma c_{\sigma C^2}^2}{\mpl^6} \bigg( \\
&& \hspace{-4cm} \frac{s^2 t^2 \left(s^6+s^4 t^2+s^2 t^4+t^6\right)}{u^8} I_4(s,t) + s^2 I_4(s,u) + t^2 I_4(t,u) \nonumber \\
&& \hspace{-4cm} + \frac{2 s^2 t^2 (s^6 + s^4 t^2 + s^2 t^4 + t^6)}{u^8} \Big( \frac{1}{t} I_3(s) + \frac{1}{s} I_3(t) \Big) - 4 u I_3(u) \nonumber \\
&& \hspace{-4cm} + \frac{67 s^5 + 469 s^4 t + 1066 s^3 t^2 + 1394 s^2 t^3 + 951 s t^4 + 341 t^5}{15u^7} \Big( s^2 I_2(s) + t^2 I_3(t) \Big) + I_2(u) \nonumber \\
&& \hspace{-4cm} + \frac{1}{2880 \pi^2} \frac{469 s^6+1876 s^5 t+5023 s^4 t^2+6512 s^3 t^3+5023 s^2 t^4+1876 s t^5+469 t^6}{u^6} \bigg) \,. \nonumber
\eea
We note that this amplitude exhibits a soft divergence, given by 
\beq
\amp_\loop^{\gr \times \nm}|_{1/\epsilon} = \frac{\amp^\nm}{8 \pi^2 \mpl^2 \epsilon} \left( s \log\Big(\frac{-s}{\mu^2}\Big) + t \log\Big(\frac{-t}{\mu^2}\Big) + u \log\Big(\frac{-u}{\mu^2}\Big) \right) \,,
\eeq
with the non-minimal tree-level amplitude $\amp^\nm$, mediated by $u$-channel scalar exchange, given in \Eq{eq:ANMtreeh}.

\section{Details on smearing}\label{sec:details}

Positivity of the UV representation of the dispersion relation, \Eq{eq:smeareduv}, must be enforced over the entire $(J,s)$ plane. As disscused in \cite{Caron-Huot:2022ugt, Henriksson:2022oeu}, the plane can be divided in four regions: the large-$s$ and large-$J$ region, already discussed in \Sec{sec:smear} and leading to \Eq{eq:psi}; the large-$s$ and finite-$J$ region; the finite-$s$ and large-$J$ region, and the finite-$s$ and finite-$J$ region. 
As our optimization condition, we will require that the smeared prefactor of $c_2$, $\alpha_2$, and $g_4$ be maximized, for scalar, photons, and gravitons, respectively.

For finite $s$ and $J$, one requires \Eq{eq:positive} in a grid of finite $(J,s)$ values via a numerical solver \texttt{SDPB} \cite{Simmons-Duffin:2015qma,Landry:2019qug}. 
Since we do not find any relevant differences between solving for $d=4$ and $d=4-2\epsilon$, we restrict ourselves to $d=4$.
Our grid contains 20 values of $s$ for $s/\bar{s} \in [1.0,1.5]$ and 80 values for $s/\bar{s} \in [1.5,250]$.
The values of $J$ are taken to be $J=0,1,2,\dots,100,110,120,\dots,300$. For $\tilde{d}_{0,0}^J$, we evaluate only even $J$; for $\tilde{d}_{2,2}^J$, $J\geq2$; and for $\tilde{d}_{4,4}^J$, $J\geq4$. 
We find that this set of values is sufficient for our purposes, with $s/\bar{s} > 250$ and $J > 300$ already in the asymptotic regimes. 
We assume that any negativity that could arise at points not contained within the grid can be efficiently dealt with by increasing the resolution while slightly perturbing the functional.

The large-$s$ limit for finite $J$ is also implemented as positivity conditions in \texttt{SDPB}. In this case, one expands the integrand of \Eq{eq:positive} for large $s$ up to $1/s^m$ (we use $m=4$) and rescales the expansion in order to express it as a polynomial in $s$ plus subleading terms. After parametrizing $s/\bar{s} = 250 (1+x)$ and smearing, we obtain positivity conditions on the resulting polynomials for any $x > 0$, which \texttt{SDPB} can solve.

The finite-$s$, large-$J$ regime cannot be treated as in \cite{Caron-Huot:2022ugt, Henriksson:2022oeu}, since improved dispersion relations involving $t$-derivatives at $t = 0$ are ill-defined in the presence of graviton loops.
In \cite{Dong:2024omo}, it is argued that, due to the rapid oscillations of the Wigner $d$-functions at large $J$, the smearing between their first and last zeros in the integration domain of $p$ can be neglected, such that only the integral near the integration limits contributes, that is,
\beq
\Psi_J(s) \simeq n_J 
\left( \int_0^a dp + \int_b^q dp \right) \frac{\psi(p) \, \tilde{d}^J_{h,h}\left(1-2p^2/s\right)}{\left(1-p^2/2s\right)^{2n+1}} \,,
\eeq
where $a$ and $b$ are the locations of the zeros of the $d$-function nearest to $p=0$ and $p=q$, respectively.
Such an approximation, in the case that $s = q^2 = \bar{s}$, yields that the functional \Eq{eq:psi} leads to positivity in the large-$J$ limit as long as $a_2$ is negative, and that $\Psi_J(s) \sim 1/J^3$. Specifically, the condition $a_2 < 0$ arises from requiring positivity in the region $p \leqslant a$, thus $\Psi_J(s)$ remains positive if $q^2 < \bar{s}$.
In our case, since we mainly consider $q^2 \ll \bar{s}$, which effectively corresponds to the eikonal regime (from where we extracted $a_2 < 0$), there is no actual need to explore additional conditions from the finite-$s$, large-$J$ regime.

\begin{figure}[t]
\centering
\includegraphics[scale=0.5]{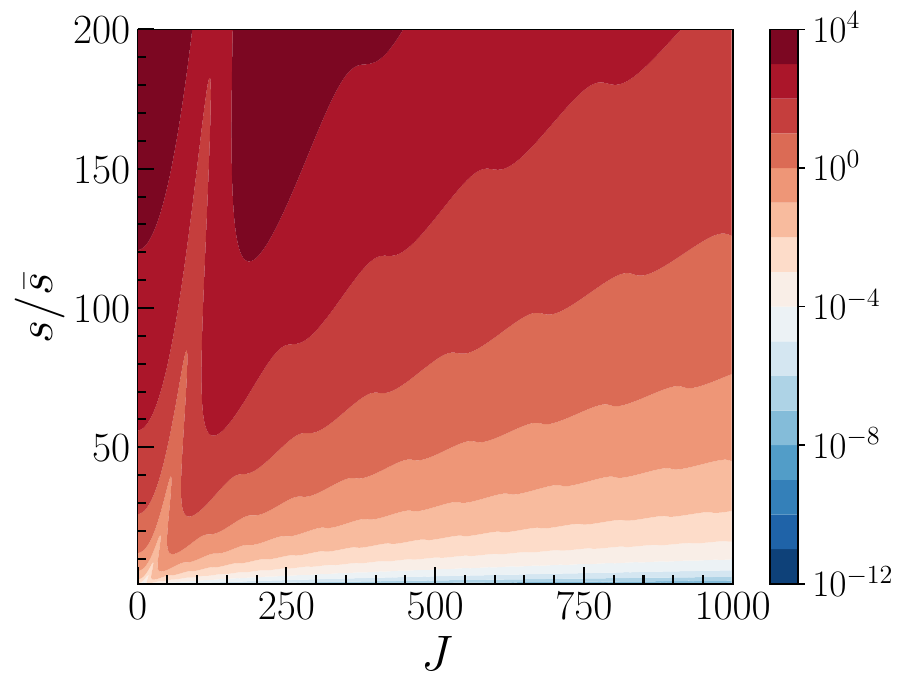}
\hspace{0.2cm}
\includegraphics[scale=0.5]{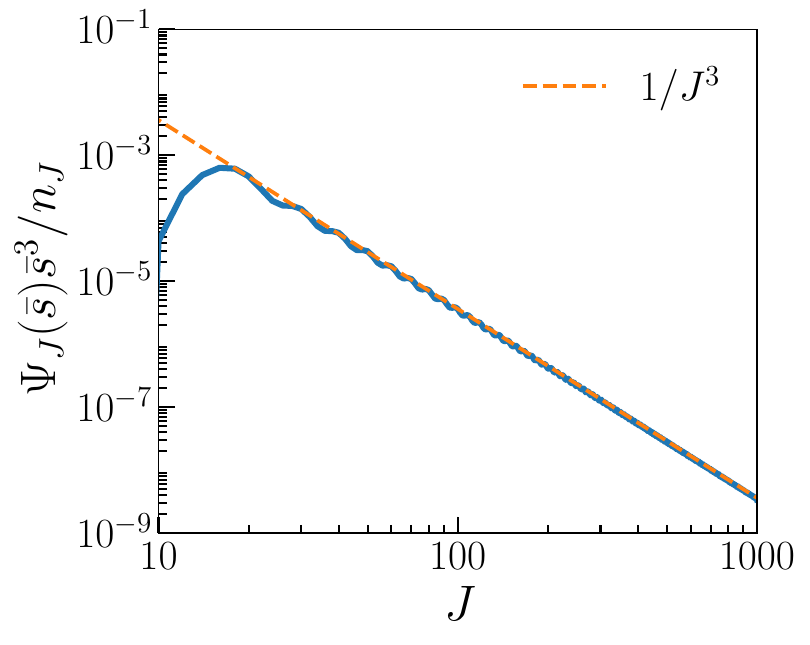}
\caption{\emph{$\Psi_J(s)$ (times $s^3/n_J$) in the $(J,s)$ plane (left) and for a fixed, finite $s = \bar{s}$ and large $J$ (right). The latter agrees with the expected $1/J^3$ behavior (orange dashed line).}}
\label{fig:positivity}
\end{figure}

In \Fig{fig:positivity} we show the value of $\Psi_J(s)$ for the four-scalar case (times a factor $s^3/n_J$). The left panel displays it over a patch of the $(J,s/\bar{s})$ plane, while the right panel compares it with the expected $1/J^3$ behaviour at large $J$. Similar plots are found for photon and graviton scattering. 
The asymptotic regions are clearly positive, while any potential negativity is confined near the origin. We stress that this can always be remedied by increasing the resolution and finely tuning the smearing functional.

\section{IR-finite sum rules}\label{sec:IRsafe}

In \cite{Bellazzini:2025bay}, analytic, crossing-symmetric, and Lorentz-invariant amplitudes free of gravitational soft divergences have been constructed,
\beq
\amp_{\Er} \equiv \lim_{\epsilon \to 0^-} \frac{\amp}{\mathcal{W}} \,, \quad
\mathcal{W} = \exp \left[ \frac{(\Er/\mu)^{-2\epsilon}}{8 \pi^2 \mpl^2 \epsilon} \left( s \log\Big(\frac{-s}{\mu^2}\Big) + t \log\Big(\frac{-t}{\mu^2}\Big) + u \log\Big(\frac{-u}{\mu^2}\Big) \right) \right] \,,
\label{eq:stripped}
\eeq
where $\mathcal{W}$ is the soft exponential introduced by Weinberg \cite{Weinberg:1965nx}.

The pattern of IR divergences in gravity is universal. At one-loop, these can be written as
\beq
\amp_\loop|_{1/\epsilon} = \frac{\amp_\tree}{8 \pi^2 \mpl^2 \epsilon} \left( s \log\Big(\frac{-s}{\mu^2}\Big) + t \log\Big(\frac{-t}{\mu^2}\Big) + u \log\Big(\frac{-u}{\mu^2}\Big) \right) \,,
\eeq
and are precisely cancelled by the $1/\epsilon$ term of the one-loop piece of $\mathcal{W}$.

As discussed in \Sec{sec:infrared}, the leading IR divergence at one loop arises from the terms of $\amp_\loop$ featuring a $1/t$ pole. In the case of the stripped amplitude \Eq{eq:stripped}, the analogous terms are given by
\bea
\amp_{\Er, \loop} &=& \frac{\amp_\tree}{8 \pi^2 \mpl^2} \left( s \log\Big(\frac{-t}{\Er^2}\Big)\log\Big(\frac{-u}{\Er^2}\Big) + t \log\Big(\frac{-s}{\Er^2}\Big)\log\Big(\frac{-u}{\Er^2}\Big) + u \log\Big(\frac{-s}{\Er^2}\Big)\log\Big(\frac{-t}{\Er^2}\Big) \right) \nonumber \\
&& + \dots \, 
\eea
where the ellipses contain other terms without a graviton pole.

In the limit considered in \cite{Bellazzini:2025bay},
\beq
\mpl \to \infty \,, \quad \Er \to 0 \,, \quad \frac{\bar{s} \log(-t/\Er^2)}{(4\pi \mpl)^2} \,\,\, {\rm fixed} \,,
\eeq
the one-loop  gravitational contribution to the scalar, photon, and graviton dispersion relations, instead of the first line in Eqs.~(\ref{eq:sumrulephi}), (\ref{eq:sumrulepho}), and (\ref{eq:sumruleh}), respectively, reads
\bea
B_{1}^{4\phi}(\bar{s},t)|_{\ir,\gr}^{\Er, \loop} &=& \frac{\bar{s}\log(-t/\Er^2)}{4 \pi^2 \mpl^4 t} \,, \\
B_{0}^{4\gamma}(\bar{s},t)|_{\ir,\gr}^{\Er, \loop} &=& \frac{\bar{s}\log(-t/\Er^2)}{4 \pi^2 \mpl^4 t} \,, \\
B_{0}^{4h}(\bar{s},t)|_{\ir,\gr}^{\Er, \loop} &=& -\frac{\log(-t/\Er^2)}{4 \pi^2 \mpl^4 \bar{s} t} \,.
\eea
After smearing the sum rules, in $d = 4$ dimensions yet with momentum transfer $p \in [\Er,q]$, with the functional \Eq{eq:psi}, we obtain IR-finite bounds on the running coefficient of the leading scalar, photon, and graviton EFT amplitudes,
\bea
c_2(\bar{s}) &\geqslant& - \frac{\zeta_\phi}{q^2 \mpl^2} \log(q/\Er) \left( 1 - \frac{4\bar{s} \log(q/\Er)}{(4\pi\mpl)^2} \right) + \dots \, \\
\alpha_2(\bar{s}) &\geqslant& - \frac{2\zeta_\gamma}{q^2 \mpl^2} \log(q/\Er) \left( 1 - \frac{4\bar{s} \log(q/\Er)}{(4\pi\mpl)^2} \right) + \dots \, \\
g_4(\bar{s}) &\geqslant& - \frac{8\zeta_h}{q^2 \bar{s}} \frac{\log(q/\Er)}{(4\pi\mpl)^2} + \dots \,.
\eea

\bibliography{draft}
\bibliographystyle{jhep}

\end{document}